\documentclass{emulateapj}

\newcommand{\obj}{NGC628}

\def\msun{\hbox{M$_\odot$}}

\def\ga{\hbox{$\Gamma$}}
\newcommand{\dr}{{\rm d}}
\def\mstar{\hbox{M$_\star$}}

\slugcomment{Accepted in ApJ, April 30th, 2017 }

\shorttitle{Cluster analysis in LEGUS}
\shortauthors{Adamo et al.}

\begin{document}


\title{Legacy ExtraGalactic UV Survey with The Hubble Space Telescope.\\ Stellar cluster catalogues and first insights into cluster formation and evolution in NGC 628\altaffilmark{1}}


\author{A. Adamo\altaffilmark{2}, J.E. Ryon\altaffilmark{3}, M. Messa\altaffilmark{2}, H. Kim\altaffilmark{4}, K. Grasha\altaffilmark{5}, D.O. Cook\altaffilmark{44}, D. Calzetti\altaffilmark{5}, J.C. Lee\altaffilmark{3, 7}, B.C. Whitmore\altaffilmark{3}, B.G. Elmegreen\altaffilmark{8}, L. Ubeda\altaffilmark{3},  L.J. Smith\altaffilmark{9}, S.N. Bright\altaffilmark{3}, A. Runnholm\altaffilmark{2}, J.E. Andrews\altaffilmark{10},  M. Fumagalli\altaffilmark{11},  D.A. Gouliermis\altaffilmark{12,13}, L. Kahre\altaffilmark{14}, P. Nair\altaffilmark{15}, D. Thilker\altaffilmark{16}, R. Walterbos\altaffilmark{14},  A. Wofford\altaffilmark{17}, A. Aloisi\altaffilmark{3}, G. Ashworth\altaffilmark{11}, T.M. Brown\altaffilmark{3}, R. Chandar\altaffilmark{18}, C. Christian \altaffilmark{3}, M. Cignoni\altaffilmark{19,20}, G.C. Clayton\altaffilmark{21},  D.A. Dale\altaffilmark{6},  S.E. de Mink\altaffilmark{22}, C. Dobbs\altaffilmark{23},  D.M. Elmegreen\altaffilmark{24}, A.S. Evans\altaffilmark{25,26}, J.S. Gallagher III\altaffilmark{27},  E.K. Grebel\altaffilmark{28}, A. Herrero\altaffilmark{29,30}, D.A. Hunter\altaffilmark{31}, K.E. Johnson\altaffilmark{25}, R.C. Kennicutt \altaffilmark{32}, M.R. Krumholz\altaffilmark{33}, D. Lennon\altaffilmark{34}, K. Levay\altaffilmark{3}, C. Martin\altaffilmark{35}, A. Nota\altaffilmark{9}, G. \"Ostlin\altaffilmark{2}, A. Pellerin\altaffilmark{36}, J. Prieto\altaffilmark{37}, M.W. Regan\altaffilmark{3},   E. Sabbi\altaffilmark{3},  E. Sacchi\altaffilmark{38,39}, D. Schaerer\altaffilmark{40}, D. Schiminovich\altaffilmark{41}, F. Shabani\altaffilmark{28}, M. Tosi\altaffilmark{39}, S.D. Van Dyk\altaffilmark{39},  E. Zackrisson\altaffilmark{43}}
\altaffiltext{1}{Based on observations obtained with the NASA/ESA Hubble Space Telescope, at the Space Telescope Science Institute, which is operated by the Association of Universities for Research in Astronomy, Inc., under NASA contract NAS 5-26555. } 
\altaffiltext{2}{Department of Astronomy, Oskar Klein Centre, Stockholm University, AlbaNova University Centre, SE-106 91 Stockholm, Sweden; adamo@astro.su.se}
\altaffiltext{3}{Space Telescope Science Institute, Baltimore, MD}
\altaffiltext{4}{Gemini Observatory, La Serena, Chile}
\altaffiltext{5}{Dept. of Astronomy, University of Massachusetts, Amherst, MA 01003}
\altaffiltext{6}{Dept. of Physics and Astronomy, University of Wyoming, Laramie, WY}
\altaffiltext{7}{Visiting Astronomer, Spitzer Science Center, Caltech. Pasadena, CA}
\altaffiltext{8}{IBM Research Division, T.J. Watson Research Center, Yorktown Hts., NY}
\altaffiltext{9}{European Space Agency/Space Telescope Science Institute, Baltimore, MD}
\altaffiltext{10}{Dept. of Astronomy, University of Arizona, Tucson, AZ}
\altaffiltext{11}{Institute for Computational Cosmology and Centre for Extragalactic Astronomy, Department of Physics, Durham University, Durham, United Kingdom}
\altaffiltext{12}{Zentrum f\"ur Astronomie der Universit\"at Heidelberg, Institut f\"ur Theoretische Astrophysik, Albert-Ueberle-Str.\,2, 69120 Heidelberg, Germany}
\altaffiltext{13}{Max Planck Institute for Astronomy,  K\"{o}nigstuhl\,17, 69117 Heidelberg, Germany}
\altaffiltext{14}{Dept. of Astronomy, New Mexico State University, Las Cruces, NM}
\altaffiltext{15}{Dept. of Physics and Astronomy, University of Alabama, Tuscaloosa, AL}
\altaffiltext{16}{Dept. of Physics and Astronomy, The Johns Hopkins University, Baltimore, MD}
\altaffiltext{17}{Instituto de Astronom\'ia, Universidad Nacional Aut\'onoma de M\'exico, Unidad Acad\'emica en Ensenada, Km 103 Carr. Tijuana-Ensenada, Ensenada 22860, M\'exico}
\altaffiltext{18}{Dept. of Physics and Astronomy, University of Toledo, Toledo, OH}
\altaffiltext{19}{Department of Physics, University of Pisa, Largo B. Pontecorvo 3, 56127, Pisa, Italy }
\altaffiltext{20}{INFN, Largo B. Pontecorvo 3, 56127, Pisa, Italy}
\altaffiltext{21}{Dept. of Physics and Astronomy, Louisiana State University, Baton Rouge, LA}
\altaffiltext{22}{Astronomical Institute Anton Pannekoek, Amsterdam University, Amsterdam, The Netherlands}
\altaffiltext{23}{School of Physics and Astronomy, University of Exeter, Exeter, United Kingdom}
\altaffiltext{24}{Dept. of Physics and Astronomy, Vassar College, Poughkeepsie, NY}
\altaffiltext{25}{Dept. of Astronomy, University of Virginia, Charlottesville, VA}
\altaffiltext{26}{National Radio Astronomy Observatory, Charlottesville, VA}
\altaffiltext{27}{Dept. of Astronomy, University of Wisconsin--Madison, Madison, WI}
\altaffiltext{28}{Astronomisches Rechen-Institut, Zentrum f\"ur Astronomie der Universit\"at Heidelberg, M\"onchhofstr.\ 12--14, 69120 Heidelberg, Germany}
\altaffiltext{29}{Instituto de Astrofisica de Canarias, La Laguna, Tenerife, Spain}
\altaffiltext{30}{Departamento de Astrofisica, Universidad de La Laguna, Tenerife, Spain}
\altaffiltext{31}{Lowell Observatory, Flagstaff, AZ}
\altaffiltext{32}{Institute of Astronomy, University of Cambridge, Cambridge, United Kingdom}
\altaffiltext{33}{Research School of Astronomy and Astrophysics, Australian National University, Canberra, ACT Australia}
\altaffiltext{34}{European Space Astronomy Centre, ESA, Villanueva de la Ca\~nada, Madrid, Spain}
\altaffiltext{35}{California Institute of Technology, Pasadena, CA}
\altaffiltext{36}{Dept. of Physics and Astronomy, State University of New York at Geneseo, Geneseo, NY}
\altaffiltext{37}{Department of Astrophysical Sciences, Princeton University, Princeton, NJ }
\altaffiltext{38}{Dept. of Physics and Astronomy, Bologna University, Bologna, Italy}
\altaffiltext{39}{INAF -- Osservatorio Astronomico di Bologna, Bologna, Italy}
\altaffiltext{40}{Observatoire de Gen\`eve, University of Geneva, Geneva,Switzerland}
\altaffiltext{41}{Dept. of Astronomy, Columbia University, New York, NY}
\altaffiltext{42}{IPAC/CalTech, Pasadena, CA}
\altaffiltext{43}{Department of Physics and Astronomy, Uppsala University, Box 515, SE-751 20 Uppsala, Sweden}
\altaffiltext{44}{California Institute of Technology, Pasadena, CA}

\begin{abstract}
We report the large effort which is producing comprehensive high-level young star cluster (YSC) catalogues for a significant fraction of galaxies observed with the Legacy ExtraGalactic UV Survey (LEGUS) Hubble treasury program. We present the methodology developed to extract cluster positions, verify their genuine nature, produce multiband photometry (from NUV to NIR), and derive their physical properties via spectral energy distribution fitting analyses. We use the nearby spiral galaxy NGC628 as a test case for demonstrating the impact that LEGUS will have on our understanding of the formation and evolution of YSCs and compact stellar associations within their host galaxy. Our analysis of the cluster luminosity function from the UV to the NIR finds a steepening at the bright end and at all wavelengths suggesting a dearth of luminous clusters. The cluster mass function of NGC628 is consistent with a power-law distribution of slopes $\sim -2$ and a truncation of a few times $10^5$ \msun. After their formation YSCs and compact associations follow different evolutionary paths. YSCs survive for a longer timeframe, confirming their being potentially bound systems. Associations disappear on time scales comparable to hierarchically organized star-forming regions, suggesting that they are expanding systems. We find mass-independent cluster disruption in the inner region of NGC628, while in the outer part of the galaxy there is little or no disruption. We observe faster disruption rates for low mass ($\leq$ $10^4$ \msun) clusters suggesting that a mass-dependent component is necessary to fully describe the YSC disruption process in NGC628. 
\end{abstract}

\section{Introduction}
Young star cluster (YSC) populations, commonly detected in local star-forming galaxies, can be powerful tracers of the star formation process in their host galaxies. YSCs form in and interact with their host galaxies, and as bound objects they allow us to study the star formation histories (SFHs) of their parent galaxies \citep[e.g.][]{2010A&A...517A..50G, 2010MNRAS.407..870A}. It is intriguing to find very ancient gravitationally bound stellar objects, i.e. the globular clusters (GCs). Potentially GCs and YSCs could share the same formation process, although, the former have most likely formed under different interstellar medium (ISM) physical conditions \citep[e.g.][]{2010Natur.464..733S}. Both similarities and differences between the young and ancient cluster populations are numerous and it is not clear yet whether YSCs could evolve into GCs \citep[e.g.][]{2015MNRAS.454.1658K}. 

Through the access to multiband HST datasets it has been possible to conduct several studies of YSC populations in local galaxies. YSCs appear to be a common product of star formation in local galaxies. They form in the densest regions of the giant molecular clouds (GMCs), nested at the bottom of the hierarchically structured interstellar medium \citep[ISM,][]{1997ApJ...480..235E}. Therefore, they can be used to probe star formation processes in local galaxies. 

Investigating properties like the mass function (or the luminosity function) of YSC populations can help to constrain their formation mechanism and how they are linked to the overall star formation process in galaxies. Throughout the literature there is consensus and evidence that both the initial cluster mass and luminosity functions (CMF and CLF, respectively) are well described by a power-law slope of approximately $-2$ \citep[][among many others]{1999AJ....118.1551W, 2002AJ....124.1393L, 2003A&A...397..473B, 2006A&A...450..129G, 2009A&A...501..949M, 2010AJ....140...75W, 2014ApJ...787...17C, 2014AJ....147...78W}. It has also been observed that the range for the recovered CLF slopes in several nearby galaxies is quite large, $-2.8 <\delta< -1.5$ \citep[e.g.][]{2011MNRAS.415.2388A, 2011AJ....142..129A, 2014AJ....147...78W}. Blending effects, important in crowded fields and in galaxies at large distances (above 80 Mpc), have the tendency to flatten the slope of the CLF \citep{2013MNRAS.431..554R}. However, in some dwarf starburst galaxies \citep{2010MNRAS.407..870A, 2011MNRAS.415.2388A, 2011AJ....142..129A}, the CLF slopes appear to be flatter than those of rich YSC populations in, e.g., M51 and the Antennae over the same luminosity range. On the other hand, steeper slopes have been observed as a function of wavelength \citep[e.g.][]{2008A&A...487..937H}, and at brighter magnitude ranges \citep[e.g.][]{2010ASPC..423..123G, 2012MNRAS.419.2606B} in some local spiral galaxies.  This steepening suggests that we find a smaller number of luminous clusters than expected if the luminosity function results from an underlying mass function described by a power-law with slope $-2$ and no upper mass limits. The dearth of very luminous (thus, massive) clusters could be explained if the CMF were not a pure power-law, but took the form of a Schechter function, which includes an exponential truncation at masses above \mstar\, \citep[e.g.][]{2006A&A...450..129G, 2009A&A...494..539L}. However, the true shape and universality of the CMF remains still under debate and requires the investigation of a significantly larger sample of galaxies.

Another key aspect not yet fully understood is whether or not there is a change in the cluster formation efficiency (\ga, the mass of star formation that is locked into star clusters) as a function of galactic environment \citep[e.g.][]{aa..nb..2015Spr}. Observational evidence suggests an increase in the cluster formation efficiency as a function of SFR density, $\Sigma_{\rm SFR}$ \citep[e.g.][]{2010MNRAS.405..857G, 2011MNRAS.415.2388A, 2015MNRAS.452..246A}. A model proposed by \cite{2012MNRAS.426.3008K} suggests that this trend is produced by the link between the cluster formation efficiency and the properties of the hierarchically structured ISM. In this model YSCs form in regions which reach gas densities above a critical value. Higher gas pressures (thus, higher gas surface densities) will favour the formation of clusters \citep{2015MNRAS.452..246A}. Since gas surface density is directly linked to the $\Sigma_{\rm SFR}$ via the Schimdt-Kennicutt \citep{2012ARA&A..50..531K} relation, it can explain the observed increase of cluster formation efficiency. Evidence has also been reported \citep{2015ApJ...810....1C} that \ga\, does not strongly correlate with total SFR, suggesting that when SFR is used instead of $\Sigma_{\rm SFR}$, the environmental dependency becomes a second order effect.

The nature of this type of relation is fundamental for understanding the clustering properties of star formation. For example, blue compact galaxies dominated by a recent starburst appear to have a cluster formation efficiency above 40\% \citep{2011MNRAS.415.2388A}. Since the majority of the massive stars (the dominant source of ionising photons) is forming in clusters, cluster feedback has a large impact on the ISM of these galaxies \citep[e.g.][]{2015A&A...576L..13B}. The \ga\, vs. $\Sigma_{\rm SFR}$ relation can also explain why dwarf irregulars living in galaxy clusters host significantly larger samples of GCs than their counterpart dwarf systems which have spent most of their time in the field. The former, entering the over-dense regions, have experienced a starburst event which has produced numerous clusters \citep{2015arXiv150900030M}.
\begin{figure*}
\centering
	\includegraphics[scale=0.85]{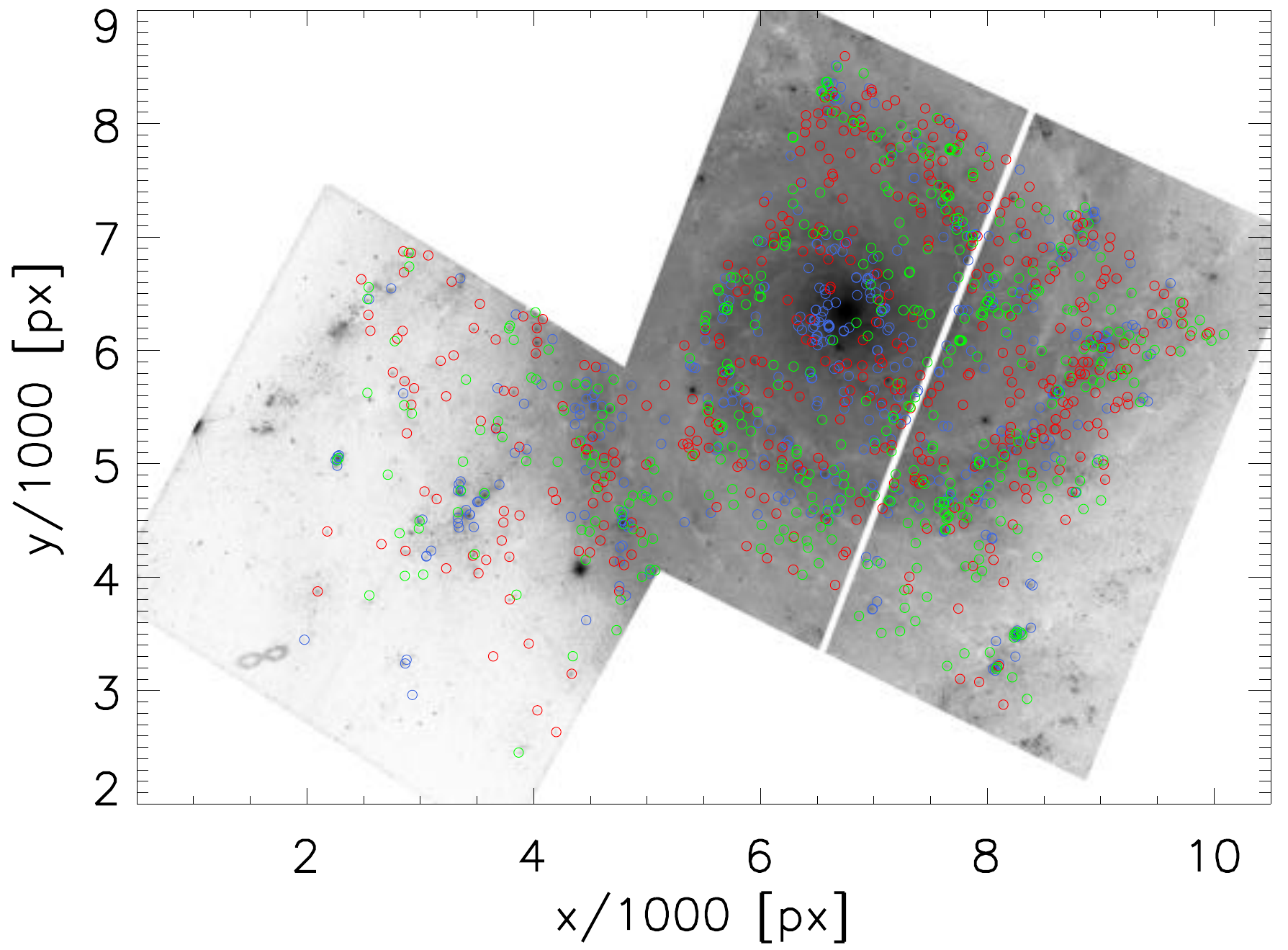}
    \caption{A mosaic image of the two F555W pointings of NGC 628, covering the inner part of the galaxy and a portion of the outer disk in the South-East (image rotated with North-up). The circles show the position of class 1 (red), class 2 (green), and class 3 (blue) cluster candidates. See section 2.2.2 for a description of our classification used here. Detected objects are covering the portions of the field of view that are in commune among the imaging taken in the 5 standard LEGUS filters.}
    \label{fig:fig1}
\end{figure*}

A full description of YSC populations in local galaxies also requires a clear understanding of their evolution in the host galaxies. Two main scenarios are currently considered, and analyses based on the YSC populations of the same galaxies have not reached an agreement. \citep[e.g.][for a summary]{aa..nb..2015Spr}. The disruption model put forward by \citet[e.g.][]{2005ApJ...631L.133F} is historically based on the disruption rate recovered from the age distributions of YSCs in the Antennae galaxy \citep[see][for the latest analysis]{2010AJ....140...75W}. It proposes that YSCs rapidly dissolve (80-90 \% each age dex) first because of internal evaporation (e.g. two- body relaxation, stellar evolution) and on long time scales because of external (e.g. tidal fields) processes. Fall, Chandar \& Whitmore (2009) have suggested that these processes happen over different time-scales and are independent of the cluster masses. The other scenario is described in \citet{2005A&A...441..117L} and observationally supported by recovered disruption rates in the solar neighbours, SMC, M33,  M51 \citep[e.g.]{2003MNRAS.338..717B, 2005A&A...441..949G}. It suggests that YSCs do not disrupt so rapidly, but their mass losses become significant after some time because of interactions with the GMCs, host tidal fields, stellar evolution \citep{2010ARA&A..48..431P}. Disruption, in this latter scenario, is dependent on the mass of the YSCs, with low mass clusters disrupting more rapidly. There is another important difference between the two empirical scenarios described above, i.e. the role of the galactic environment. The mass independent scenario proposes a ``quasi-universal" disruption rate independent of the cluster mass and galactic environment where clusters are formed indicating that the primary disruptive influences may be due to internal processes. The other scenario is clearly anchored to the differences in the properties of the host galaxies. Theoretical models \citep{2010ApJ...712..604E, 2011MNRAS.414.1339K} suggest that interactions with dense ISM (clusters form in giant molecular clouds complexes which will exert tidal forces on the clusters) can also disrupt clusters of any mass. Therefore, we should see that the disruption rate should change as a function of environment \citep[see][for a compilation of observational evidence]{aa..nb..2015Spr}. 

Many fundamental questions related to cluster formation and evolution remain still unanswered. The Legacy ExtraGalactic UV survey (LEGUS) is a Hubble treasury program designed to provide a homogeneous imaging dataset in five bands (from the UV to the NIR) of a large sample (50) of local star-forming galaxies representative of the variety observed within the Local Volume \citep[][hereafter C15]{2015AJ....149...51C}. The homogeneous imaging coverage, including two filters below the Balmer break ($\sim4000$ \AA), is enabling us to recover high-quality cluster photometric and physical properties for a large number of YSC populations in local galaxies. In the pre-LEGUS era, only a handful of galaxies can claim comparable data and products, like M31 \citep{2015ApJ...802..127J}; the Antennae \citep{2010AJ....140...75W}; M83 \citep{2014ApJ...787...17C, 2015MNRAS.452..246A}. The access to a large sample of galaxies with high-quality cluster catalogues produced with the most up-to-date techniques and homogeneous approaches will provide a common ground on which to investigate and try to answer all the open questions addressed above.

The aim of this work is to present the LEGUS cluster analysis and provide guidelines to the numerous cluster catalogues that will be released to the astronomical community in 2017. We use as a test-bench the nearby spiral galaxy \obj\, (also known as M74, distance $\sim$ 9.9 Mpc from C15, see Fig.~\ref{fig:fig1}). Morphologically this galaxy has been classified as a Hubble type SAc galaxy\footnote{value taken from NED}, a multiple spiral-arm system. Previous studies of \obj\, \citep[e.g.][]{2000AJ....120.1306L, 2002AJ....124.3118T} find a H{\sc ii}-region luminosity function slightly shallower than $-2$ and no significant change of slope as a function of galactocentric distances but only between arm and inter-arm regions \citep{1989ApJ...337..761K}. \citet{2006ApJ...644..879E} report luminosity and mass distributions of increasing stellar aggregate boundaries compatible with a power-law index of $-2$, suggesting that star formation in the \obj\, inner and outer disk proceeds in a scale-free hierarchical fashion as result of a turbulence-dominated ISM. The LEGUS dataset (see Fig.~\ref{fig:fig1}) now available offers us the possibility to continue this investigation at the smallest and yet densest star-forming scales, at the bottom of this hierarchical process. 

In the first part of this paper we provide a thorough description of the methodology developed to extract cluster candidate positions, produce final photometric tables, investigate completeness limits of our dataset (Section 2). In Section 3 we describe the fitting methods used to derive cluster candidate physical properties. The color  properties of the cluster candidates of our test-galaxy \obj\, are analysed in Section 4. In Section 5 we present the CLF, CMF, disruption rate analysis. In the final sections, we discuss our results in the framework of cluster formation and evolution and summarise the main result in the Conclusions.

\section{The photometric cluster catalogues}

\subsection{Dataset description}
We refer the reader to C15 for a descriptions of the standard data reduction of the LEGUS imaging datasets which are currently released at the webpage {\em https://archive.stsci.edu/prepds/legus/}.
Each LEGUS target has a standard and homogeneous filter coverage provided by archival and newly acquired imaging with either the Wide Field Camera 3 (WFC3) or the Advanced Camera for Surveys (ACS).  
All the galaxies have WFC3 imaging in the F275W and F336W filter. Three other bands cover the blue optical (ACS/F435W or WFC3/F438W), visual (ACS or WFC3 F555W or F606W), NIR (ACS or WFC3 F814W). Hereafter, the conventional Johnson passband naming $UV$, $U$, $B$, $V$, and $I$ band will be used for the five filters, although the filter throughput is not converted to the standard Johnson system.  We refer hereafter to the $V$ band as the reference frame. The photometry provided by the LEGUS analysis is in Vega magnitude system.
Reduced science frames have been drizzled to a common scale resolution, which corresponds to the native WFC3 pixel size (0.04 $\arcsec$/px). The frames have all been aligned and rotated north-up. 

A description of the LEGUS imaging available for \obj\, is given in Table~\ref{tab1}. The dataset consists of a mixture of archival and newly acquired data which have been reduced accordingly to the standard LEGUS approach (see C15 for details).

\begin{deluxetable*}{cccccccc}
\tablecolumns{8}

\tabletypesize{\small}
\tablecaption{The LEGUS dataset of NGC628.\label{tab1}}
\tablewidth{0pt}
\tablehead{
\colhead{Filters} & \colhead{Program number} & \colhead{PI} & \colhead{exptime} & \colhead{ZP(Vega)} & \colhead{aver apcor\tablenotemark{a}}&
 \colhead{det limits\tablenotemark{b}} &  \colhead{det threshold}
\\
\colhead{ } & \colhead{} & \colhead{} & \colhead{sec} & \colhead{mag} & \colhead{mag} &
\colhead{mag} & electron/sec
\\
\colhead{(1)} & \colhead{(2)} & \colhead{(3)} & \colhead{(4)} & \colhead{(5)} & \colhead{(6)} &
\colhead{(7)} & \colhead{(8)}
}
\startdata
\multicolumn{8}{c}{{\bf  Inner pointing (NGC628c)}}\\
\hline
WFC3/F275W & 13364 & Calzetti & 2481.0 & 22.632 & -0.817$\pm$ 0.066 &   23.29 & 0.009\\
WFC3/F336W & 13364 & Calzetti & 2361.0 & 23.484 & -0.750 $\pm$ 0.060 &  23.91 & 0.010\\
ACS/F435W & 10402 & Chandar & 1358.0 & 25.784 & -0.656 $\pm$ 0.034 & 24.93 & 0.013\\
ACS/F555W & 10402 & Chandar & 858.0 & 25.731 & -0.634 $\pm$ 0.034 &  25.05 & 0.021\\
ACS/F814W & 10402 & Chandar & 922.0 & 25.530 & -0.751 $\pm$ 0.037 & 24.27 & 0.030\\
\hline
\multicolumn{8}{c}{{\bf  Outer pointing (NGC628e)}}\\
\hline
WFC3/F275W & 13364 & Calzetti & 2361.0 & 22.632 & -0.795  $\pm$ 0.097&   23.38 & 0.009\\
WFC3/F336W & 13364 & Calzetti & 1119.0 & 23.484 & -0.706  $\pm$ 0.059&  23.48 & 0.018\\
ACS/F435W & 10402 & Chandar & 4720.0 & 25.784 & -0.695  $\pm$ 0.039&  25.26 & 0.010\\
WFC3/F555W & 13364 & Calzetti & 965.0 & 25.816 & -0.740  $\pm$ 0.038&  25.22 & 0.024\\
ACS/F814W & 10402 & Chandar & 1560.0 & 25.530 & -0.843  $\pm$ 0.050&  24.42 & 0.029
\enddata
\tablenotetext{a}{Averaged aperture corrections used to produce the final AV\_APCOR cluster catalogues.}
\tablenotetext{b}{The listed values correspond to the 90\% completeness limits  at the detection thresholds listed in column 8. Completeness limits have been estimated using synthetic clusters with sizes larger than 1 pc. See details about the meaning of the recovered completeness values in the main text.}

\end{deluxetable*}

\subsection{Extraction and selection of cluster candidates}

\subsubsection{Automated steps \label{autcat}}

A custom pipeline, {\bf  legus\_clusters\_extraction.py} (version 4.0) has been developed to produce initial cluster candidate catalogues which contain, for each source, an ID number, the position in pixel coordinates, the final photometry including errors (in the Vega system), the concentration index (CI), and the number of filters in which the source was detected with a photometric error $\le 0.3$ mag. A readme file includes all the key information about the galaxy, the parameters used to build the catalogue, the content of the columns.

\begin{figure*}
	\includegraphics[scale=0.34]{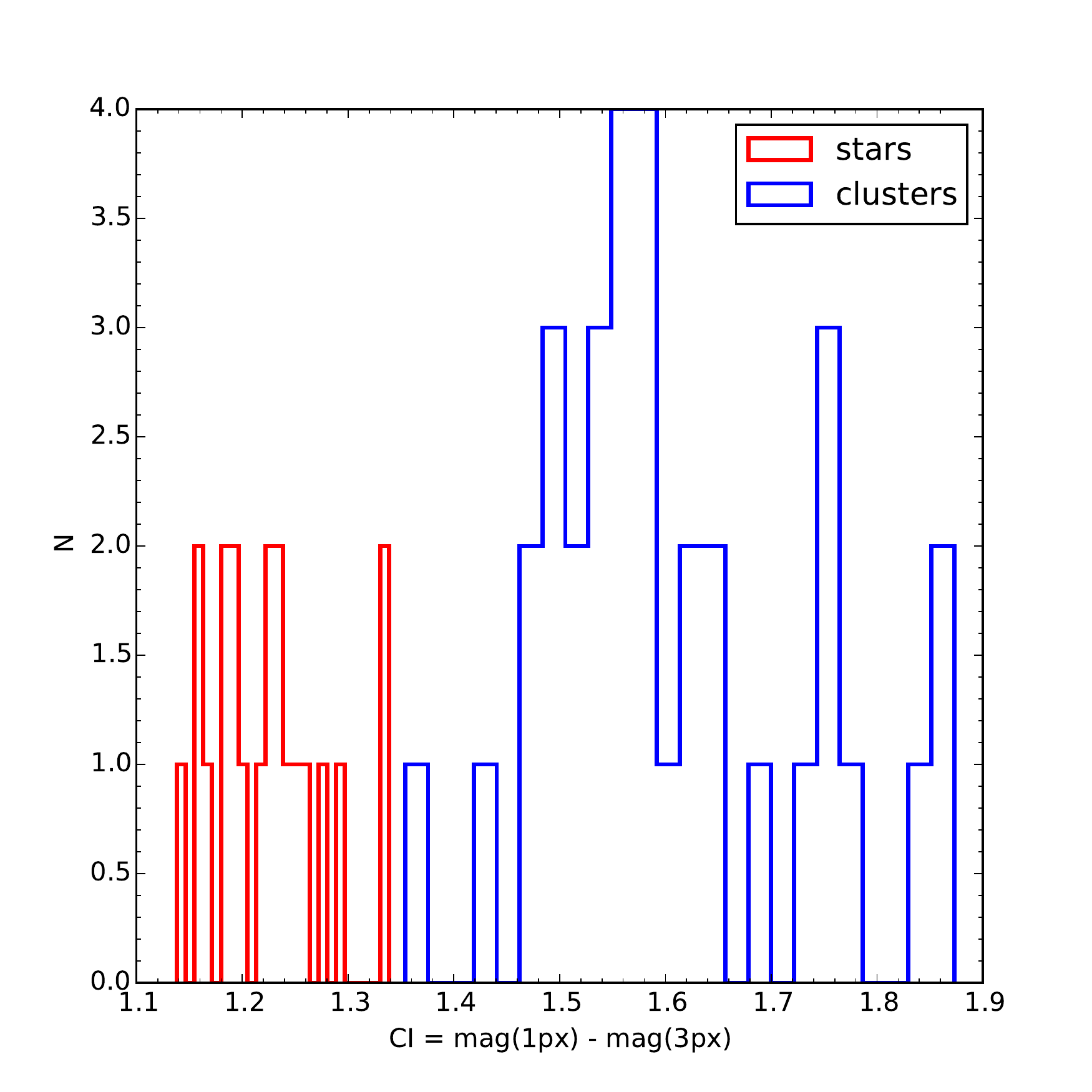}
	\includegraphics[scale=0.34]{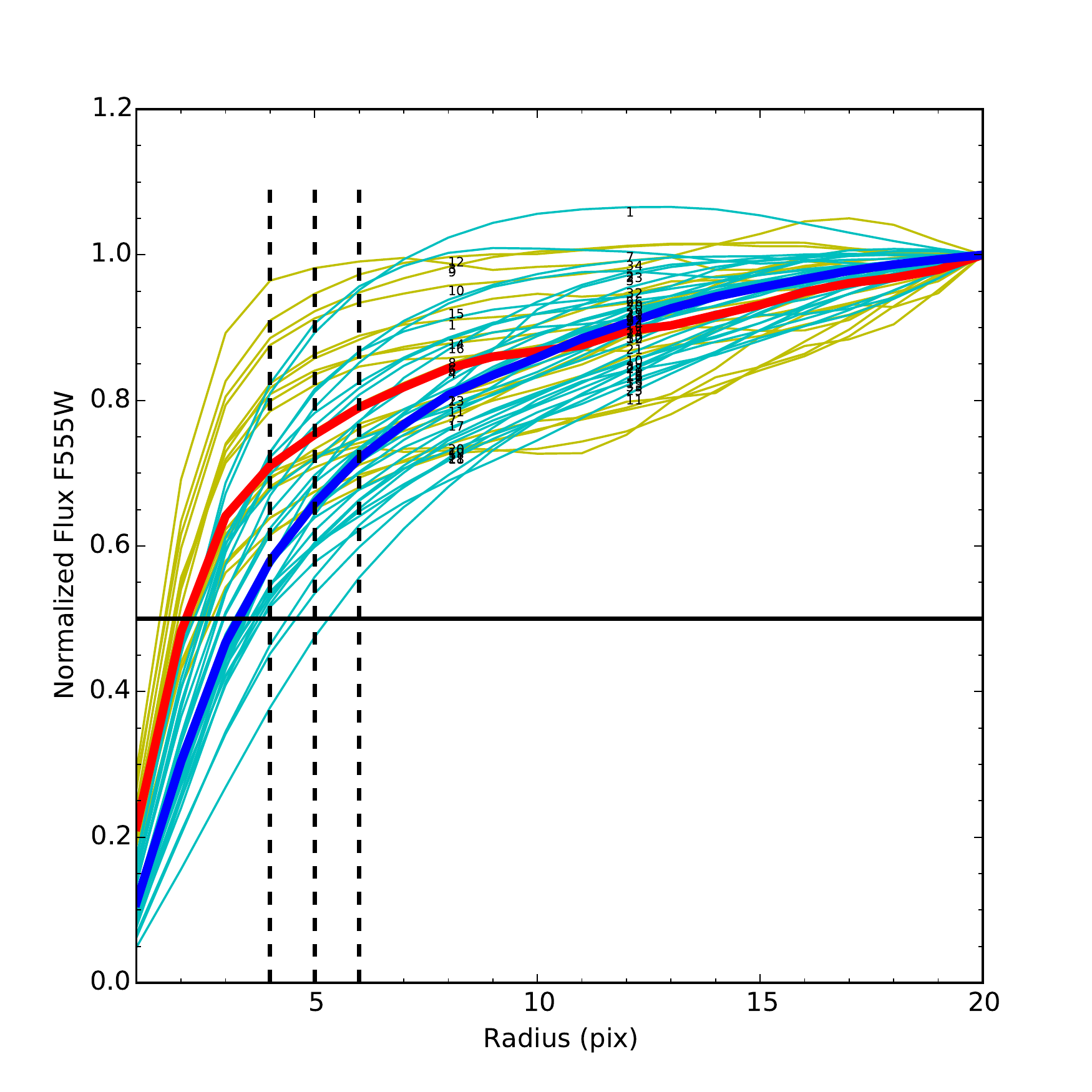}
	\includegraphics[scale=0.34]{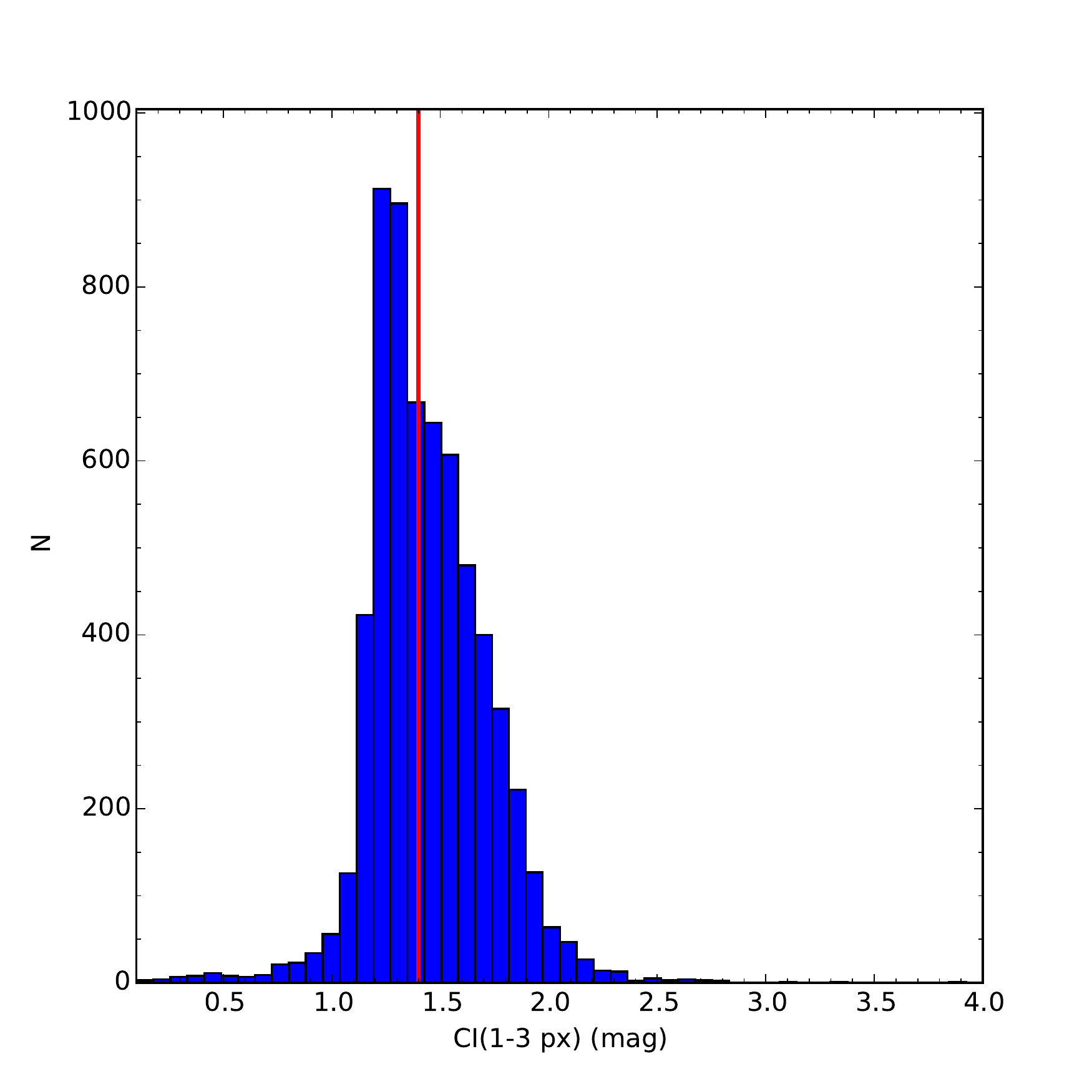}\\
	\includegraphics[scale=0.235]{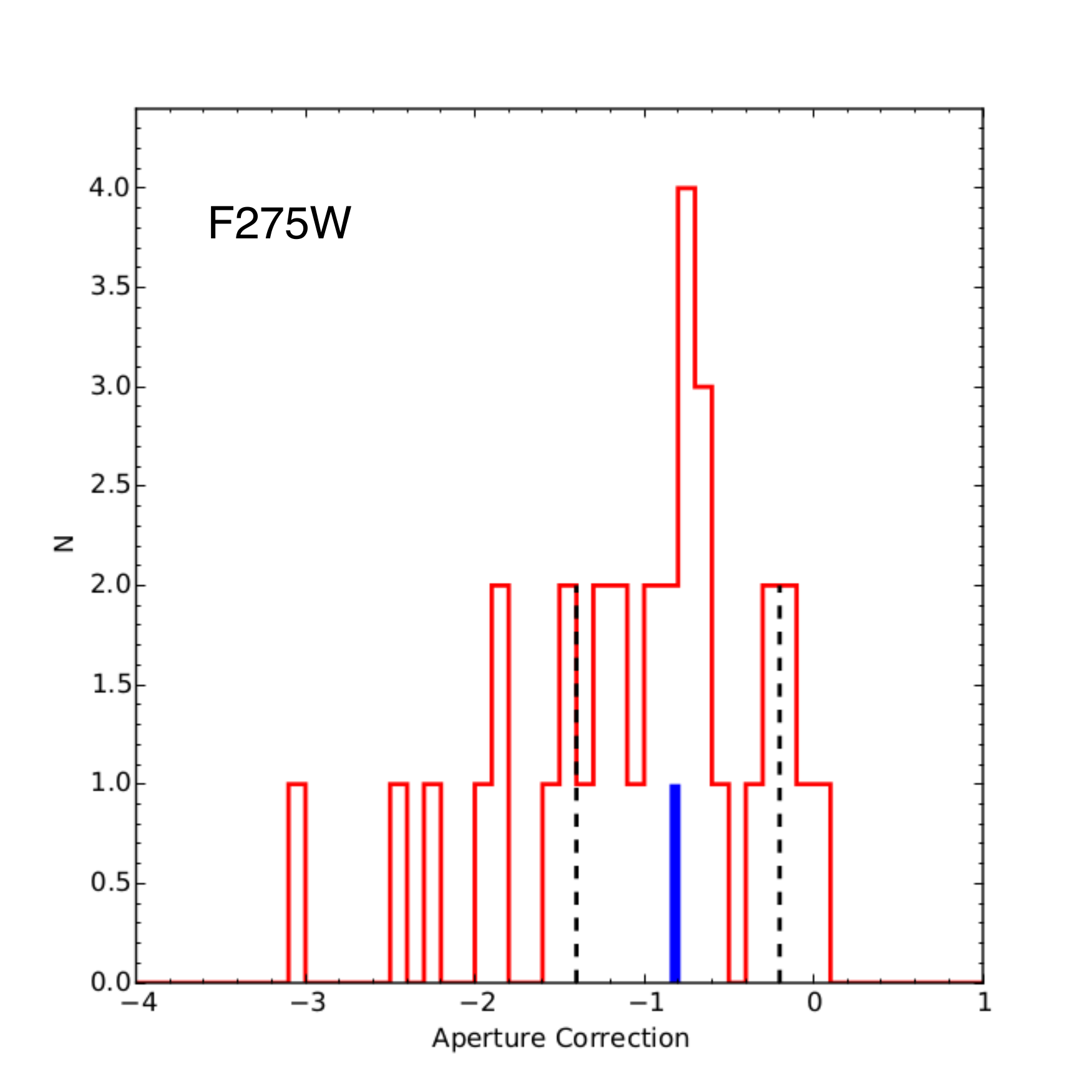}
	\includegraphics[scale=0.235]{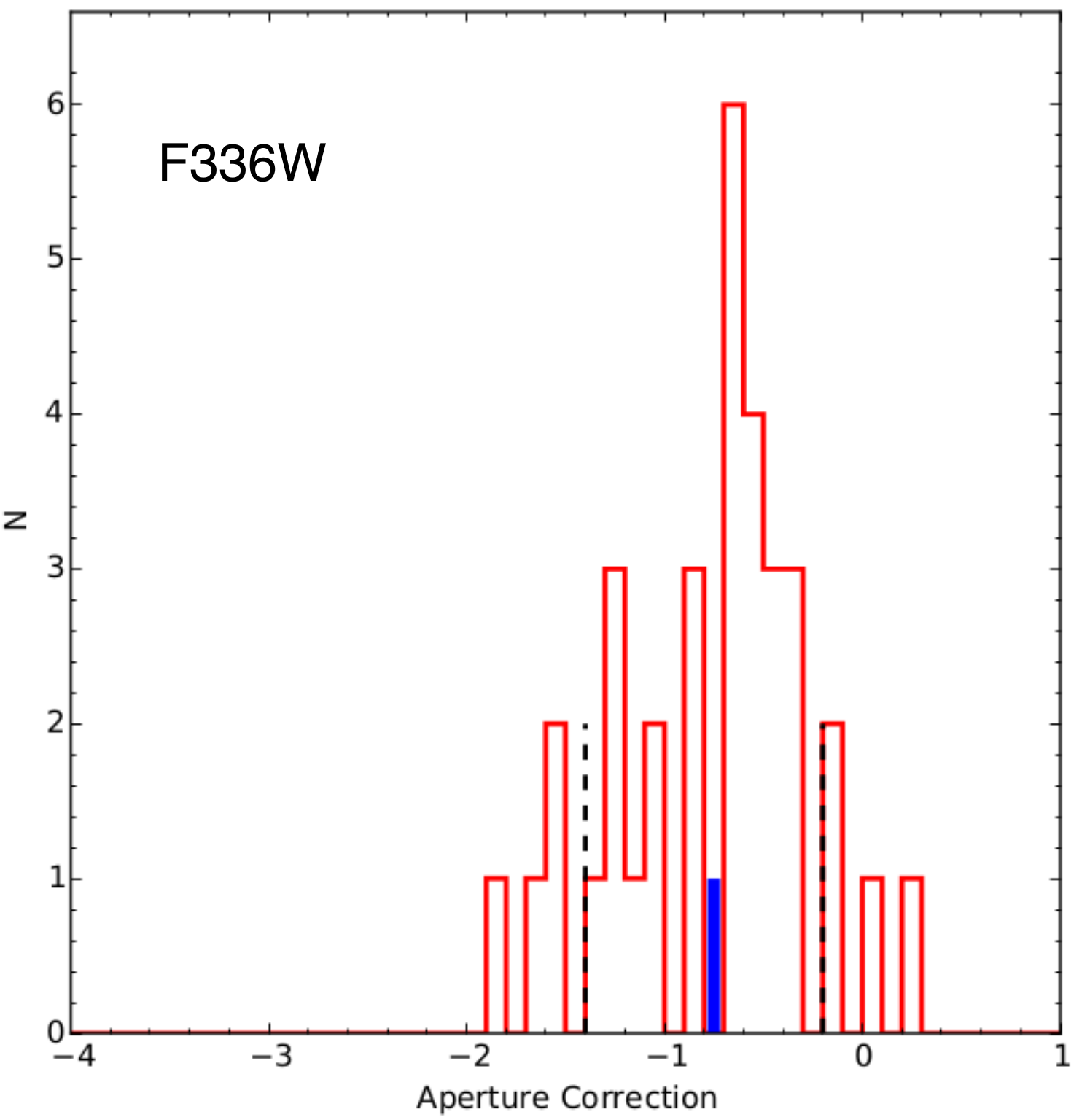}
	\includegraphics[scale=0.235]{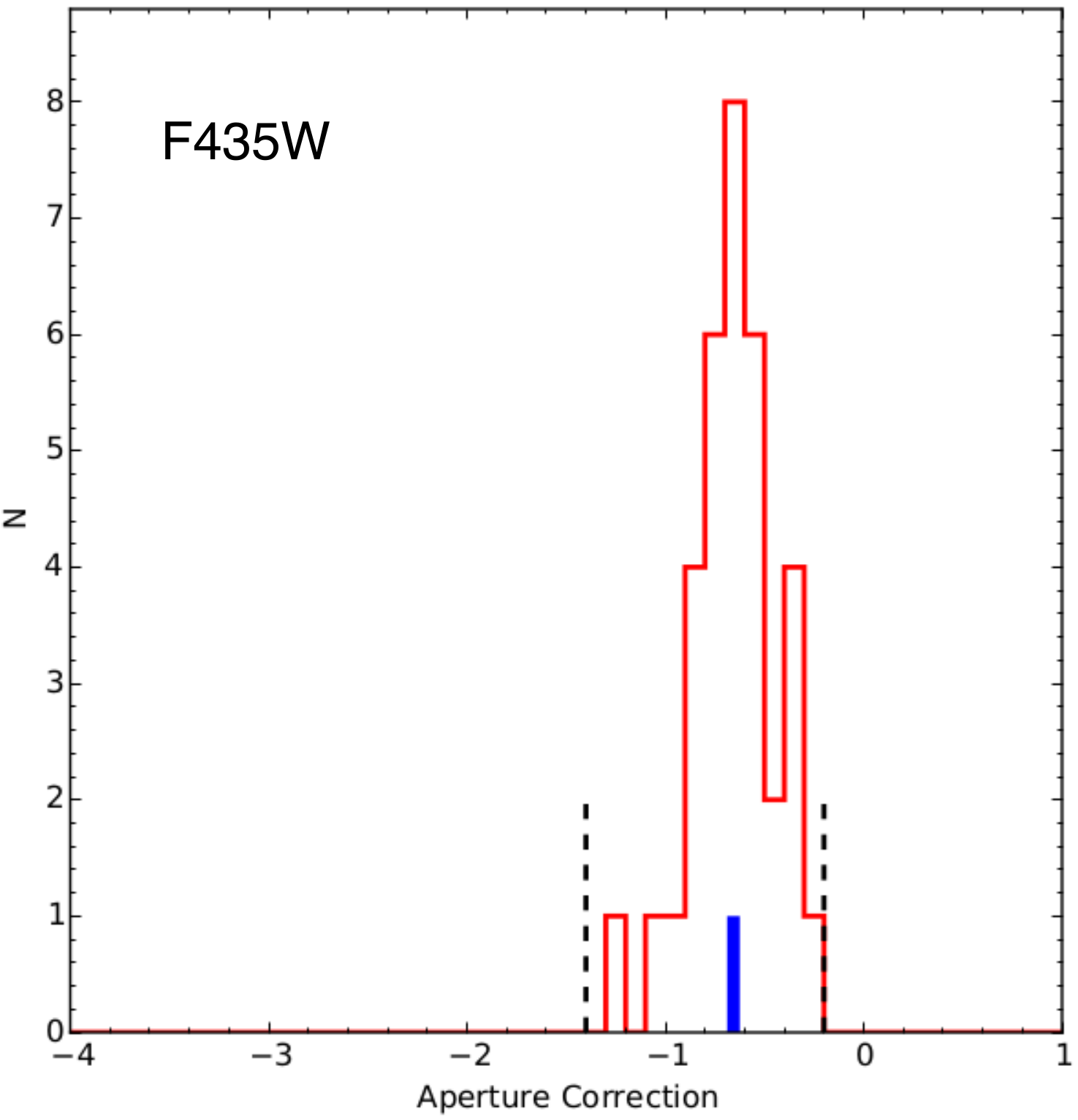}
	\includegraphics[scale=0.235]{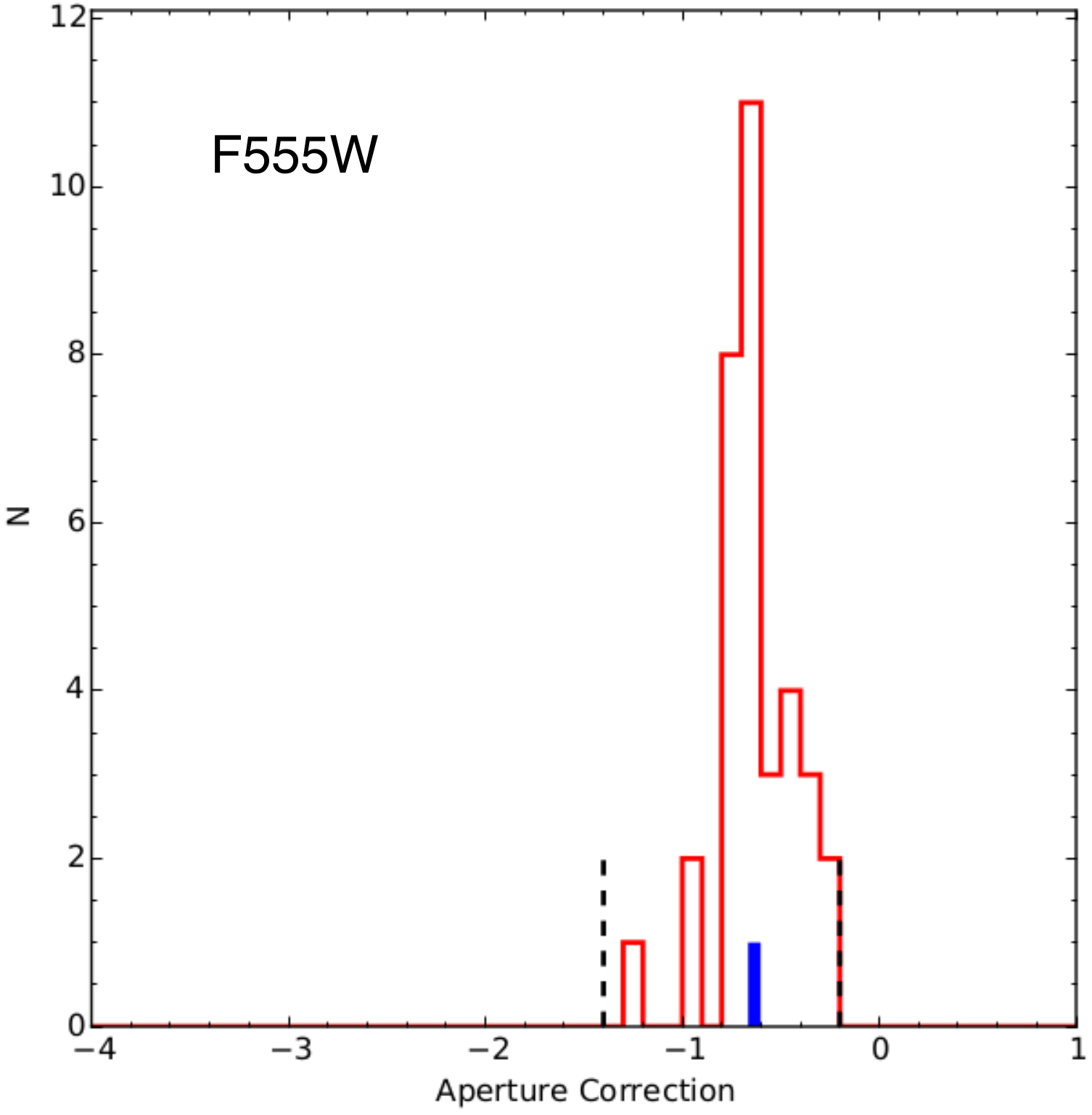}
	\includegraphics[scale=0.235]{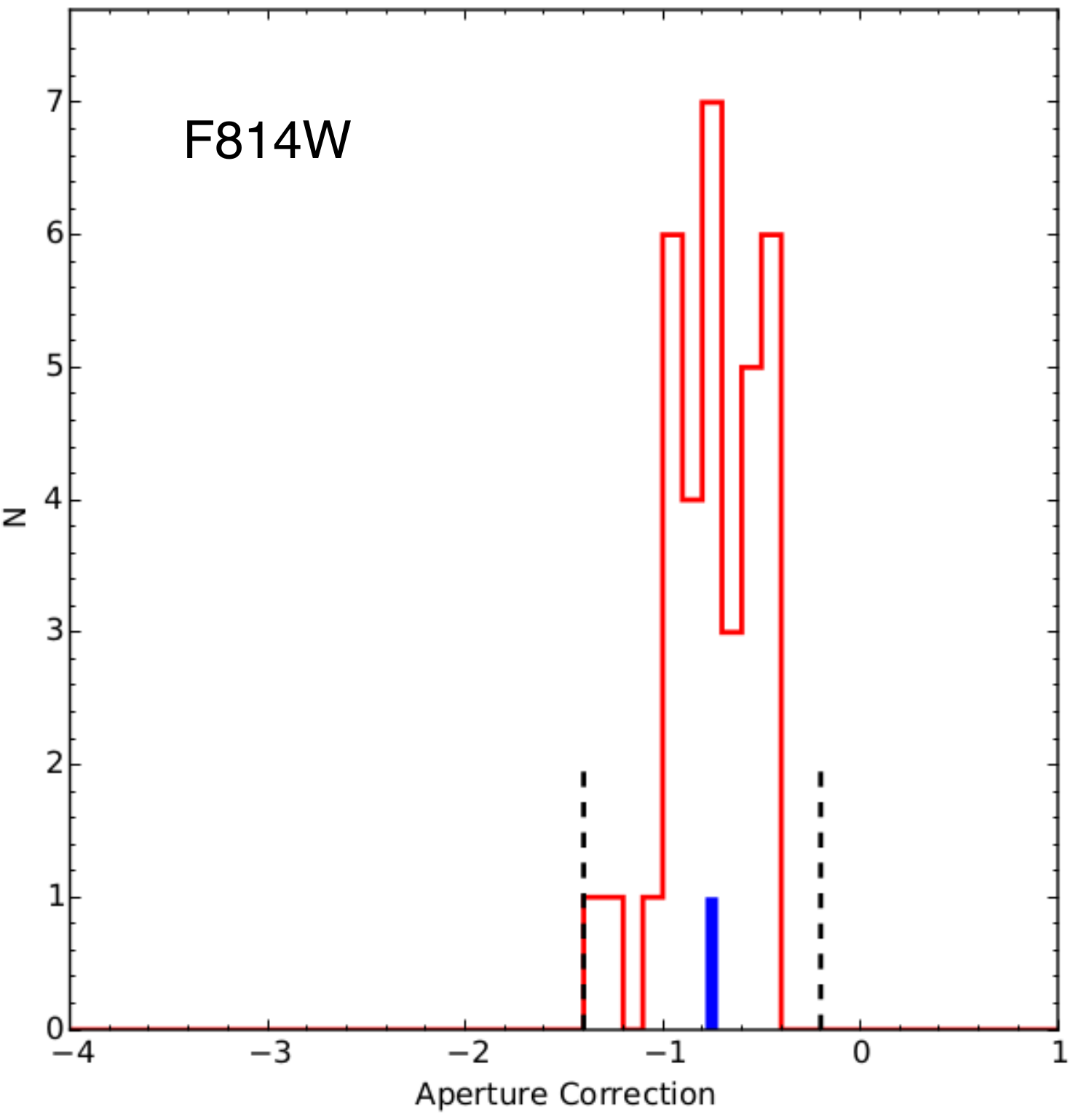}
    \caption{Plots produced by the {\bf  legus\_clusters\_extraction.py} pipeline and used to decide important photometric parameters, like CI cut, aperture radius, and range of allowed aperture corrections for each galaxy. The plots included here are for the NGC628c pointing. {\bf  Top row}. On the left, the recovered CI distributions of the star and cluster control samples are shown. In the middle, the growth curves (normalised flux in $V$ band versus aperture) of stars (dark yellow solid lines) and clusters (cyan solid lines) contained in the control samples are plotted. Median curves of both samples are overplotted with red (stars) and blue (clusters) solid lines. The black horizontal solid line shows where the 50\% flux is reached. The vertical dashed lines show which fraction of flux is contained in apertures of 4, 5, and 6 px. IDs are included for each single curve, and are used to remove sources that show extreme deviations from the general trends. In the right, the distribution of CI of all the sources extracted in Step 1. The red vertical line shows the CI reference value used to distinguish between unresolved sources (stars) and resolved objects (candidate clusters). {\bf  Bottom row}. The aperture correction recovered using the cluster control-sample. The black vertical dashed lines show the limits within which the average values are estimated for each band. The blue line shows the average value (see Table~\ref{tab1}).}
    \label{fig:fig2}
\end{figure*}

The pipeline comprises six consecutive steps and produces several diagnostics that help to fix key parameters to produce the cluster candidate catalogues for each galaxy. We present here a detailed description of each step in the pipeline, and its application to NGC 628. This galaxy has been observed in two different pointings, NGC628c (inner region) and NGC628e (outer region). The two datasets have been analysed separately, because of different combinations of cameras and filters (see Table~\ref{tab1}).

\begin{itemize}
\item {\bf  Step 1} -- Source extraction. The pipeline uses SExtractor \citep{1996A&AS..117..393B} to extract source positions from the white-light image produced with the five standard LEGUS bands (see C15 for the method used to produce white-light images). In cases where the white-light image is not optimal (e.g. it shows numerous artefacts at the edges of the single frames, or the $I$ band in low stellar density fields is dominated by red giants) we replace it with the reference $V$ band frame. The SExtractor input parameters are optimised to extract objects with at least a 3 $\sigma$ detection in a minimum of 5 contiguous pixels. These numbers change when the white-light or the $V$ band filter is used for detection. We do not apply any filtering to the image but we use small background cells to make sure that the strong gradient in the background between arms and inter-arm regions does not affect our capability of detecting sources. These initial catalogues are visually screened to understand whether potential clusters have been missed and any possible improvement in the SExtractor configuration file can be made. The configuration file used for the source extraction will be released together with the final LEGUS cluster catalogues for each galaxy. At the end of this first step we detect 6272 (NGC628c) and 4539 (NGC628e) candidate clusters.

\item{\bf  Step 2} -- Determination of CI and aperture radius parameters. In this step, the user selects a training sample of objects that are clearly stars (i.e., isolated, bright, and with steep luminosity profiles) and another sample that are clearly clusters (i.e., isolated, relatively bright, and with shallow luminosity profiles) using the reference frame ($V$ band). The pipeline performs aperture photometry on these sources using radii of increasing size (from 1 px to 20 px with a step of 1 px). The local sky background annulus is set at 21 px and is 1 px wide. From this photometry, the pipeline calculates the CI (the magnitude difference between apertures of radius 1 pix and 3 pix) and the curves of growth for the input lists of stars and clusters. In Figure~\ref{fig:fig2} we show an example of the plots produced by the pipeline in Step 2 and 3. The top left panel shows the CI distribution recovered for the star and cluster control sample. From this plot, the user selects a CI value that separates the visually-selected stars and clusters as cleanly as possible. In the case of NGC628c the value we apply is 1.4. This value is an initial guess that can be iteratively adjusted in Step 3.  As we will discuss and further investigate in Step 3 and in the completeness analysis, a selection criterion based on the CI cut allows us to remove a significant amount of stars from the catalogue, but it will also remove compact clusters which appear unresolved on the frame (see Section 2.3 for a detailed discussion).  The middle panel of Figure~\ref{fig:fig2} shows the growth curve analysis on the stellar and cluster reference sample. The median stellar growth curve (red thick line in the plot) shows how the flux increases for an averaged unresolved source on this specific reference frame. This curve is linked to the stellar PSF produced by a given combination of detector and filter and will change from galaxy to galaxy. The median cluster growth curve shows (blue thick line in the plot) how the flux is distributed in a resolved cluster spread function (CSF)\footnote{Since the PSF is the spread function of a point-like source, we will refer to the spread function of clusters as cluster spread function. Notice that a CSF does not mean a cluster resolved in their stellar content. Clusters in the LEGUS galaxies cannot be resolved in their single stars. This is also true for the closest targets, where an unresolved cluster core is typically surrounded by partially resolved single stars.}. The increase in flux is slower in the resolved cluster than in the averaged stellar one, explaining why the CI is larger for resolved (clusters) than unresolved (stellar-like) sources \citep[the CI method was first introduced by][]{1992AJ....103..691H, 1993AJ....106.1354W}. Inspection of the growth curves allows the user to fix the size of the aperture radius used to perform photometry. For each galaxy we fix the aperture radius to do cluster photometry to the smallest integer number of pixels containing at least 50\% of the flux within the aperture (aperture corrections for the missing flux are discussed in Step 4).  We select the smallest aperture to reduce as much as possible the risk of contaminations from neighbouring sources. We will refer to this value as the science aperture, to distinguish it from other apertures used in the analysis.

\item {\bf  Step 3} -- CI selection and multi-band photometry. The CI distribution of all the extracted sources in Step 1 are plotted together with the CI cut value selected in Step 2. The top right plot of Figure~\ref{fig:fig2} shows the distribution for NGC628c. As already discussed in the literature \citep[e.g][]{2010ApJ...719..966C, 2015MNRAS.452..246A}, the CI distributions of point sources which are unresolved (stars) will show a narrow peaked distribution (e.g. peak in the right panel of Figure~\ref{fig:fig2}). On the other hand, clusters are resolved and they show a broad range of sizes \citep[e.g.][]{2015MNRAS.452..525R}, thus, their CI distribution appears significantly broader than the stellar one (this can be seen in the distribution as the prominent wing extended to large CI values). The goodness of the chosen CI value for the cut is then directly tested on the reference frame ($V$ band) via visual inspection of the objects in the catalogue that satisfy this condition. We check whether a smaller CI introduces a large contamination of stars. The chosen best CI value is, eventually, indicated as a selection criterion. All sources with a CI smaller than the reference value (1.4 for the ACS $V$ band of NGC628c and 1.3 for the WFC3  $V$ band of NGC628e) are removed from the initial catalogue. We perform then multiband aperture photometry on the sources that pass this selection step, using the science aperture fixed in Step 2 and a local sky annulus located at 7 px with a width of 1 px. The same science aperture and local sky annulus is used in all the 5 filters.

\item {\bf  Step 4} -- Averaged and CI-based aperture correction. We apply two different aperture correction methods both widely used in the literature.

In the first case, we produce averaged aperture corrections using the cluster control-sample produced in Step 2. The correction is estimated as the difference between the magnitude of the source recovered at 20 px (sky annulus at 21 px, 1 px wide) minus the magnitude of the source obtained using the science aperture (sky annulus at 7 px, 1 px wide). Not all the sources in the control samples are detected in all the filters. To avoid clusters not detected in some filters from skewing the averaged values of the aperture corrections we reject sources that have corrections outside a certain range. We show for example the aperture correction distributions recovered for the cluster control-sample of NGC 628c in the 5 LEGUS band in the bottom panels of Figure~\ref{fig:fig2}. The vertical dashed lines show the limits within which we use the values to produce the average correction. These limits can be adjusted by the user for a given galaxy. In the case of NGC 628c some of the sources are very faint or not detected in the bluest filters, therefore, their corrections become extreme. In Table~\ref{tab1}, we list the final averaged aperture corrections and the errors (standard deviations) for each filter of each pointing in NGC 628. These values are added to 
the photometry produced in Step 3. The standard deviation of the aperture correction recovered from the sample is added in quadrature to the photometric error, to take into account the uncertainties introduced by the averaged correction. With this method the differences in sizes of the clusters are not taken into account. However, as one can see from the recovered values in Table~\ref{tab1}, the differences in the corrections as a function of waveband are quite small. Therefore, averaged aperture corrections do not change the shape of the spectral energy distribution (SED) of the source, they only change the normalisation. This will slightly affect the mass estimates (not the derived ages and extinction values) but it will be well within the average 0.1 dex uncertainties that the SED fitting methods produce.

In the second approach, we produce aperture corrections based on the CI of the source in each band. The method to derive the CI versus aperture correction relation will be described in a forthcoming paper (Cook et al. 2016). This method has initially been developed to produce an aperture correction for very extended clusters (partially resolved in their stellar components) detected in the LEGUS dwarf galaxies and afterward extended to the rest of the LEGUS targets. The CI-based aperture correction for each source is calculated by measuring the CI in each band and applying the CI-aperture correction relation determined for the appropriate instrument and science aperture size. While this method takes into account the size (indirectly measured via the CI) of the source as a function of waveband, it can not only change the normalisation of the SED but also the shape (so it will affect the mass, age, and extinction). This effect will be particularly large in faint sources and in crowded regions where the estimate of the CI become more uncertain. 

At the end of Step 4 we have two sets of aperture corrections calculated. Within the LEGUS naming convention, the two final photometric catalogues produced in Step 5 with these two correction methods and derived analyses are referred to as  AV\_APCOR and CI\_BASED.

\item {\bf  Step 5} -- Final products. The quality of the photometry is checked in this final step. We remove all the sources which have not been detected in at least two filters (the reference $V$ band and either $B$ or $I$ band) with a photometric error $\sigma_{\lambda} \leq 0.3$ mag. In each band, we correct the photometry of these sources by the foreground Galactic extinction \citep{2011ApJ...737..103S}. Two automatic final photometric catalogues are produced, e.g., the \emph{automatic\_catalog\_ngc628c\_avgapcor.tab} and the \emph{automatic\_catalog\_ngc628c\_cibased.tab}. Both catalogues contain the ID of the sources, positions in pixels and in RA and DEC, multiband photometry and errors, CI, and a flag indicating the number of filters in which each source was detected. In the case of NGC628c and e, the automatic catalogues produced at the end of this step contain 3086 and 593 cluster candidates, respectively. 

These sources, however, are likely not only star clusters. Among them there are still interlopers, i.e. pairs and multiple stars in crowded regions, background galaxies, bright stars in the galaxy which have CI slightly larger than the limit used here, foreground stars, objects and artefacts at the edge of the science frames. To minimise the contaminations of our final catalogues, we visually inspect a fraction of the sources found in the automatic catalogues that satisfy some extra selection criteria. We select for visual inspection all the sources which are detected in at least 4 bands (in the \emph{automatic\_catalog\_ngc628c\_avgapcor.tab}) with a photometric error below 0.3 mag and have an absolute $V$ band magnitude brighter than $-6$ mag. The $V$ band magnitude cut is applied to the $V$ band photometry obtained with the CI based aperture correction. This allows us to include slightly fainter, very diffuse clusters since they have large values of the CI and therefore get a larger aperture correction.

\item {\bf  Step 6} -- Missed clusters. During the visual classification of the sources it is also possible to add sources that have not been included in the visual inspection catalogue. If such extra sources are not found in the automatic catalogues then we use Step 6 to produce their final photometry in the same homogeneous way as done for the bulk of the other sources. The number of added sources is typically below 1 \% but it can change from galaxy to galaxy.
  
\end{itemize}

\subsubsection{Classification of cluster candidates}

\begin{figure*}
\centering
	\includegraphics[scale=0.7]{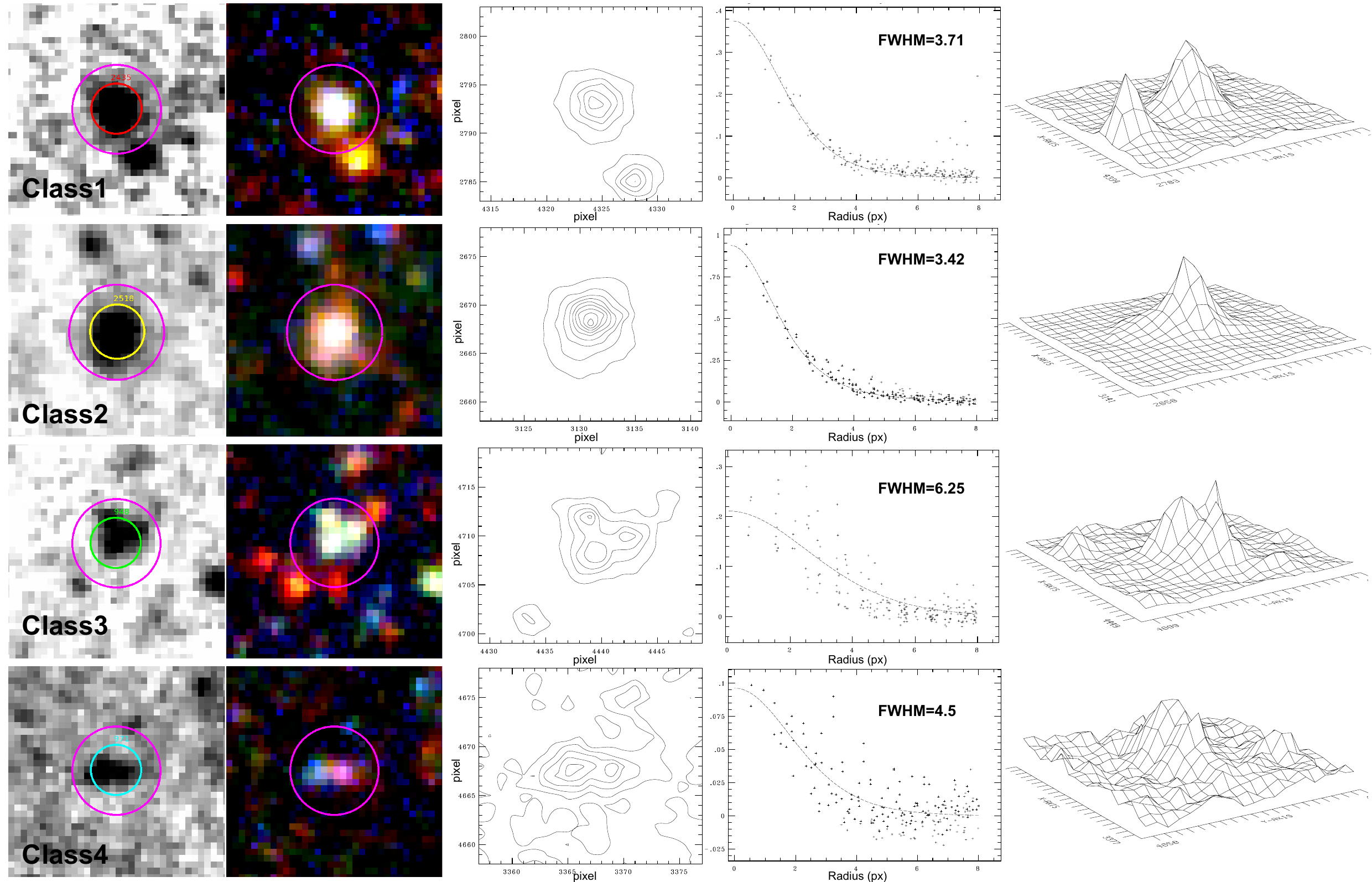}
    \caption{Panels inspected for the visual classification of the sources. An example for each identified class is listed. The first two panels of each row show each object in the reference frame $V$ (logarithmic scale) and in a 3-color composition. The outer purple ring has a radius of 7 px ( $=0.28 \arcsec\sim$13.4 pc) and shows the position of the local sky annulus. The inner ring shows the aperture radius used to do photometry (4 px $=0.16 \arcsec\sim$ 7.7 pc). The middle and right plots show the contour, the radial profile, and the surface plot of the object. The FWHM reported in the radial profile is estimated using a MOFFAT light distribution. Detailed comments on each object are provided in the text. }
    \label{fig:fig3}
\end{figure*}

At this stage of the reduction three or more independent classifiers from within the LEGUS team visually inspect the cluster candidates. A tool has been made available within the LEGUS collaboration to perform this task. This tool visualises the object to be classified in two frames, in the reference $V$ band and in a combined 3-color image. Using an interactive window the user can also visualise the contour, the radial profile, and the surface plot of each source in the reference band (see  Figure~\ref{fig:fig3}) using the IRAF task IMEXAMINE. Two circles on the images show the aperture used to perform photometry (inner) and the location of the local sky (outer circle). Each source is inspected in the single band image to check whether the light is extended (cluster) or has a stellar PSF (star). The difference in the light distribution of a cluster and a star at the distance of NGC 628 can be easily discerned in the top  row panels of  Figure~\ref{fig:fig3}. The source at the centre of the inset is a cluster and it has a FWHM$=$3.71 px. A star is visible in the same cutout and its light is much less extended (FWHM$= 2.4$ px) than the one of the cluster. The contour and surface plots (middle and right panels) confirm the different behaviour in the two objects. 

Based on this inspection each object in the catalogue gets assigned one of four defined classes. In Figure~\ref{fig:fig3} we show an example object for each class taken from the central pointing of NGC 628. Class 1 clusters are compact and centrally concentrated objects, with a FWHM more extended than the stellar one. They show a homogeneous self consistent color. With respect to class 1,  class 2 systems are clusters with slightly elongated density profiles and less symmetric light distribution. Class 3 can potentially be less compact and homogeneous clusters, more likely compact associations\footnote{Typical size of stellar associations are between 50 and a few hundreds parsecs. Class 3 objects that we will call hereafter as associations, or compact associations, have a projected size of a few tens of parsecs and, in many cases, these compact associations are part of much larger stellar associations.}, and they show asymmetric profiles and multiple peaks on top of diffuse underlying wings which suggest the presence of a possible concentration of low-mass stars. Finally, class 4 objects are single stars or artefacts, or any other interlopers (e.g. background galaxies, foreground stars) that can affect the cluster analysis. The bottom  row of Figure~\ref{fig:fig3} shows an example of an object detected with extended CI, morphology and CSF profile similar to a class 3 object. However, the stars belonging to this system have clearly different colors. We consider these objects as a chance superposition along the line of sight of two stars, thus, they are excluded from the cluster catalogue. 

Within the LEGUS collaboration, we are also testing a \emph{machine-learning} (ML) recognition of the sources. The visually inspected catalogues are fed to the algorithm as training sets for the subsequent classification of sources. Testing is in progress and the results will be published in a forthcoming paper (Grasha et al. in prep). The cluster catalogues of a third of the LEGUS galaxies have currently been inspected by humans. We are providing visual classification of a fraction of the catalogues of another third of galaxies, that are used as training set by the ML algorithm.  The goal is to perform the visual classification of the remaining galaxies using a purely based ML approach.

\subsection{Testing our approach and completeness limits}

\begin{figure}
	\includegraphics[scale=0.46]{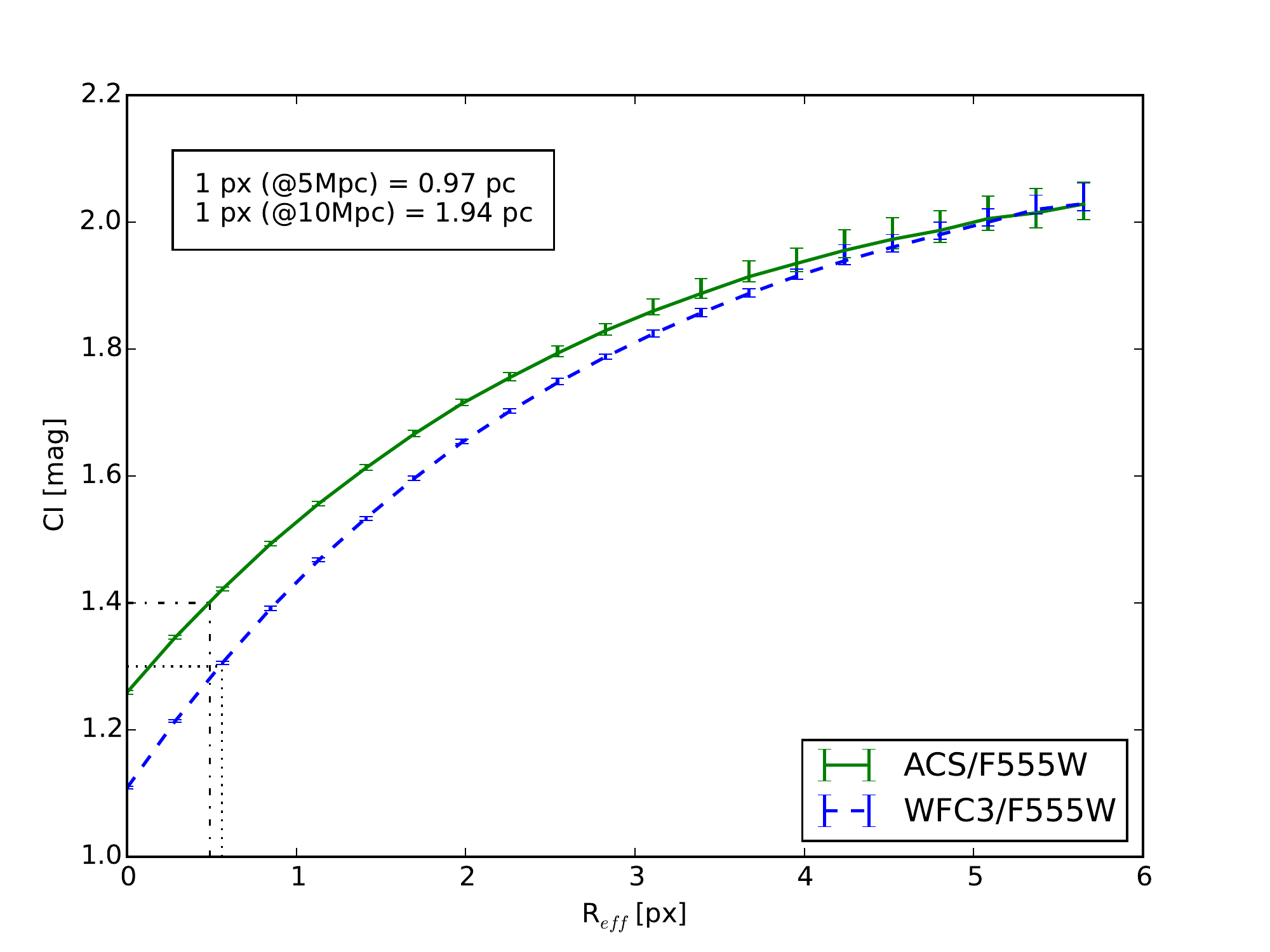}
    \caption{Relation between CI and $R_{eff}$. The two curves are the median values obtained from artificial clusters simulated with increasing $R_{eff}$. The relation has been tested for the reference $V$ band frame in both the ACS and WFC3 instruments. The error bars are the 25 and 75 \% quartiles of the measurement. An inset shows the pixel resolution corresponding to the distance range typically covered by the LEGUS galaxies. The black dotted lines show the CI values used for the cluster candidate selection in the inner (ACS/F555W) and outer (WFC3/F555W) of \obj. The applied cuts approximately correspond to $R_{eff} \sim 1$ pc.  }
    \label{fig:fig4}
\end{figure}

To investigate the impact of the assumptions and the selection criteria adopted to produce the LEGUS cluster catalogues we have performed several completeness tests. The tests are run with a custom-made pipeline, {\bf  legus\_cct.py} (LEGUS Cluster Completeness Tool)  available within the LEGUS collaboration.
The pipeline runs in five consecutive steps, i.e. creation of synthetic sources, source extraction, photometry, recovery fractions, final outputs. One of the input files is the same used to run the standard {\bf  legus\_clusters\_extraction.py}. The second input file specifies quantities necessary to create the synthetic frames containing the simulated clusters, e.g. input parameters for the BAOlab software \citep{1999A&AS..139..393L}. A stellar PSF generated with the IRAF task PSF using selected stars in the science frames is also provided. This PSF is then convolved with MOFFAT15 profiles of specified effective radii, $R_{eff}$. The resulting extended sources are then randomly distributed in a blank frame of the same dimension as the science one. A magnitude range is provided as well. To overcome crowding problems we only insert a thousand clusters per loop. The frame containing synthetic clusters is then added to the real science frame. In the next step, all the sources are then extracted using the same procedure as in the {\bf  legus\_clusters\_extraction.py}. For all the sources extracted it estimates the CI and the photometry (including averaged aperture correction) in the same way as for the real clusters. Using the known position of the simulated clusters, the software estimates a source recovery fraction before and after the CI cut is applied. 

\begin{figure}
	\includegraphics[scale=0.26]{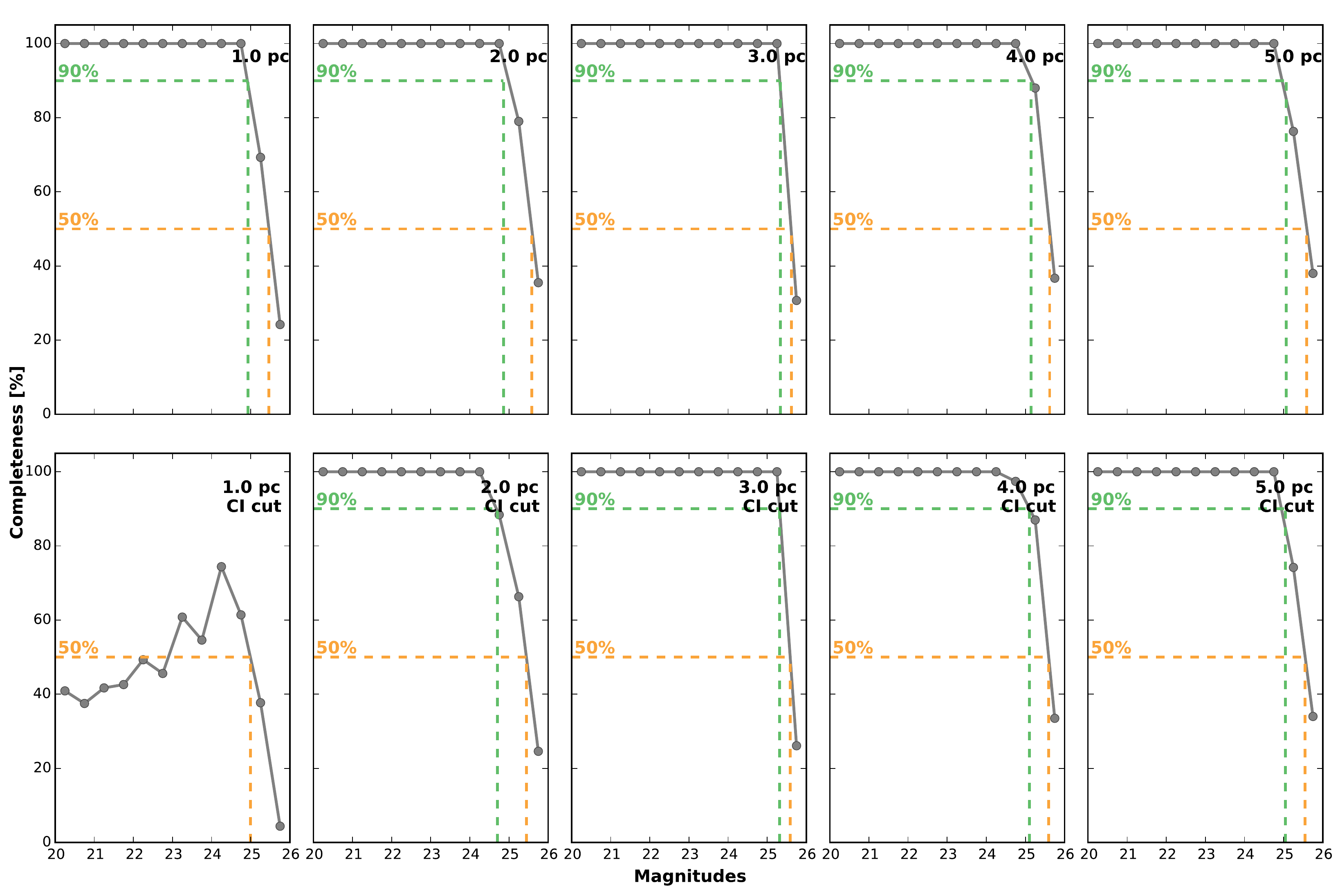}
    \caption{Recovered completeness limits in the $V$ band frame of \obj c as a function of $R_{eff}$ (annotated in each panel). The top  row shows the recovery rates before, the bottom  row after the CI cut is applied. The CI cut is only affecting the recovery of very compact and partially unresolved clusters with $R_{eff} \leq 1$ pc.}
    \label{fig:fig5}
\end{figure}

As a first result of these simulations we have investigated the relation between CI and $R_{eff}$ of increasingly extended sources, with particular emphasis on the impact that the CI cut has in removing compact clusters as a function of distance of the galaxy. In Figure~\ref{fig:fig4} we show the recovered relation between CI and $R_{eff}$ for the ACS and WFC3 F555W science frames, used as reference frames for the CI selection. As an example, we refer to the study-case \obj. The applied CI cuts of 1.4 mag and 1.3 mag for the ACS and WFC3 F555W correspond to a cluster $R_{eff}\sim1$ pc. This does not imply that all clusters of 1 pc size are systematically removed from the catalogues. We investigate the recovery fraction of sources in the $V$ band frame before and after the CI cut is applied. In Figure~\ref{fig:fig5} we illustrate the results for \obj c. Before the CI cut is applied we have 100\% recovery for sources of 1 pc as well as more extended ones. However the recovery of very compact sources goes below 50\% after the CI cut is applied. Only half of the sources with  $R_{eff}\sim1$ pc are extended enough to make into the selection. A smaller CI will include too many stellar objects, so we conclude that at the distance of \obj\, the CI cut applied removes a fraction of unresolved clusters with sizes of 1 pc and below. This choice does not introduce biases in our cluster analysis because observationally, no clear relation between the size and mass or luminosity of star clusters has been found \citep[see][and references therein]{2015MNRAS.452..525R}. Moreover, an ongoing analysis of cluster sizes in \obj\, shows a clear log-normal distribution which is well above the detection limits of 1 pc (Ryon et al. submitted).

\begin{figure}
		\includegraphics[scale=0.48]{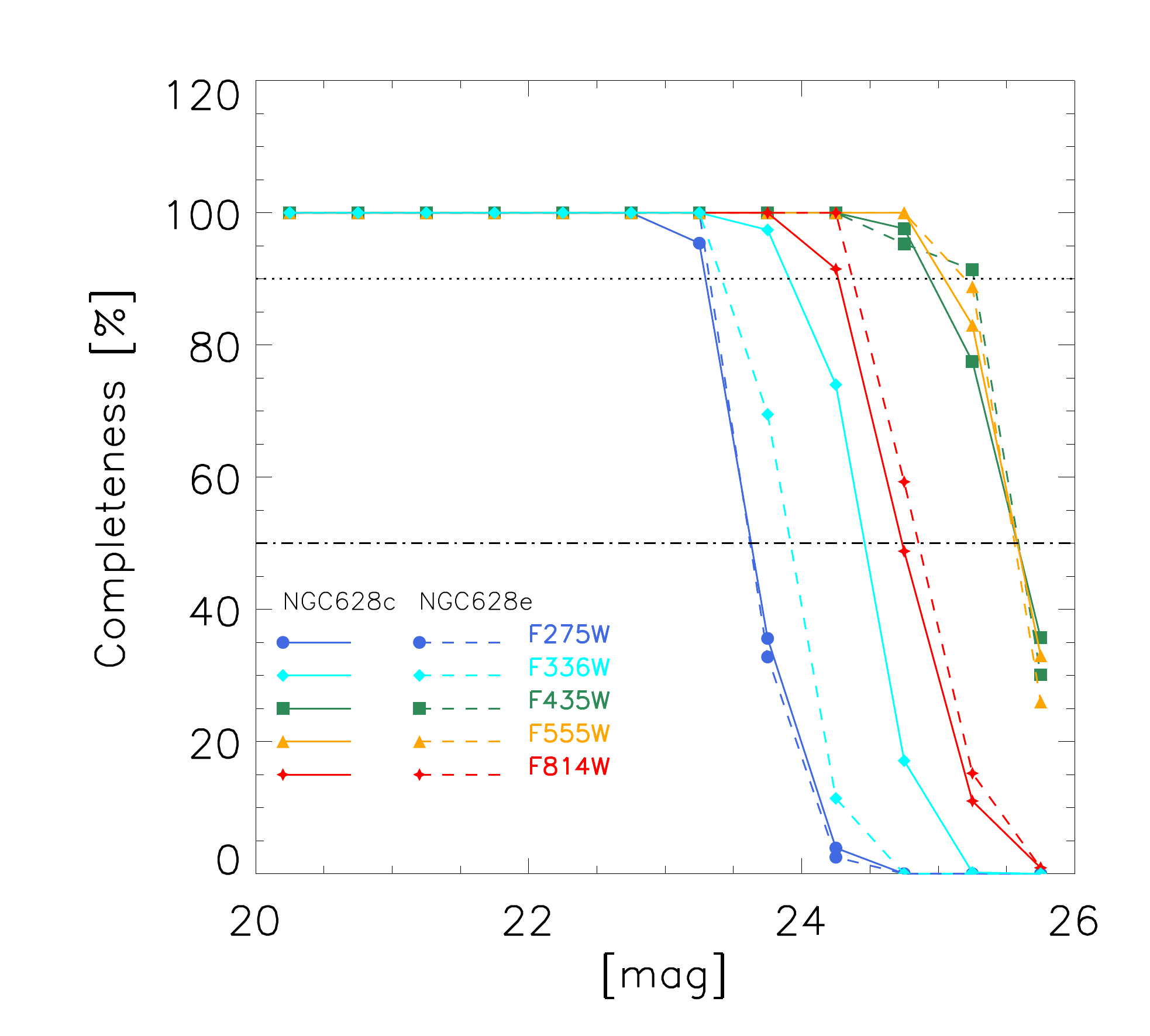}
    \caption{Recovered completeness limits as a function of different bands for clusters with sizes larger than 1 pc. Solid and dashed lines associated with different symbols have been used to show the recovery rates in the inner and outer field respectively (see inset).}
    \label{fig:fig6}
\end{figure}

In Figure~\ref{fig:fig6} we illustrate the final completeness limits of the two pointings in \obj. In Table~\ref{tab1} we list the magnitude limits corresponding to 90\% recovery of sources with a detection threshold above 3 $\sigma$ (column 8 Table~\ref{tab1}) within a minimum of 5 contiguous pixels. For the same band, the differences in the exposure times (longer exposures) have a larger impact on the recovery fractions than the differences in crowding between the inner and outer region (i.e. compare the recovery magnitude for the F336W and F435W for the inner and outer pointing). The completeness software has been run independently for each band. Therefore, these values must be considered as indicative of the detection limits intrinsic to the science frames. However, we are not taking into account that we apply to our catalogues additional selection constraints (detection in at least 2 filters, or in case of visually inspected sources in 4 filters), visual inspection, and that real clusters have different luminosities at each wavelength depending on their age, mass, and extinction. 

\subsection{Comparison with catalogues available in the literature}

\begin{figure*}
\centering
		\includegraphics[scale=0.48]{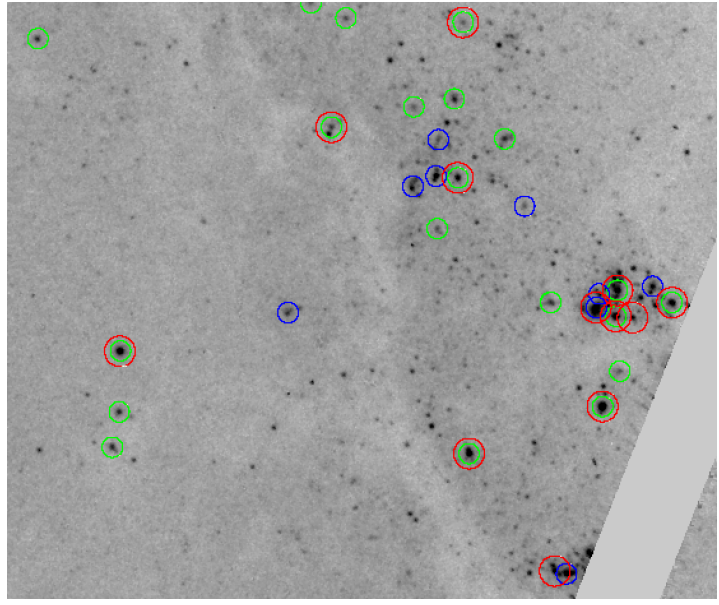}
    \caption{A zoom-in image of a region of NGC628 ($560 \times 473$ px). The LEGUS final catalogue of objects classified as class 1 \& 2 is illustrated by the green circles, class 3 by blue circles. We compare the LEGUS catalogue to the automatic catalogue (red circle sources) based on HLA observations as described in \citet{2014AJ....147...78W}. See text for differences and similarities in the selection criteria.}
    \label{fig:fig6b}
\end{figure*}

Figure~\ref{fig:fig6b} shows a comparison between the LEGUS final catalogue (green circles are class 1 \& 2, blue circles class 3) and a catalogue from \citet[][shown as red circles]{2014AJ....147...78W} for a small region in NGC 628. The Whitmore et al. catalogue was
produced automatically based on Hubble Legacy Archive (HLA)
observations, including measurements of the concentration index and
algorithms that remove close pairs and multiple hits in very crowded
regions. A relatively bright limit is used ($M_V = -8$) for the HLA-based
catalogue since no manual inspections are made.

The correspondence is fairly good, with 77 \% of the HLA-based sources
being included in the LEGUS catalog. Making a similar magnitude cut in
the LEGUS catalogue (which goes roughly 2 mag deeper) results in a
recovery rate of 73 \% of the HLA-based catalogue being found in the
bright end of the LEGUS catalog.  There are two primary differences:
1) the LEGUS catalogue includes approximately five more (out of about 100 in
the region of overlap used for the comparison) bright class 1 objects
that are clearly good cluster candidates and were missed in the HLA
(probably due to using too conservative a limit for the concentration
index), and the LEGUS catalogue includes about 15 more compact
associations - i.e., type 3) than found in the HLA-based
catalog. While selection effects are an important consideration in any
study, we note that \cite{2014ApJ...787...17C} compared completely manual,
hybrid (like with LEGUS), and completely automatic catalogs made by two
different studies and found that the selection did not result in major
changes to the primary results of the studies (i.e., the CLF, CMF, and
age distributions).

\section{Final cluster catalogues \& SED fitting procedures}
\begin{figure*}
\centering
		\includegraphics[scale=0.28]{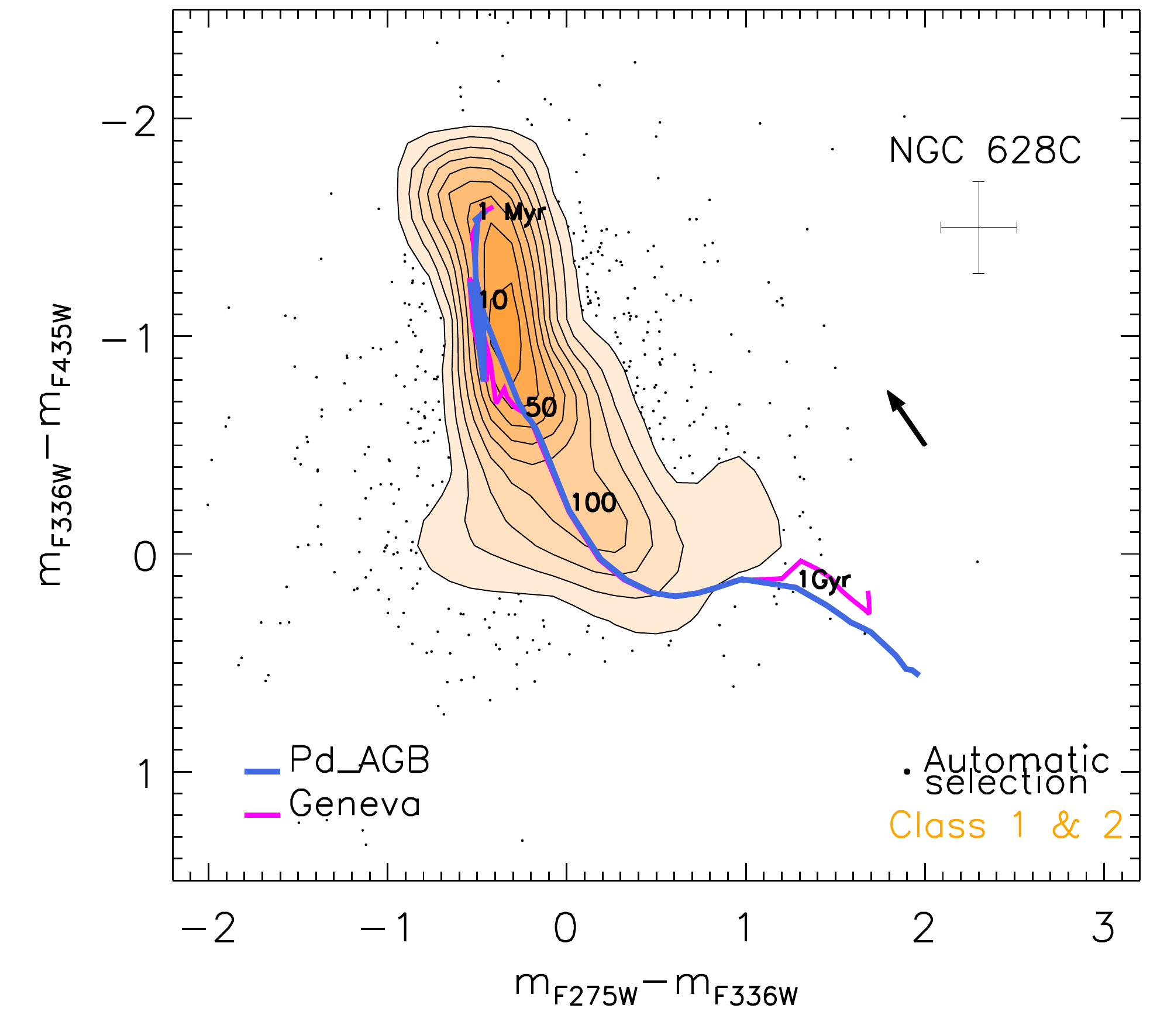}
		\includegraphics[scale=0.28]{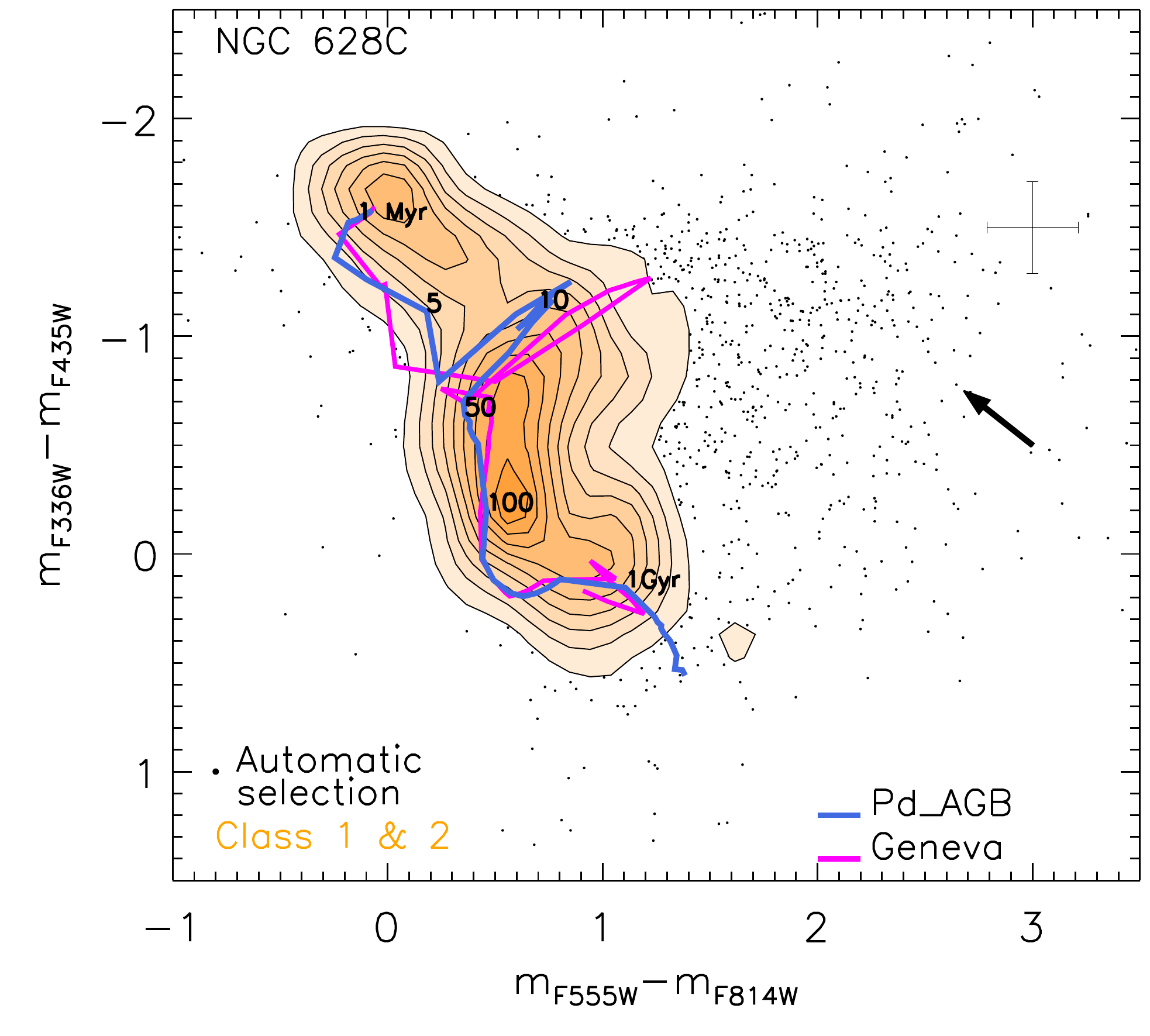}
		\includegraphics[scale=0.28]{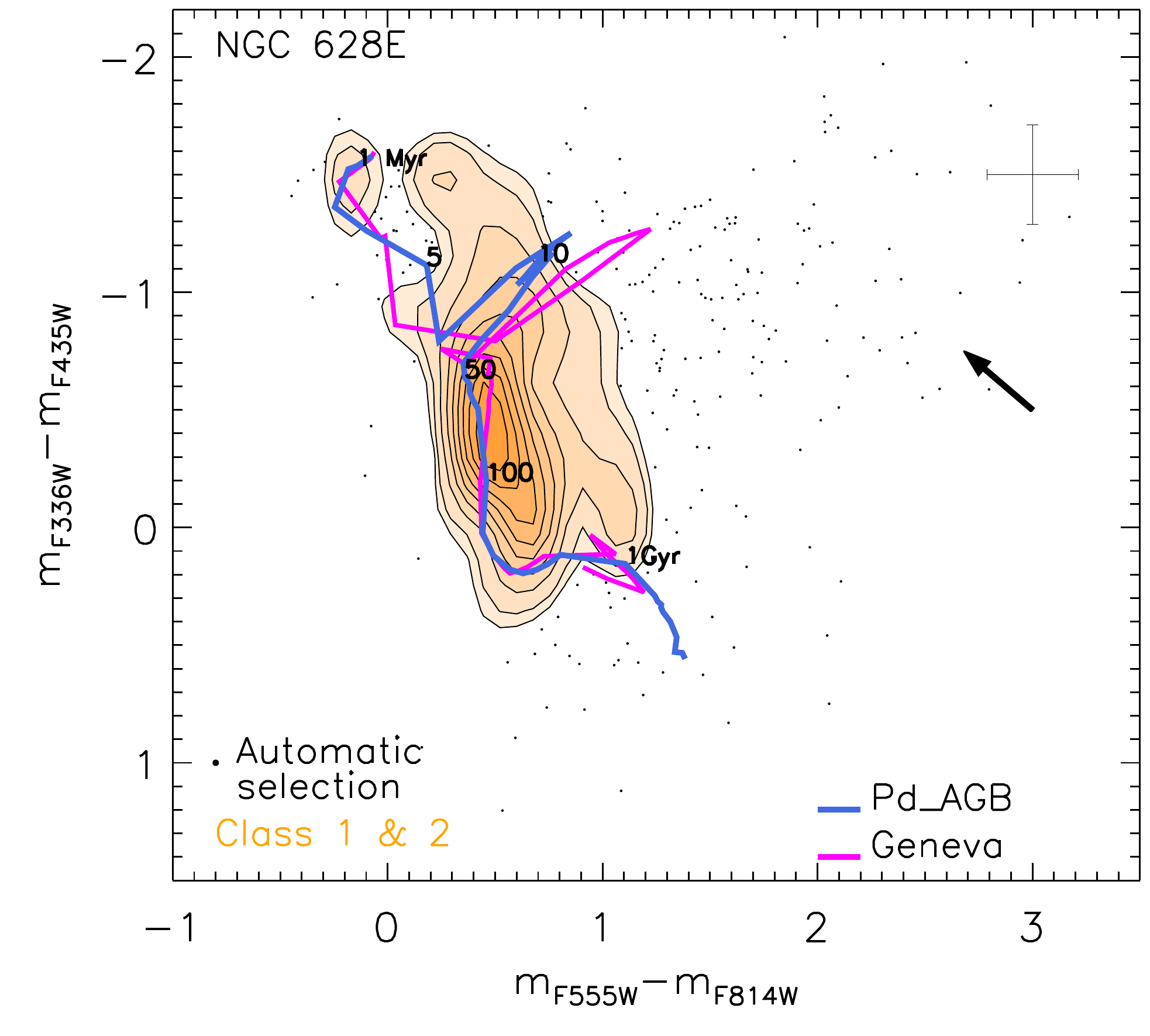}\\
		\includegraphics[scale=0.28]{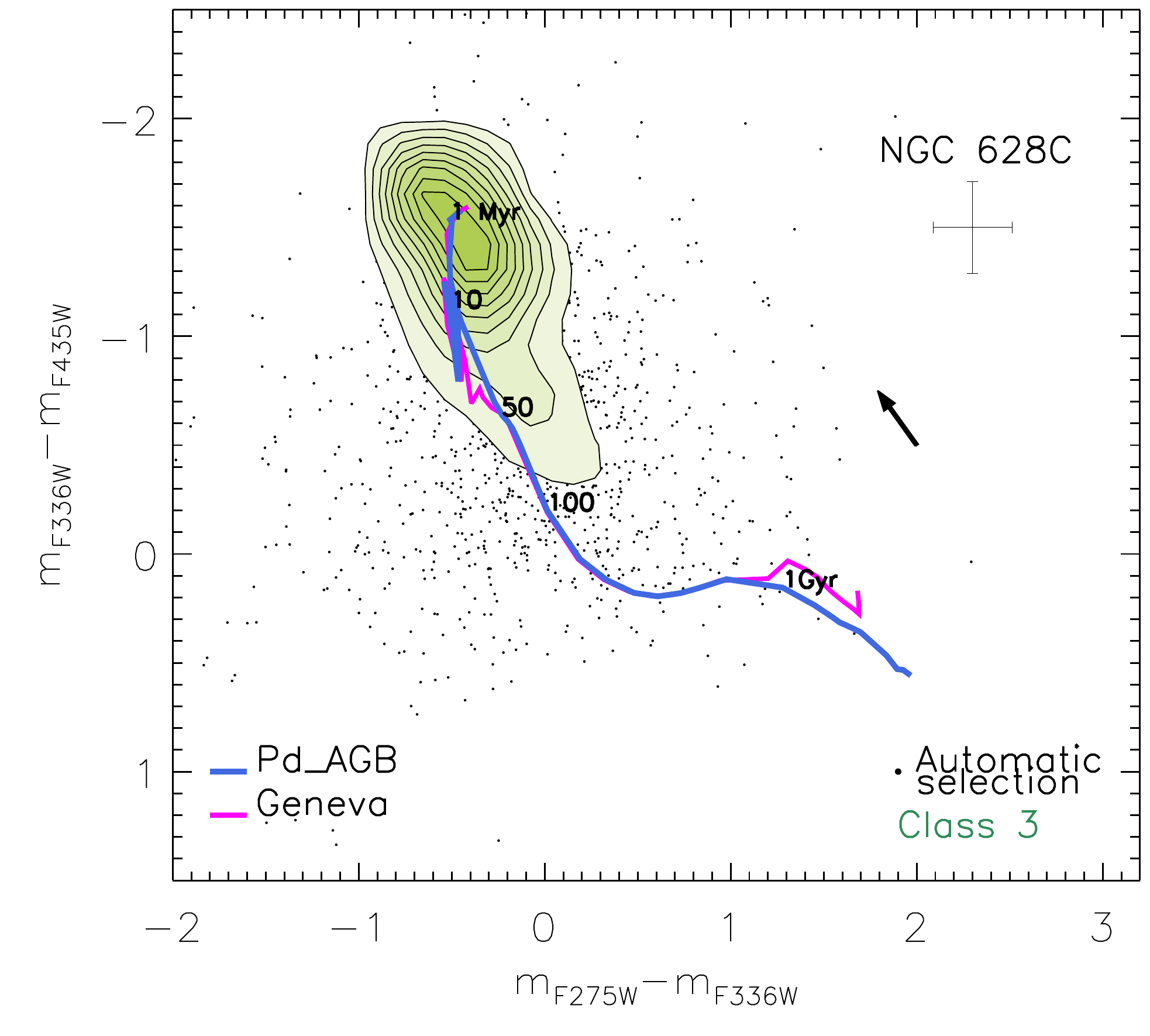}
		\includegraphics[scale=0.28]{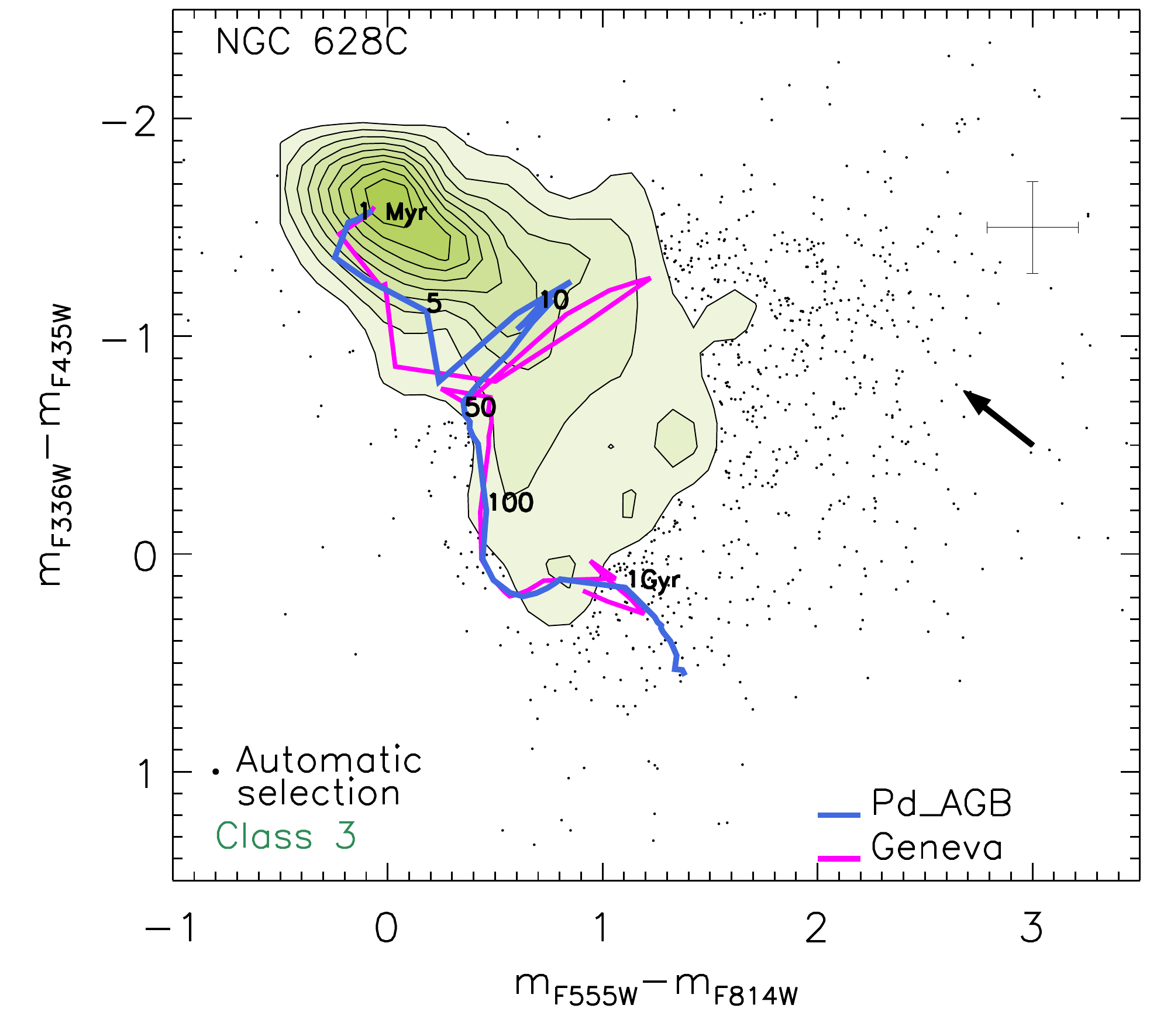}
		\includegraphics[scale=0.28]{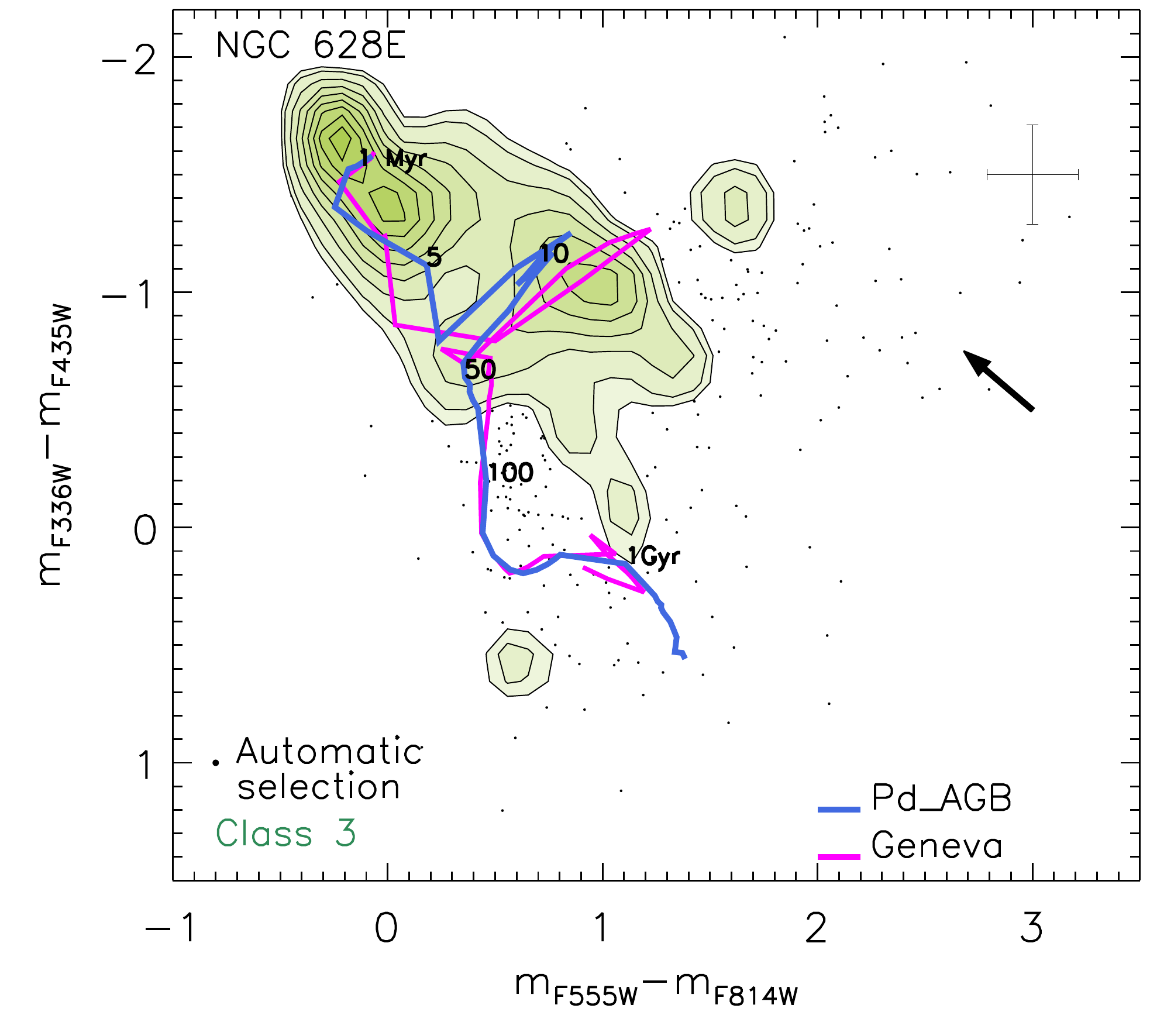}

    \caption{UV (left) and optical (middle and right panels) color-color diagrams of the cluster population of NGC628c (left and central) and NGC628e (right). The dots show the location of all the sources contained in the final reference (AV\_APCOR) catalogue. The overlaid density number contours show the distribution of class 1 \& 2 clusters (top), and class 3 objects (bottom). The contours include regions with densities equal or larger than 10\% of the 2D--histogram peak. Padova (blue) and Geneva (magenta) evolutionary tracks are included (see legend). The arrow shows in which direction the objects will move if corrected for a reddening $E(B-V) = 0.2$ mag. The error bar on the top right corner shows the average error in color. }
    \label{fig:fig7}
\end{figure*}

The photometric catalogues produced in Step 5 of the pipeline, as described in Section~\ref{autcat}, are fed into two different SED fitting algorithms. 

The analysis is performed with two different approaches that reflect the most common methods used in the literature. 

In the first case, we fit the cluster SEDs with \emph{Yggdrasil} SSP models \citep{2011ApJ...740...13Z}. We use two stellar libraries to create two sets of SSP models, Padova-AGB and Geneva tracks without rotation, available in Starburst99 \citep{1999ApJS..123....3L, 2005ApJ...621..695V}. They assume a \cite{2001MNRAS.322..231K} universal initial mass function (IMF), with stellar masses between 0.1 and 100 \msun. The IMF is assumed to be fully sampled, therefore, we will refer to these models as deterministic. These stellar models are then used as input to run Cloudy \citep{2013RMxAA..49..137F} and obtain a realistic evolution of the nebular emission line and continuum, produced by the ionised left-over gas surrounding the very young clusters. All the metallicity steps available in Starburst99 for both stellar libraries are accessible. For the Cloudy simulations we assume that the gas and the stars form in material with the same metallicity. For each galaxy, we adopted the present-day metallicity of its young populations as derived from nebular abundances in the literature and listed in C15. We use spherical solutions for the gas distribution around the ionising sources. The hydrogen number density, $n_H$, is set to typical values measured in H{\sc ii} regions, $n_H\sim10^2$ cm$^{-3}$. The covering factor, $c$ is assumed to be 0.5, i.e. roughly 50\% of Lyman continuum photons, produced by the central source, ionise the surrounding gas. The gas filling factor, $f$ is assumed to be 0.01. While the Yggdrasil interface is able to provide several combinations of Cloudy assumptions, we decided to fix the parameters to average values. Our dataset is limited to wide passbands, thus, the total integrated fluxes of these very young clusters are sensitive to the presence of emission lines and continuum from the ionised gas. The changes produced by different assumptions in the gas phases are secondary and difficult to disentangle. Including the nebular treatment in the models is fundamental to estimate the cluster physical parameters of very young star clusters \citep[e.g.][]{2001A&A...375..814Z, 2010ApJ...725.1620A, 2010ApJ...708...26R}, however, we do not have enough information to disentangle the gas conditions. The model grid used in the fitting procedure includes a combination of progressive age steps and increasing internal reddening, i.e. $E(B-V)=[0.0; 1.5]$ and steps of 0.01 mag. The models are reddened prior to be fitted to the observed photometry. We provide to the fit three extinction and/or attenuation laws. First, the Milky Way extinction law from \cite{1989ApJ...345..245C}. Second, the starburst extinction law by \citep{2000ApJ...533..682C}, assuming the stars and gas suffer the same reddening, and third, the same Calzetti et al. starburst extinction law, but instead, we assume the gas emission suffers higher extinction than the stars. The fitting algorithm is based on a traditional $\chi^2$ approach described in \citet{2010MNRAS.407..870A}. The software provides also error analyses as described in \cite{2012MNRAS.426.1185A}. Only sources that are detected with a photometric error $\sigma \leq 0.3$ mag in at least 4 bands are analysed. This condition will ensure that all the sources selected for visual inspection and a significant fraction of the excluded ones get a potential determination of their physical properties. 

In the second case, the cluster physical properties are obtained using a Bayesian analysis method together with stochastically sampled cluster evolutionary models presented by \citet[][hereafter K15]{2015ApJ...812..147K}. The analysis is based on the Stochastically Lighting Up Galaxies \citep[\emph{SLUG},][]{2012ApJ...745..145D, 2015MNRAS.452.1447K} code and its post-processing tool for analysis of star cluster properties, \emph{cluster\_SLUG} (K15).
The stellar libraries and extinction curves/attenuations used in \emph{cluster\_SLUG} are the same as for \emph{Yggdrasil}. In contrast to \emph{Yggdrasil}, the \emph{SLUG} treatment of the nebular emission is based on an analytical solution (see K15 for a discussion of the assumptions and possible differences rising from this different approach). \emph{cluster\_SLUG} provides posterior probability distribution functions (PDFs) for ages, masses, extinctions of the cluster candidates, assuming different priors on the cluster mass function and dissolution rate. 

This second approach provides a direct treatment of the uncertainties produced by low mass clusters affected by stochasticity combined with stellar evolutions. In K15 we present a direct comparison of the recovered cluster properties using the two methods presented here. 

Both deterministic and stochastic analyses are released, together with the photometric catalogues, on the LEGUS webpage. The data release for each galaxy contains 48 catalogues, 12 produced with deterministic fitting procedures and 36 with bayesian approaches\footnote{Twelve  catalogues come from a combination of the 2 photometric approaches for aperture correction (average based, CI based), 2 stellar libraries (Geneva and AGB-Padova), 3 extinction/attenuation curves. The bayesian analysis is produced for 3 different sets of priors but the same combination of photometric analysis, stellar libraries, extinction curves ($12 \times 3$ for a total of 36 catalogues).}.

Detailed studies on the impact of different flavours of stellar libraries and assumptions to build cluster evolutionary tracks are currently under investigation in the LEGUS team. \cite{2016mnraswofford} has tested the impacts of new evolutionary tracks in deriving the physical properties of very massive young star clusters detected within a small sample of LEGUS galaxies. More specifically, we have tested the use of recent published stellar libraries with single non-rotating stars, i.e. Padova \citep[][]{2014MNRAS.445.4287T}, Geneva \citep{2012A&A...537A.146E}, Geneva with single rotating stars \citep{2012A&A...537A.146E}, and BPASS with interacting binaries \citep{2008MNRAS.384.1109E}.

In the following sections we will analyse the cluster population of \obj.  The goal is to probe cluster formation conditions and evolution in this galaxy. We will check whether our results depend on the assumptions made to retrieve the cluster photometry and physical properties. Differences will be outlined and taken into account in our interpretations of the results. 

\begin{figure}
		\includegraphics[scale=0.40]{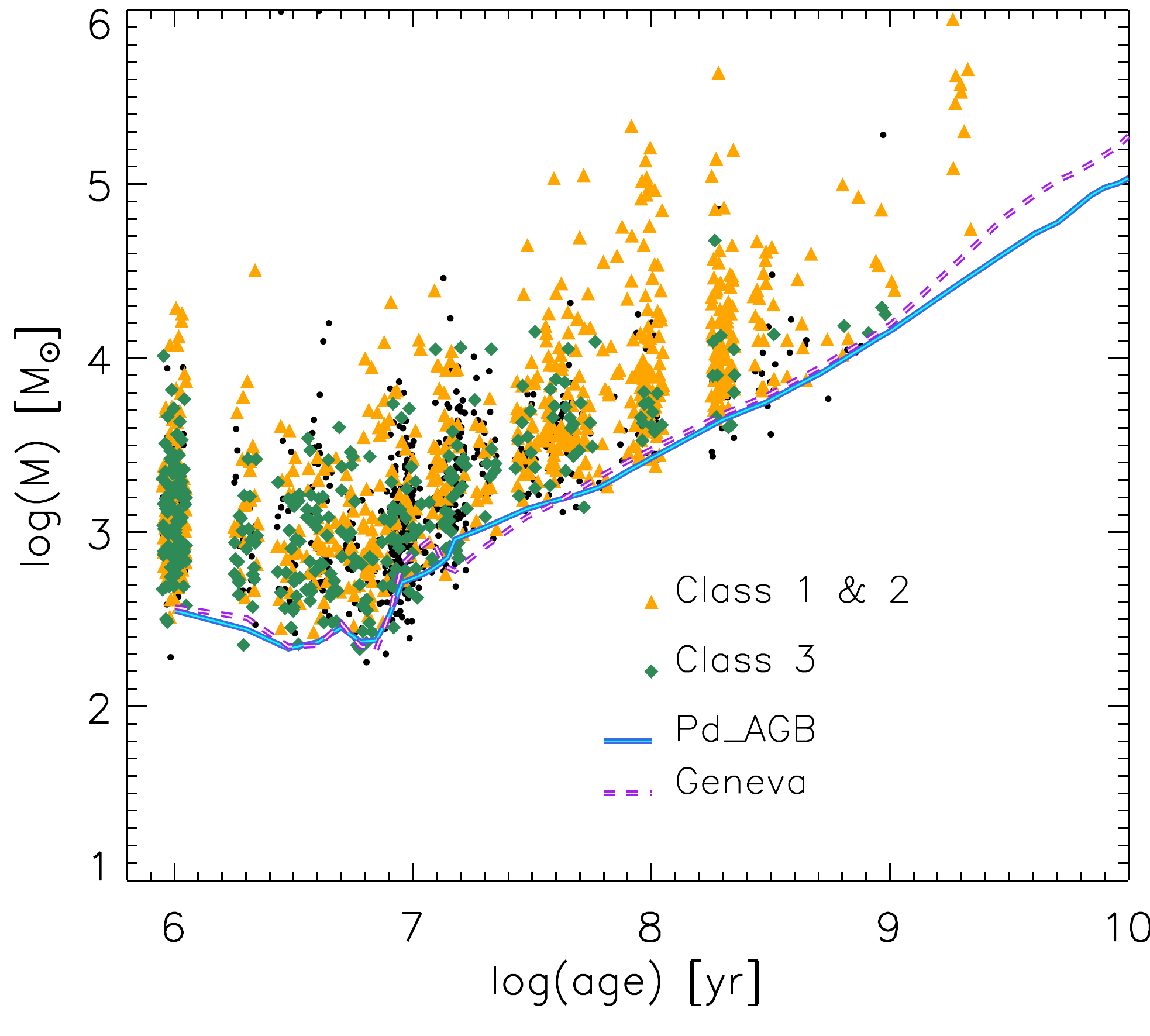}
    \caption{Age-mass diagram of the \obj c pointing. The mass limits as a function of age for both evolutionary tracks, shown by the blue and purple lines, are estimated assuming a M$_V=-6$ mag (23.98 mag), i.e. the magnitude cut used to select objects for visual inspection. The underlying black dots show the sources in the reference catalogue which have been detected in at least 4 filters, but have not been visually inspected. A small random shift is applied to the age of each source so that they do not overlap creating single columns at each age step. The discrete age steps of the models are anyway visible.}
    \label{fig:fig8}
\end{figure}

\section{The photometric properties of the NGC628 YSC population \label{ccd_phot}}

\begin{figure*}
\centering
		\includegraphics[scale=0.4]{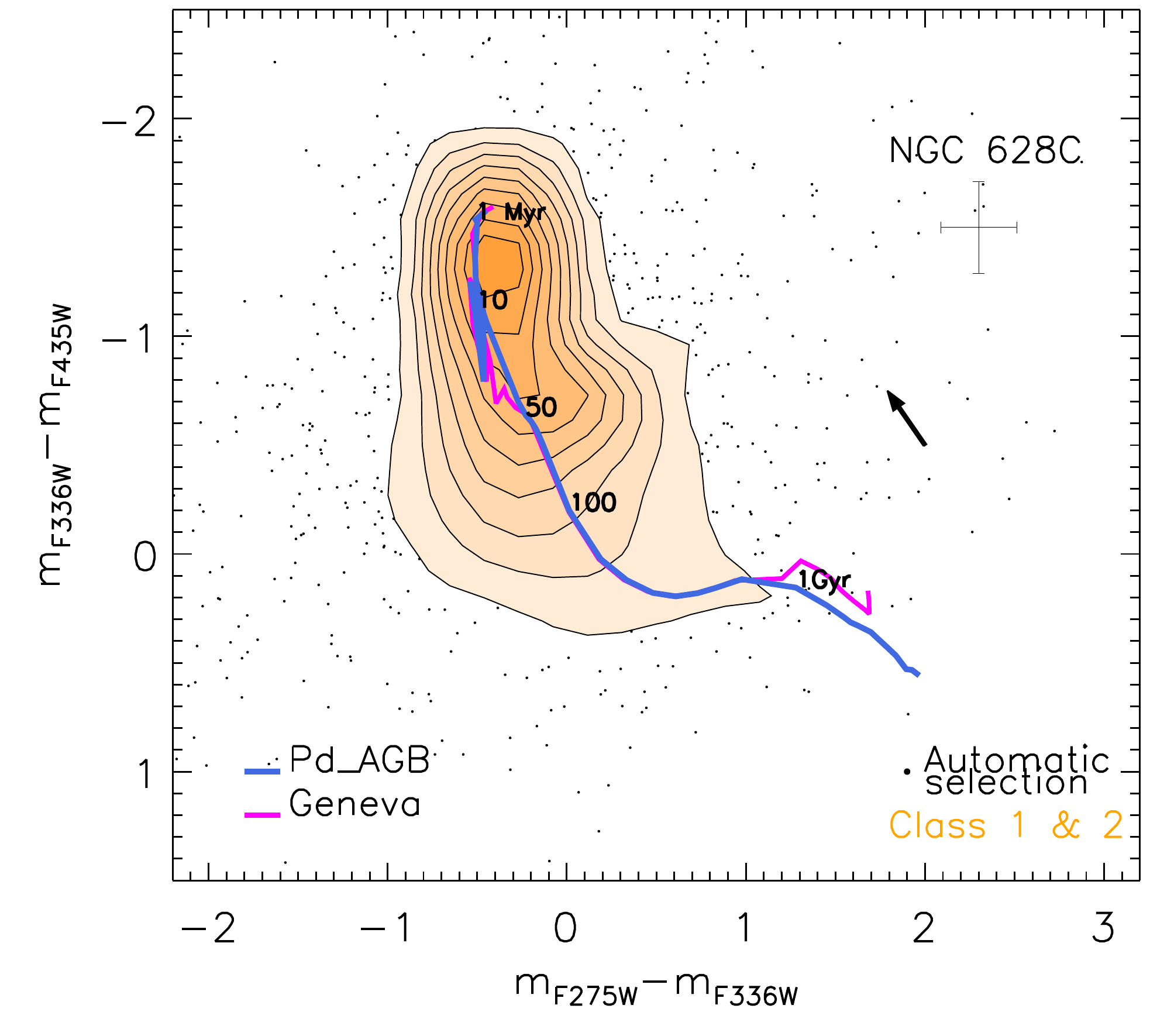}
		\includegraphics[scale=0.4]{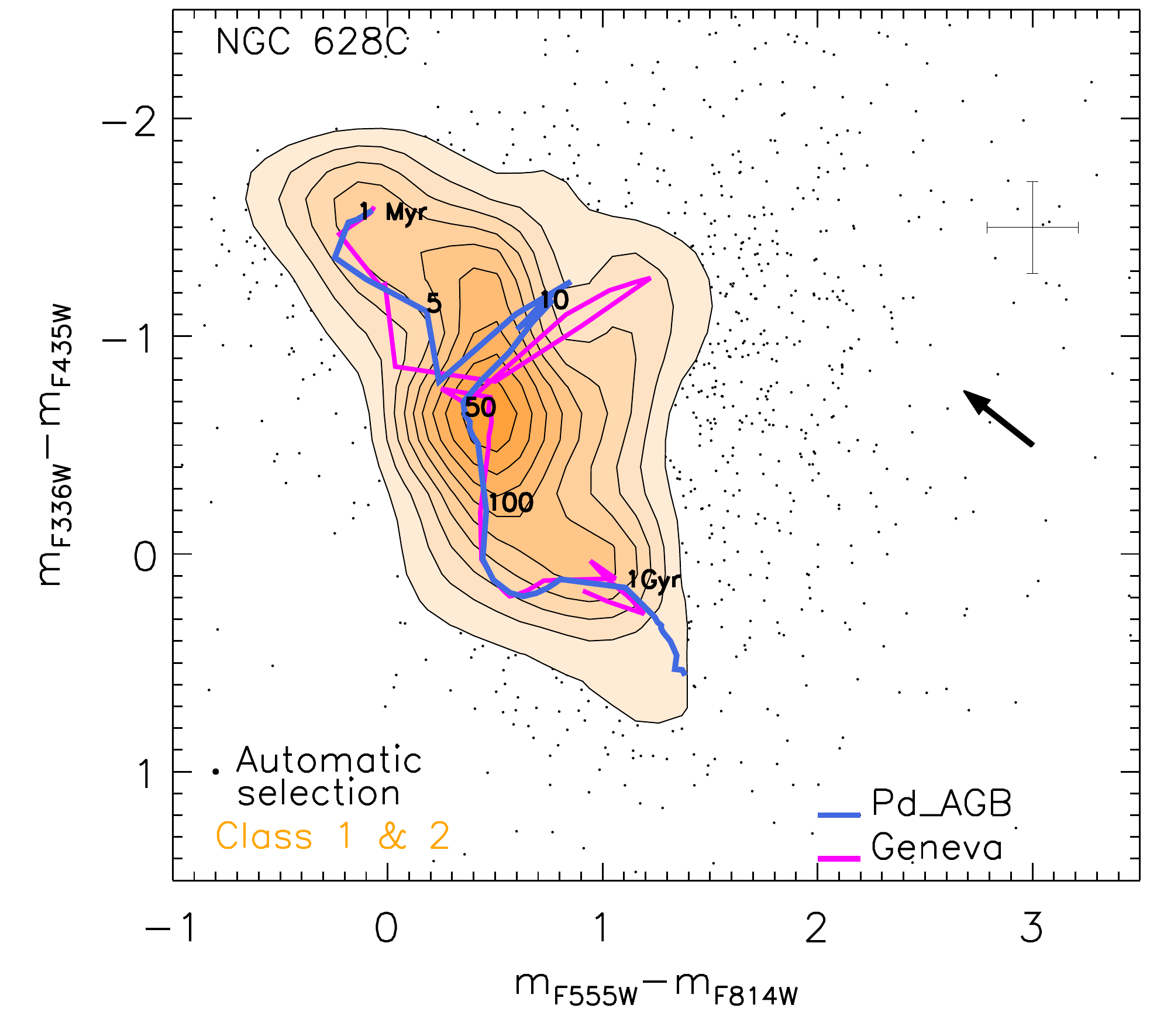}\\
		\includegraphics[scale=0.4]{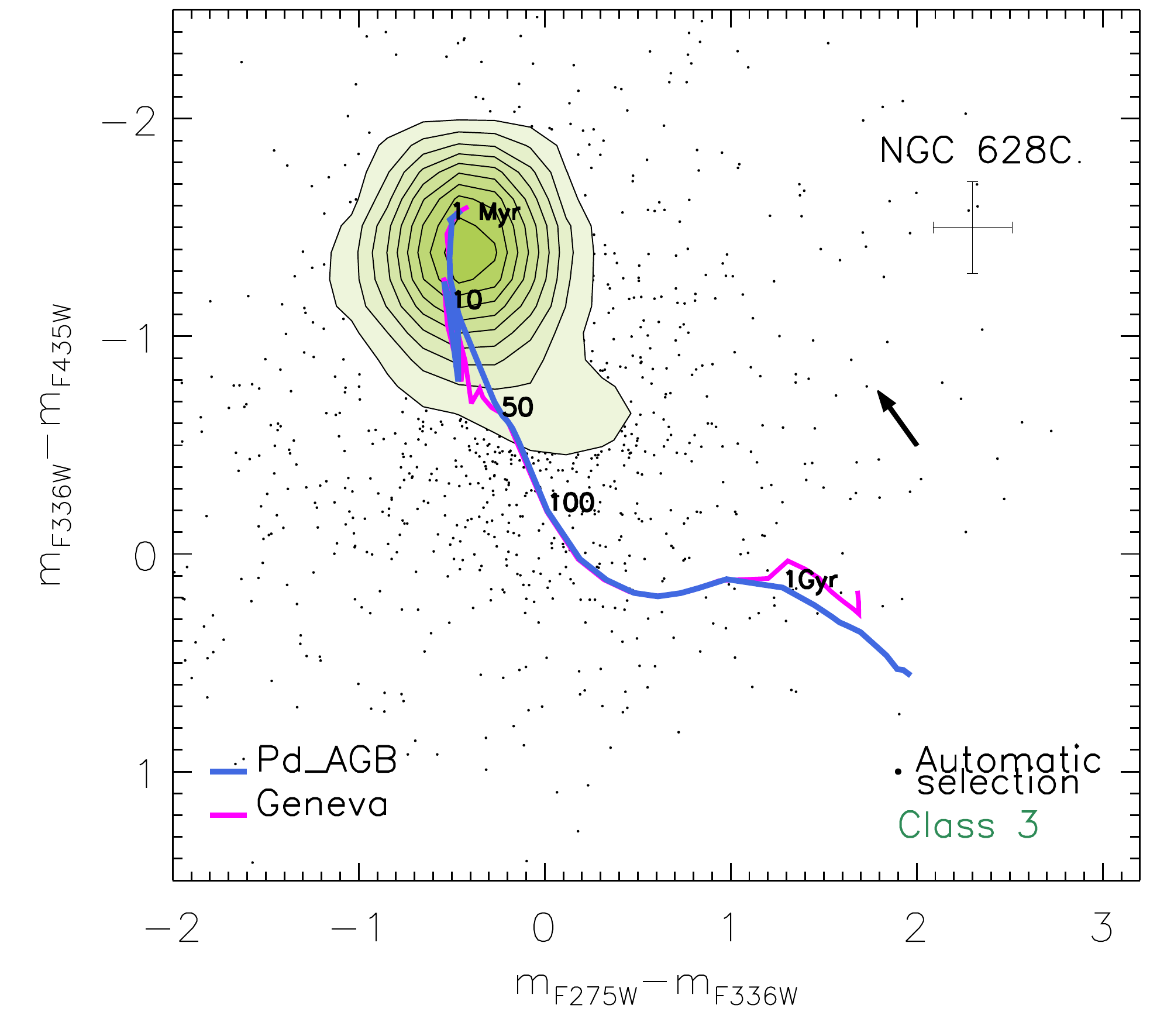}
		\includegraphics[scale=0.4]{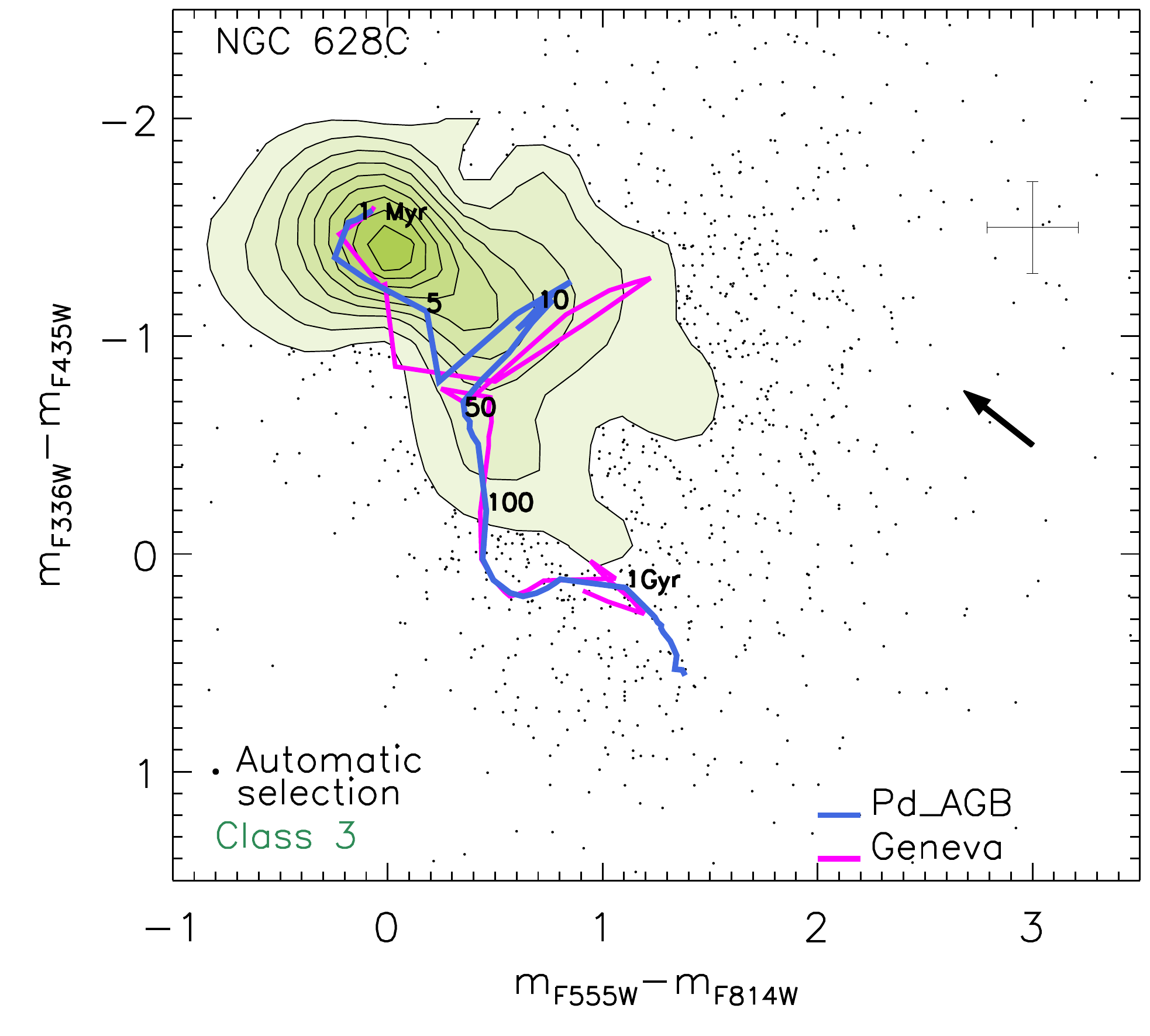}
    \caption{UV (left) and optical (right) color-color diagrams of the cluster population of NGC628c based on CI\_BASED catalogues. The dots show the location of all the sources contained in the CI\_BASED catalogue. The overlaid density number contours show the distribution of class 1 \& 2 clusters (top), and class 3 objects (bottom). See Figure~\ref{fig:fig7} for more details.}
    \label{fig:fig9}
\end{figure*}
In the upcoming analysis we use as a reference sample the YSC catalogue of NGC 628 obtained with photometry corrected by an average aperture correction, SED fits produced with the deterministic approach, assuming Padova stellar evolutionary models and solar metallicity tracks, and the Milky Way extinction law \citep{1989ApJ...345..245C} to take into account the internal reddening. The reference catalogues of NGC628c and NGC628e contain  3086 and 593 cluster candidates, respectively. Of these sources, roughly 1600 and 380 passed the selection required to move onto the next step of visual inspection (e.g., they have photometric errors $<$ 0.3 mag in all four bands and CI values $>$ than 1.4). In NGC628c (NGC628e), 334 (92) systems have been classified as class 1, 357 (80) as class 2, and 326 (87) as class 3. The remanning 583 (121) sources have been assigned class 4. 

The main difference between class 1 and class 2 clusters is linked to the apparent morphology of the systems, with the latter having elongated surface brightness. Ellipticity in the surface brightness profile of well resolved nearby YSCs has been measured in the MW and the Local Group \citep[e.g.][]{2012MNRAS.426.2427S}, but they find that it is not linked to any specific dynamical status of the cluster (e.g. expanding and/or disrupting systems). In the LEGUS galaxies, clusters are not fully resolved and our ability to recover their morphology is very much dependent of the distance of the galaxy and the crowding of the region. In \cite{2015ApJ...KG}, we compare the cluster physical properties such as ages, masses and extinctions of the three different classes (1,2,3) of systems identified in \obj. We report that the median age and mass of class 1 clusters are slightly higher than for the other two classes. This trend may be the result of the methodology we apply to classify cluster candidates. YSCs are born in cluster complexes and giant young star-forming regions. This implies that during the first $\sim 10$ Myr it is challenging to identify the real morphology of a cluster because of the complexity of the regions where clusters form (e.g. the Tarantula nebula in the LMC is a good nearby example). Conversely, we believe class 3 objects constitute a different type of system from class 1 and 2 objects. In particular, their multi-peak nature suggest these to be likely compact associations. In light of these considerations, in the rest of the analysis we will look at cluster physical properties unifying class 1 and 2 under the same group and comparing their properties to class 3. 

In Figure~\ref{fig:fig7} we show the UV and optical color-color diagrams of all the sources included in the automatic catalogue as dots, while the class 1 \& 2 clusters and class 3 systems from our reference (AV\_APCOR) catalogue are showed with contours of density numbers. The spread in colors of the automatically selected populations is clearly reduced after visual inspection. The color-color diagrams show that in the inner region of \obj, the cluster population is well distributed along the evolutionary track with peak of number densities of clusters at very young ages ($\sim$ 1-10 Myr) and between 50 and a several hundreds of Myr. We see a clear difference in the distribution of class 1 \& 2 clusters versus class 3, the latter is fairly concentrated in a region of the diagram corresponding to 1-10 Myr and it extends up to $\sim$ 50 Myr, with relatively few object at older ages. The difference between the age distributions of the classes of objects confirms that the morphological classification of class 3 objects, characterised by multiple peaks in their brightness profile and asymmetries in the light distribution, is probably selecting stellar associations that evaporate and disappear in short time scales in the galactic stellar field \citep[e.g.][]{2011MNRAS.410L...6G} and not gravitationally bound objects.

The age-mass diagram in Figure~\ref{fig:fig8} confirms the observed trend in the color properties of our clusters and associations. The majority of class 3 objects have ages below 20 Myr and smaller masses than class 1 \& 2 as already reported by \cite{2015ApJ...KG}. The maximum mass observed per age bin increases on average as a function of age, a result of the nearly constant SFR, the stochastic nature of the cluster formation process, and of the size-of-sample effect \citep[longer age intervals imply a higher chance to form more massive clusters, e.g.][]{2003AJ....126.1836H}. 

We also investigate the colors of the cluster population detected in the outer pointing of \obj. The color-color diagrams reported in the panels on the right of Figure~\ref{fig:fig7} show that the cluster population (class 1 and 2) in the outer region of NGC628 has a less significant population of very young clusters (the contour levels are very weak in the region corresponding to ages younger than 5 Myr) with respect to the inner pointing. We see that the outer field color distribution is dominated by a cluster population with ages between 50 and a few hundreds Myr. On the other hand the distribution of the class 3 associations behaves in a similar fashion in the two galactic regions, with their numbers quickly fading after 50 Myr. Similar behaviours in the YSC populations of the inner and outer regions have already been reported in the literature for M83 \citep[e.g.][]{2011MNRAS.417L...6B, 2014ApJ...787...17C} and can likely be linked to the role of the environment where YSCs are forming and evolving. The large sampling of different galactic environments provided by LEGUS will be a key to interpret these observational trends in the near future.

\subsection{A comparison between AV\_APCOR and CI\_BASED catalogues of NGC 628c}

In this section we compare the photometry and analysis performed on the CI\_BASED catalogue to our reference catalogue. For convenience we show only the analysis performed on the  NGC628c pointing, but the outcome is very similar for the outer field. 

The advantage of using average aperture corrections is that this method does not change the shape of the SEDs or the intrinsic colors of the clusters. As we see in Table~\ref{tab1}, the differences between the applied aperture corrections change only slightly between different bandpasses (i.e. it reflects the change in the PSF as a function of filter/camera). Therefore such a correction will change the normalisation of the cluster SED but not the shape. This method, however, does not take into account that, at the distance ranges of the LEGUS targets, clusters are partially resolved, that is their CSF changes as a function of the cluster size. However, since the intrinsic shape of the SED is preserved and only the normalisation is affected, this method may overestimate the mass of very compact clusters or underestimate the mass of the very extended sources, but the uncertainties will be within the 0.1-0.2 dex in logarithmic age and mass usually produced by the fitting method. 
\begin{figure}
		\includegraphics[scale=0.40]{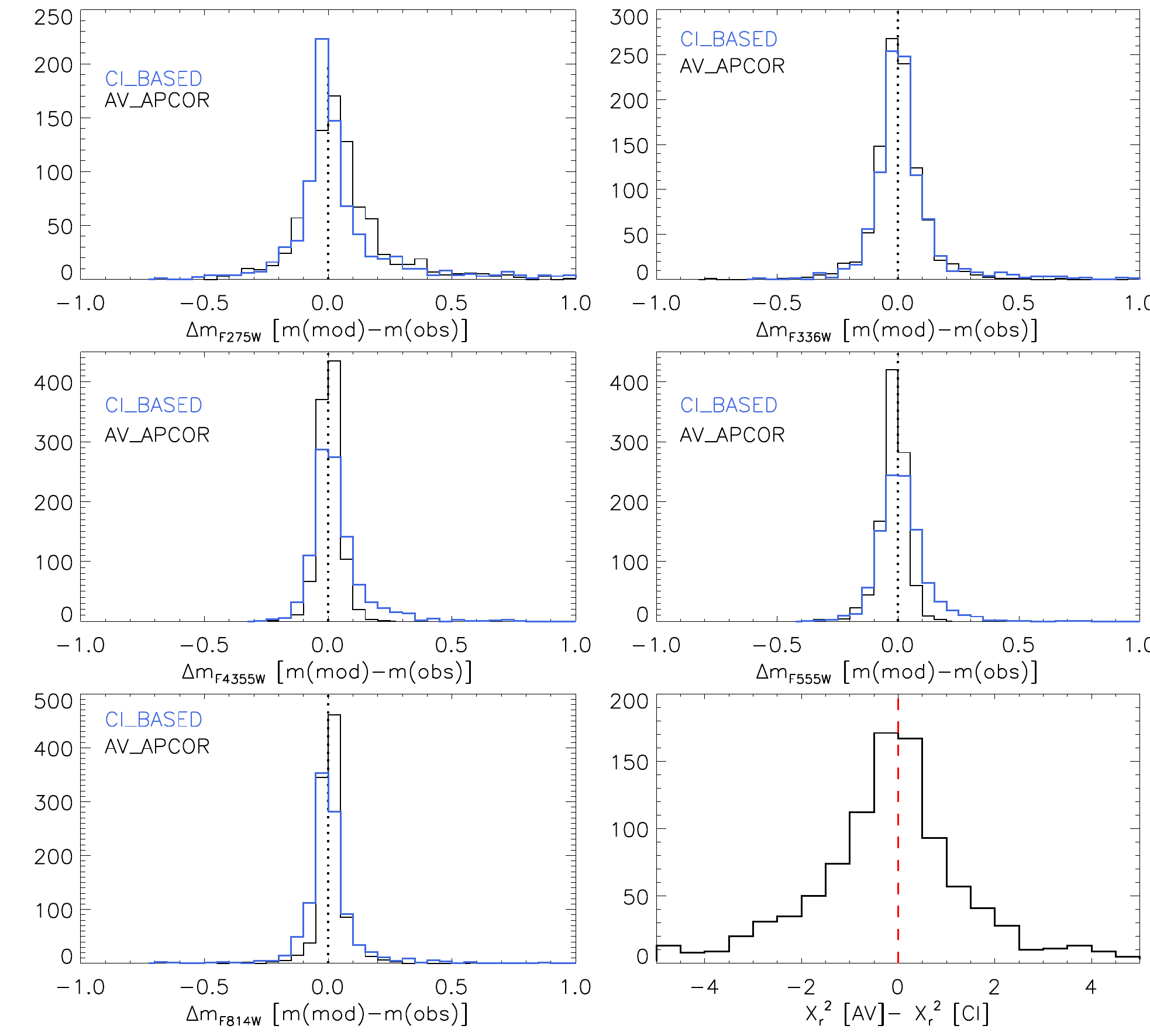}
    \caption{Residuals produced by the SED fitting analysis of both AV\_APCOR (black) and CI\_BASED (blue) catalogues of NGC 628c as a function of waveband. The bottom right panel shows the difference between the recovered reduced $\chi^2$ in the AV\_APCOR and CI\_BASED. Only class 1, 2, and 3 have been used to produce these distributions.}
    \label{fig:fig10}
\end{figure}
\begin{figure}
		\includegraphics[scale=0.40]{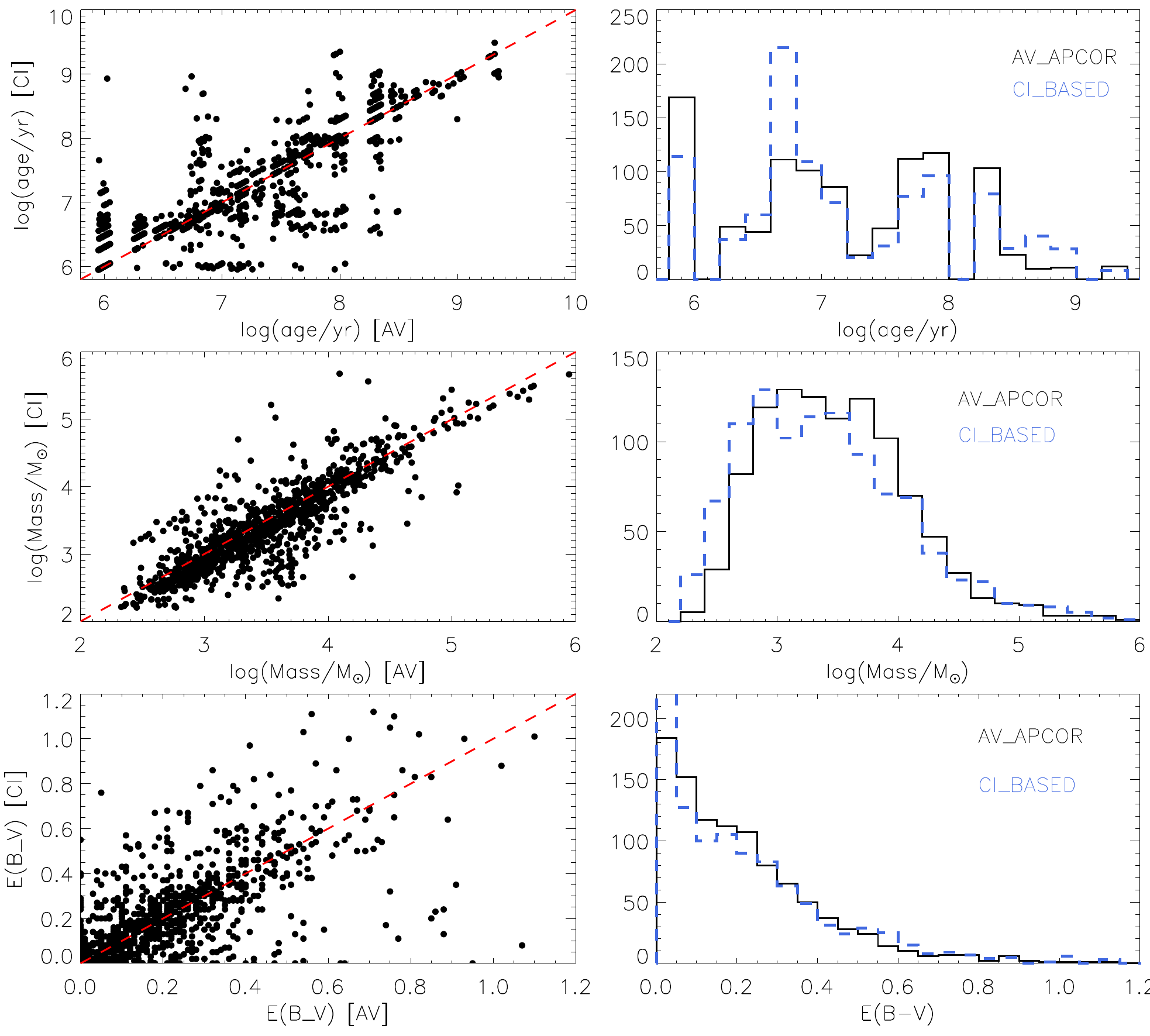}
    \caption{Recovered physical properties for class 1, 2,  and 3 for both AV\_APCOR (x-axes and black solid line histogram) and CI\_BASED (y-axes and blue dashed line histograms)  catalogues of NGC 628c. The red dashed lines in the left panels show the location of the one-to-one correlation.}
    \label{fig:fig11}
\end{figure}

The CI based aperture correction takes into account the relation between the size of the cluster and the required aperture correction. The relation has been derived in each band using simulations of YSCs of varying size (see Cook et al. in prep.). This method has the great advantage of taking into account the cluster sizes. The limitation, however, resides in the uncertainties produced by the CI estimates as a function of wavelength. This means that the CI based method not only changes the normalisation of the SED but also the shape. Therefore, the uncertainties propagate in the estimated mass, age and extinction of the source.

In Figure~\ref{fig:fig9} we show the same UV and optical color-color diagrams but derived using the photometry of the  CI\_BASED catalogue. We notice that while the overall location of the clusters and associations is similar in the two photometric catalogues, the contours of the CI\_BASED photometry are more extended on the left side of the tracks, a spread that is larger than the photometric uncertainties. In general, one expects sources identified as clusters to diffuse on the right side of the evolutionary tracks because of reddening, while the spread on the left side is mainly produced by photometric errors and uncertainties in the calibration.

In Figure~\ref{fig:fig10}, the analysis of the residuals (i.e. the difference between the observed  and the best model integrated fluxes) of class 1, 2, and 3 clusters in each band does not show significant differences in the residuals of the F336W filter in both AV\_APCOR and CI\_BASED catalogues. The residuals of the F275W filters in the CI\_BASED catalogue are more concentrated around zero (the best match) than the AV\_APCOR ones. The opposite trend is observed in the AV\_APCOR residuals of the $BVI$ filters, where they show a narrower distribution around the best match than the CI\_BASED ones. The latter are less peaked and show a tail towards positive values (i.e. the observed magnitude is brighter than the one predicted by the best model) in the $B$ and $V$ bands and negative values in the $I$ band. Overall, we see that the differences between the reduced $\chi^2$ obtained from the fit performed to the photometry of the AV\_APCOR and CI\_BASED catalogues show a more significant negative tail suggesting that the fit to the CI\_BASED photometry is slightly worse.

\begin{figure*}
\centering
		\includegraphics[scale=0.4]{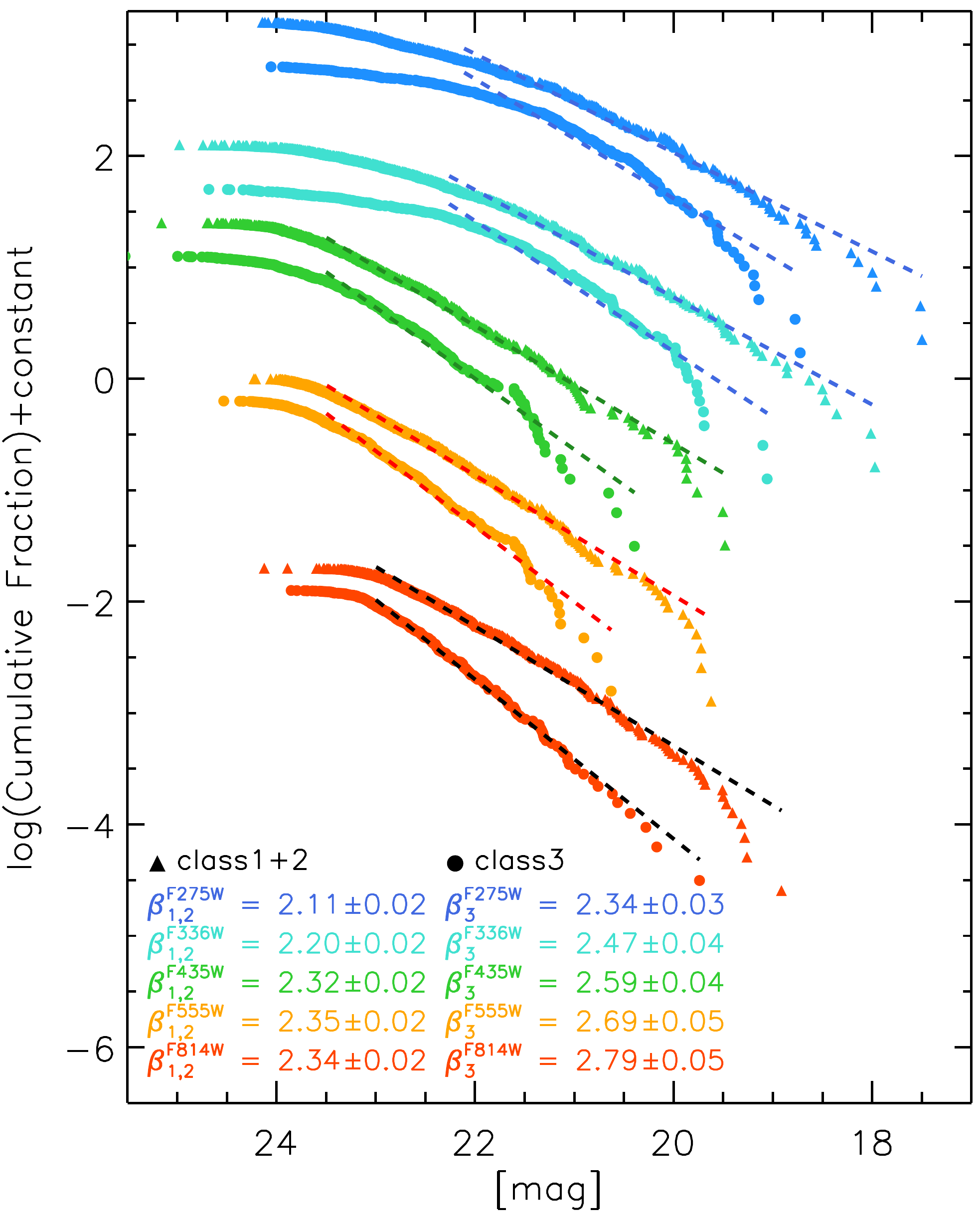}
		\includegraphics[scale=0.40]{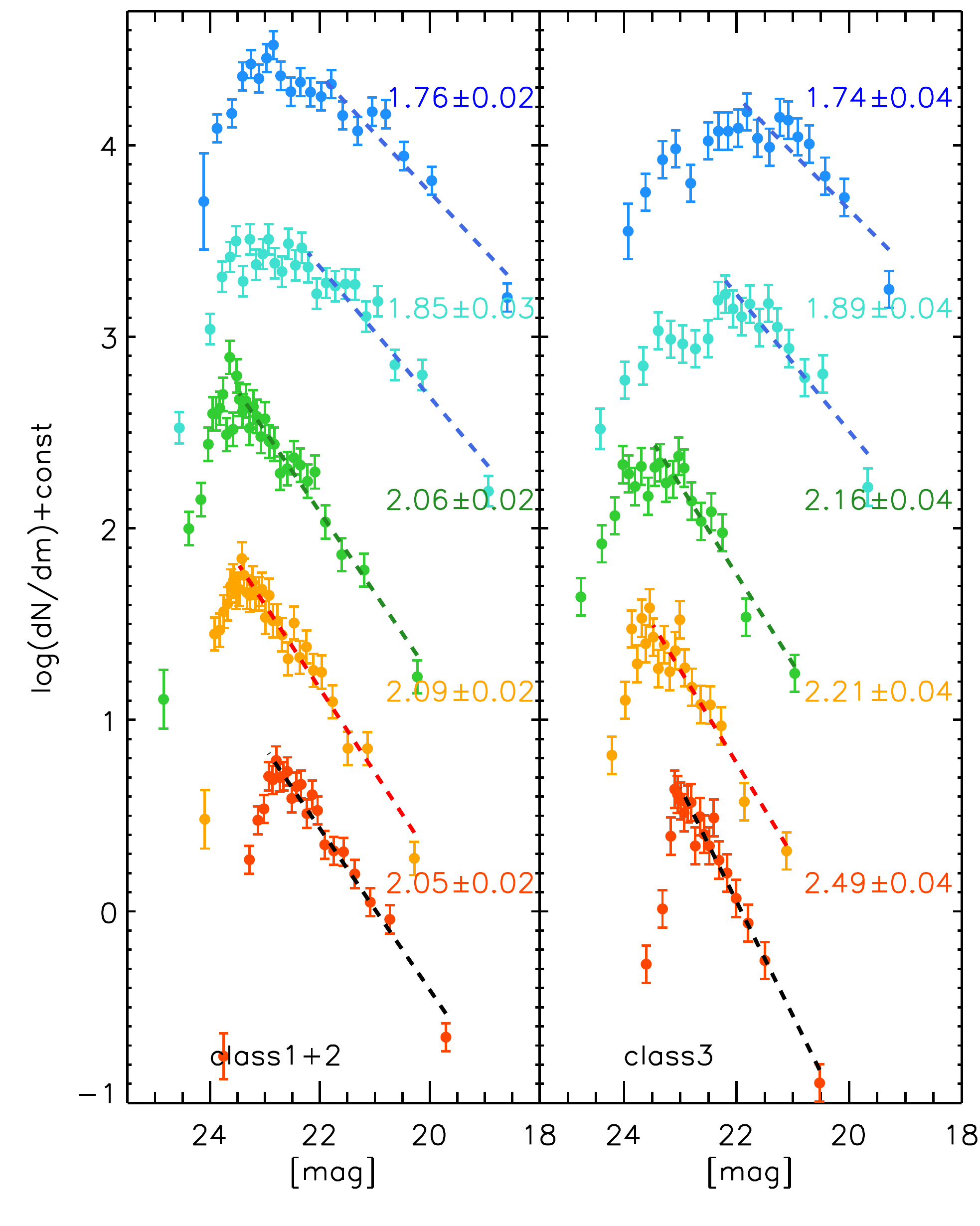}\\
    \caption{The luminosity function of the whole cluster population of NGC628 in the 5 standard LEGUS bands. From top to the bottom, we plot the CLFs obtained in the $UVUBVI$ filters. On the left panel, we fit cumulative distributions of the magnitudes of class 1 \& 2 and class 3, the recovered slopes are listed in the insets. On the right plot we fit distributions of bins containing the same number of objects. The plot consists of two panels, the left one shows the distributions for class 1 \& 2, the right one for class 3. The fit has been performed including the brightest object (bin) down to the system (bin) with magnitude comparable to the detection limits listed in Table 1.}
    \label{fig:fig12}
\end{figure*}
A direct comparison of the recovered ages, masses, and extinctions of class 1, 2, and 3 objects of the AV\_APCOR and CI\_BASED catalogues is shown in Figure~\ref{fig:fig11}. 
Although on average we see correspondence along the one-to-one line, some deviations are also significant. The most important difference in the CI\_BASED catalogue is the appearance of the pronounced peak at around 5 Myr visible in the top right distribution of Figure~\ref{fig:fig11}. According to the top left panel, these systems have been assigned in the AV\_APCOR both younger, similar, or older ages. Since the mass-to-light ratio is smaller at younger ages, the objects that become younger in the CI\_BASED analysis will also have smaller stellar mass, therefore, we can explain why the AV\_APCOR and CI\_BASED mass histograms (central right panel of Figure~\ref{fig:fig11}) differ in the mass bins between 1000 and $10^4$ \msun. In the CI\_BASED catalogue these objects have been assigned masses $\leq10^3$ \msun. In total we estimate that about 20 \% of the class 1, 2 and 3 systems that are in common in the two catalogues have differences in ages larger than 0.1 dex, which is the average uncertainty recovered for age estimates obtained with deterministic methods. 
This small fraction of deviating sources may explain why we do not see significant differences in the recovered CLFs, CMFs, and disruption rates (Section 5) of the cluster population in \obj\, when AV\_APCOR and CI\_BASED catalogues are used. Further investigation of these two approaches is presented in Cook et al. (in prep.). In the next Section we will report the results of the analysis performed on the YSC population of \obj\, using only the AV\_APCOR reference catalogue.

\section{Constraining the formation and evolution of clusters and associations in NGC628}

In Section 4, we have observed that the photometric properties of clusters (class 1 \& 2, in our analysis) and compact associations (class 3) show significant differences, with the latter class disappearing from the regions of the color-color diagrams occupied by more evolved stellar populations. In the following sections we analyse and compare physical and statistical properties of class 1 \& 2  and class 3 objects separately. The aim is to probe differences and analogies between these two types of stellar objects that can help us to understand their formation process and evolution. 

\subsection{Multiwavelength analysis of the CLF of likely bound and unbound systems \label{luminosity}}

The luminosity function of YSCs is typically described as a power-law function in the luminosity space, $dN/dL\propto L^{-\beta}$. However, we fit the function in a logarithmic space, where the luminosity is replaced by the magnitude, so that $\log [d(N)/dM ] \propto \theta\times M$, where $\beta=2.5\times\theta+1$ \citep[e.g.][]{2002AJ....124..147W, 2008A&A...487..937H}. 

We apply two different techniques, equally used in the literature, to analyse the CLF of NGC628. In Figure~\ref{fig:fig12} we report the observed luminosity functions in each band, using cumulative distributions (left panel), like in \citet{2012MNRAS.419.2606B}; and bins containing the same number of objects (right panel), following the \citet{2005ApJ...629..873M} approach. 

 The luminosity properties of the clusters are directly observable, not affected by age or mass determinations. However, their luminosity distributions depend on both the detection limits of our datasets and by the adopted extraction procedure combined with the selection criteria we impose to yield the final catalogue. To build the CLF we select only sources that  have been visually classified as class 1, 2, or 3 (thus they are brighter than -6.0 mag in $V$ band and detected in four filters with photometric error smaller than 0.3 mag) and are younger than 200 Myr. In Figure~\ref{fig:fig13} we show the age-mass diagnostic diagram including the recovered 90\% detection limits in the 4 bands required for detection and the $V$ band cut at $-6.0$ mag, all converted to limiting masses as a function of age. We observe that the $V$ band cut applied for the visual inspection is more conservative than the detection limits of our dataset. However, we notice here that the resulting flattening of the distributions at the low luminosity ends is produced by the combination of a sharp magnitude cut combined with detection limits of the science frames and the method used to produce the final position catalogue of cluster candidates. The age limit of 200 Myr enables us to directly compare the CLF to the CMF (see Section 5.3). In total we count 733 (370) class 1 \& 2 (class 3 numbers are indicated between brackets) objects in the F275W filter, 846 (404) in the $U$ band, 851 (408) in $BVI$ bands before the age cut. After the age cut is applied we are left with 703 (369), 778 (397), and 783 (401) class 1 \& 2 (class 3) objects in $UV$, $U$, and $BVI$ bands, respectively. To prevent that incompleteness affects our analysis we perform the fit of the binned and cumulative distributions from the brightest bin (object) down to the bin (object) with a magnitude brighter than 22.12, 22.26, 23.50, 23.50, 23.0 mag in $UV$, $U$, $B$, $V$, and $I$ respectively. These limiting values have been chosen to avoid the shallower regions of the distributions and are more conservative than the 90\% completeness magnitudes reported in Table~\ref{tab1}.

In the case of an equal number of object bins, the slopes and the associated uncertainties are produced by the IDL package LINFIT, which takes into account the weighted error associated with each bin. In the case of the cumulative functions the error analysis has been performed with bootstrapping techniques. To take into account how the photometric uncertainty associated with each point affects the final recovered slope, we perform Monte Carlo realisations of 1000 cumulative distributions. With each observed magnitude we associate an uncertainty extracted from a gaussian distribution with standard deviation equal to the maximum tolerated error of 0.3 mag. Each cumulative realisation is thus fitted. The final error associated with the observed slope is the standard deviation of all the recovered indexes.

In general both methods produce slopes that are close to an index of $-2$. We notice that the recovered slopes for the binned data are shallower than in the case of cumulative distributions. This is mainly the result of the differences between the two techniques (see Section 5.3), hence slopes  determined using binned and cumulative distributions cannot be directly compared. Some important features are, however, observable in both analyses. We find a clear steepening as a function of increasing wavelength, i.e. the recovered slopes become significantly steeper than $-2$ in the $BVI$ filters. This is true for both class 1 \& 2 clusters and class 3 associations  (see insets in Figure~\ref{fig:fig12}). The distributions of the two bluest filters show an extended flat peak that in the cumulative distributions appears as a significant curvature. Moreover, the cumulative distributions at all wavelengths show a clear steepening at the bright magnitude ends. Such steepening is not observed in the binned distributions  because while a variable-size binning technique mitigates biases introduced by equally-spaced binning approaches \citep{2005ApJ...629..873M}, it also tends to wash out small-scale variations  (see Section 5.3). Because of the small number of very luminous objects the brightest bin of the CLF has a width of about 1.5--2 mag encompassing the range where the steepening in the cumulative function is observed. Our simulations, in Section 5.3, show that the departures from a single power law function happens mainly at the bright (massive) end, thus the cumulative analysis in better suited to investigate the shape of the luminosity and mass functions.

The luminosity function is a direct observable of the underlying mass function integrated over time. The trends we observe in the CLF of NGC628 suggest a dearth of luminous (massive) clusters/associations thus simultaneously analysing  the CLF and CMF can tell us something important about the how clusters form and evolve in this galaxy.
 
\subsection{The analysis of the cluster mass function}

\begin{figure}
\centering
		\includegraphics[scale=0.43]{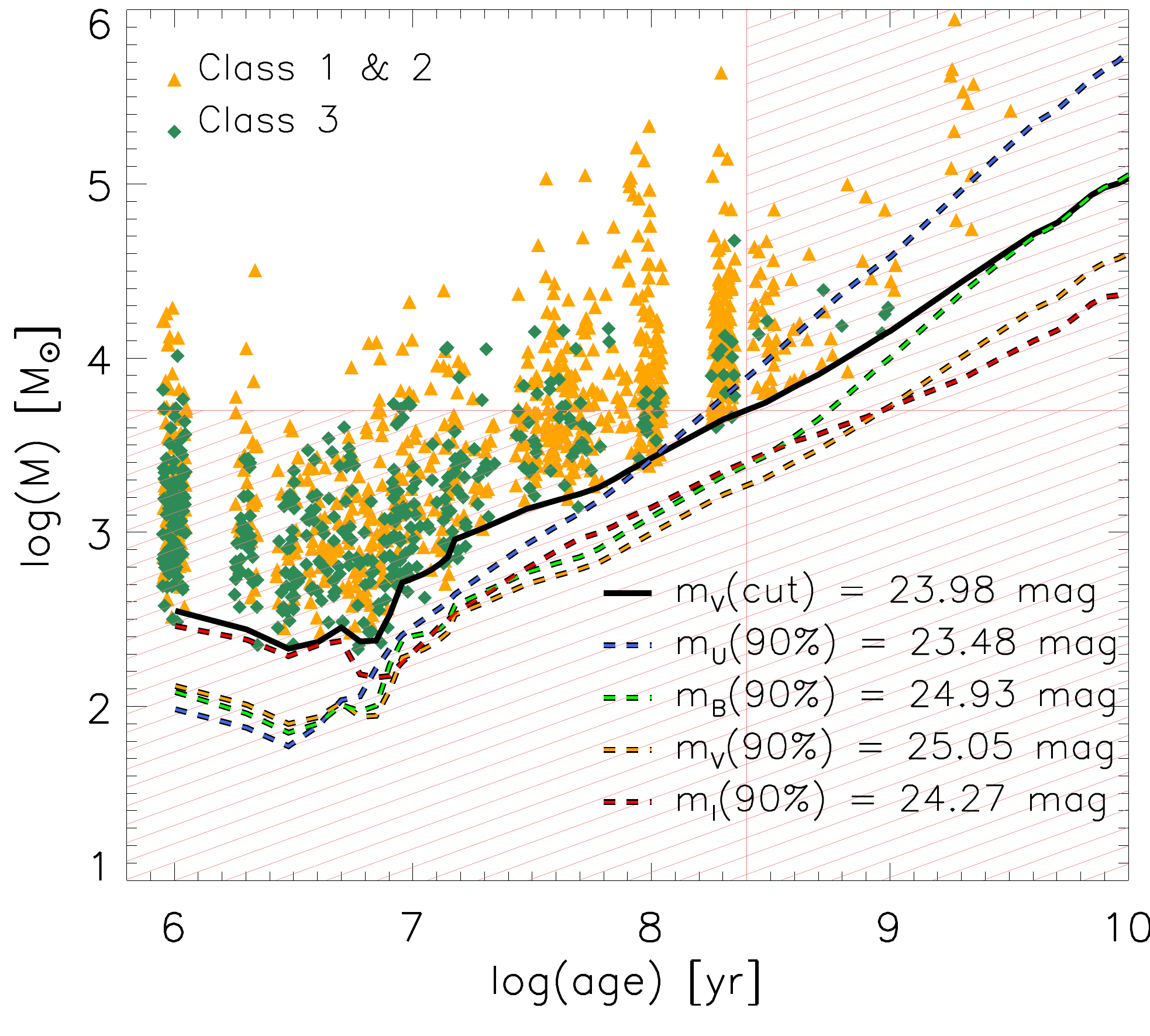}
    \caption{The age-mass diagram of the cluster and association populations of both the inner and outer pointing of NGC628. The magnitudes corresponding to 90 \% completeness limits (see Table~\ref{tab1}) in the four bands required for the analysis have been converted into mass limits as a function of age using Yggdrasil models. We also include the detection limits in age and mass imposed by the $M_V < -6$ mag selection criterion. The latter cut ensures we are above 90\% recovery in all the four bands for masses above 5000 \msun and ages up to about 200 Myr. The shadowed areas show which part of the sample has been excluded in the analysis of the CMF and disruption rates.} 
    \label{fig:fig13}
\end{figure}

To derive masses we need to make an extra step where ages and internal reddening of the sources are extracted. As discussed in K15, the stellar physical properties derived with our deterministic method are severely biased at very low masses due to stochastic variations from the small number of stars.. Figure 14 of K15 shows the one-to-one comparison between Yggdrasil deterministic and SLUG stochastically derived cluster properties suggesting that important deviations occur at cluster masses below 5000 \msun. This mass limit was, in recent years, widely adopted in the YSC analysis based on deterministic approaches. We assume the same mass limit in our current analysis. Figure~\ref{fig:fig13} shows the age-mass diagram of the clusters/associations in our two HST pointings of NGC628. A mass cut of 5000 \msun\, gives us complete detection up to a stellar age of 200 Myr. Sources falling within the shadowed areas are, thus, not included in the analysis of the CMF and disruption rates presented hereafter. In total we count 320 class 1 \& 2 clusters and 42 class 3 objects that pass this mass and age selection. 
 We notice that our mass cut is very close to the completeness limits of our catalogues at the last age bin at 200 Myr. We tested whether using a higher mass cut at 8000 \msun\, produces different outcomes. We do not see any change in the recovered CMF properties, only higher errors because of the smaller number of clusters available for the analysis. 

\begin{figure}
\centering
		\includegraphics[scale=0.45]{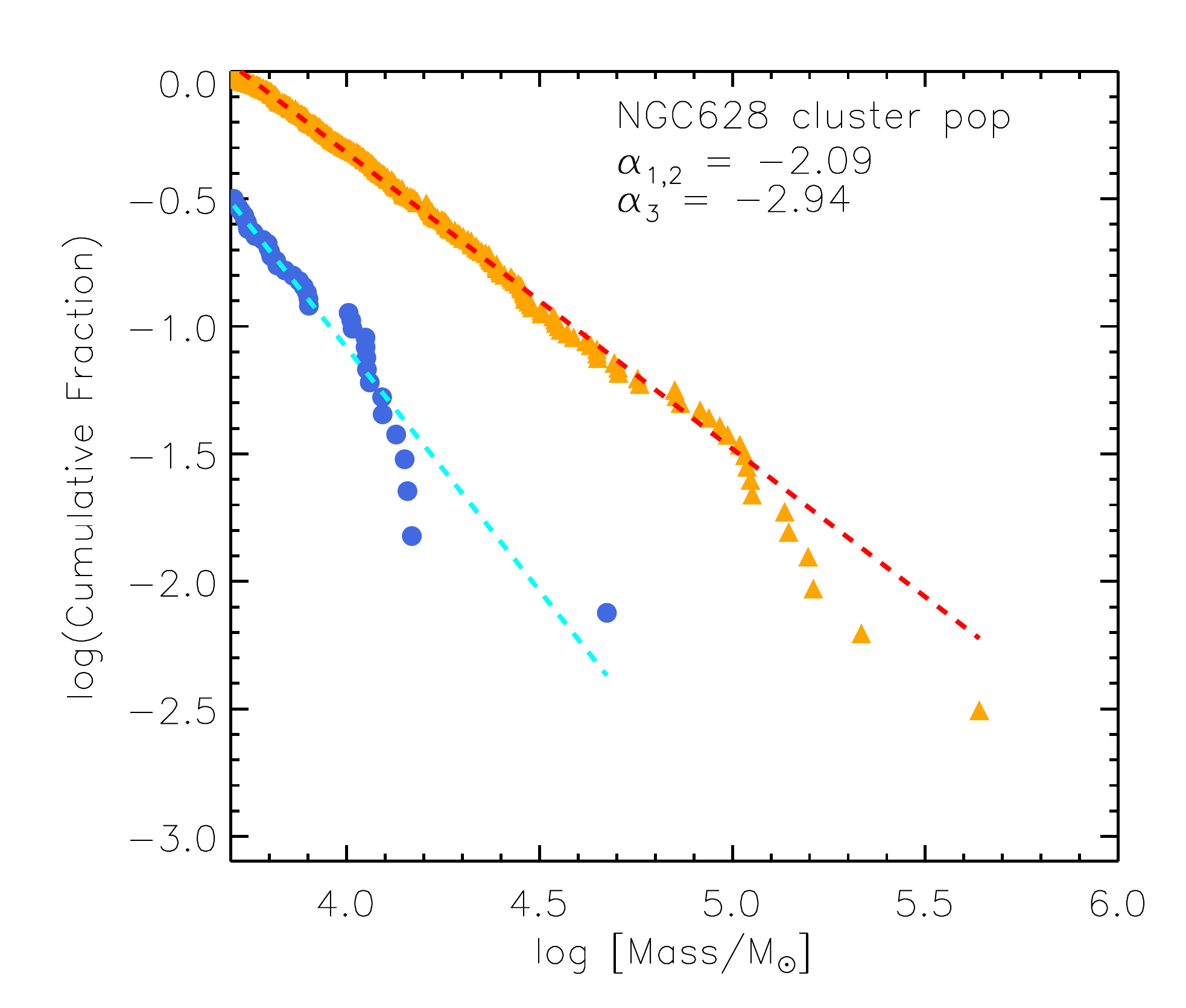}\\
		\includegraphics[scale=0.45]{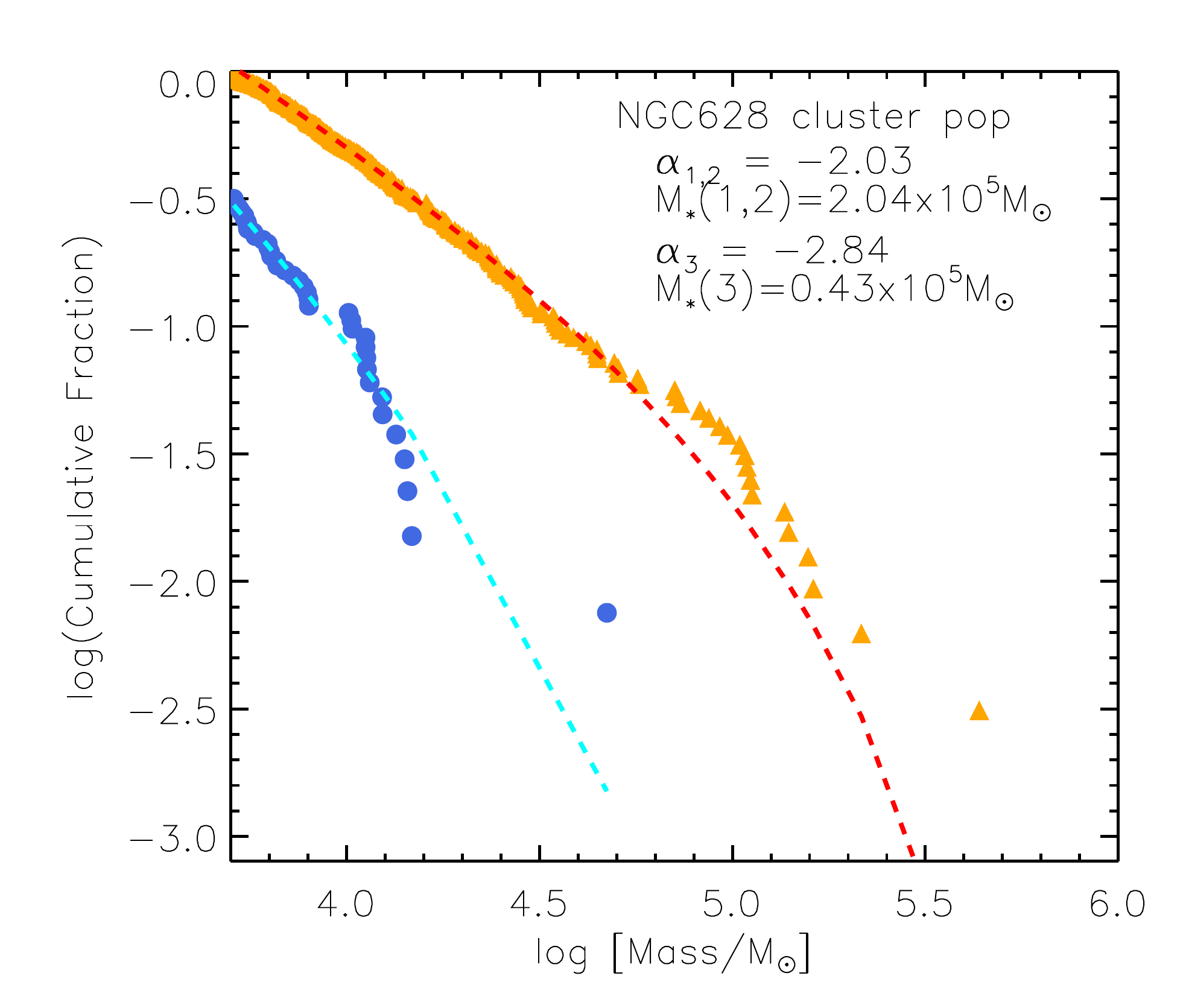}
    \caption{Cumulative mass functions of class 1 \& 2 (orange triangles) and class 3 (blue dots) systems. The distributions have been created only with objects younger than 200 Myr and the fit includes only systems more massive than 5000 \msun. The recovered slopes for the two subpopulations are reported in the inset.}
    \label{fig:fig14}
\end{figure}

The observed mass function of class 1 \& 2 (orange triangles) and class 3 (blue dots) systems are shown as cumulative distributions in Figure~\ref{fig:fig14}. The observed cumulative distributions are fitted using the IDL  maximum-likelihood fitting package {\tt MSPECFIT} \citep{2005PASP..117.1403R}. We perform two different fits to the cumulative distributions, a single power-law function, in the integral form $N(M' > M) \propto M^{\alpha}$, and a power law function with a truncation at the upper-mass end, i.e. $N(M' > M) \propto N_0[(M/M_{\star})^{\alpha}-1]$ \citep[see][and references therein for a complete discussion of the formalism]{2005PASP..117.1403R}. The resulting fitted parameters for the two functions (single power law  in the top panel, truncated power law bottom panel) are included in the insets of Figure~\ref{fig:fig14}. In Table~\ref{tab:massfunction} we list the recovered values for the class 1 \& 2 population. $M_{\star}$, the index $\alpha_{SF}$, and $N_0$, that is the number of objects more massive than $2^{1/\alpha}M_{\star}$ are determined for a truncated function, while the pure power-law fit provides the index $\alpha_{PLF}$. As described in \citet{2005PASP..117.1403R} if the resulting $N_0$ is significantly large than 1 then a truncated CMF form is preferred to the more traditional single power-law function. When $N_0 < 1$ the truncation mass is unconstrained thus a single power-law fit is sufficient. In Table~\ref{tab:massfunction} we also include errors. The errors associated with the observed maximum cluster mass, M$_{max}$, and fifth most massive cluster mass, M$_{max}^{5th}$ have been computed during the SED fitting procedure and described in Section 3. The errors associated with the best fitting parameters have been computed using deviations from  1000 iterations of bootstrap trials. 

In general, we observe that both a very steep single power-law fit and a truncated function fit with a slightly flatter index can reproduce the observed mass distribution for class 3 objects. However, the number of associations is very small (42) thus is not possible to impose any further constraint.

On the other hand, the analysis of the mass distributions of class 1 and 2 systems yields, for both a single power law and a truncated function type fits, slopes very close to $-2$.  However, as already noticed during the analysis of the CLF, the approximation of the CMF by a single power law function (see top panel of Figure~\ref{fig:fig14}) overestimates the expected number of clusters at the upper mass end of the distribution.  A fit to the observed CMF of class 1 \& 2 with a truncated power-law function (bottom panel of Figure~\ref{fig:fig14}) yields a similar slope to but it mitigates the differences at the high mass end of the CMF distribution. The resulting $N_0$ (see value listed in Table~\ref{tab:massfunction}) is larger than 1 suggesting that the latter function provides a better fit to the observed CMF.  Thus a truncated function with slope $\alpha_{SF} = -2.03$ and 
\mstar$\sim2.0 \times 10^5$ \msun\, is the statistically favoured description of the observed CMF of NGC628. However it is important to notice that the number of clusters more massive than $5\times 10^4$ \msun\, is about 22 and only half of those clusters are more massive than $10^5$ \msun\, so the constraint on \mstar\, is weak and the uncertainties on $N_0$ large. 

As exercise we try to estimate the expected number of clusters more massive than \mstar. Using the combination of far-UV and 24 $\mu$m fluxes of the area covered by the LEGUS pointings of NGC628, we estimate a SFR of about 0.59 \msun$/yr$. Assuming that the SFR was constant for the last 200 Myr we estimate that a total stellar mass of $1.18\times10^8$ \msun\, has been formed in the region. Using the cluster formation efficiency definition given in \citet{2015MNRAS.452..246A} and clusters in the age range between 1 and 100 Myr (same as the age range to which the estimated SFR is sensitive to), we derive for this region of NGC628 a cluster formation efficiency of 12\%. This means that 12\% of the total stellar mass of $1.18\times10^8$ \msun\, is in bound clusters, i.e. $1.42\times 10^7$ \msun. Using the latter amount as the total stellar mass in clusters, we can estimate the number of clusters more massive than \mstar. Observationally we find 2 clusters more massive than \mstar. Assuming a pure power-law mass function of slope $-2.09$ (with upper mass $1.\times 10^7$ \msun) we estimate that 5 clusters more massive than \mstar\, should have formed in the last 200 Myr. A Schechter type function, as described by equation 3, results in 1 cluster more massive than \mstar. The estimated total stellar mass in clusters results in cluster numbers that are consistent with the observed ones but does not produce any definitive proof that can help to discern the real shape of the upper mass function. Therefore, the solution provided by a pure power-law function with slope $-2.09$ cannot be discarded.

\begin{figure}
\centering
		\includegraphics[scale=0.43]{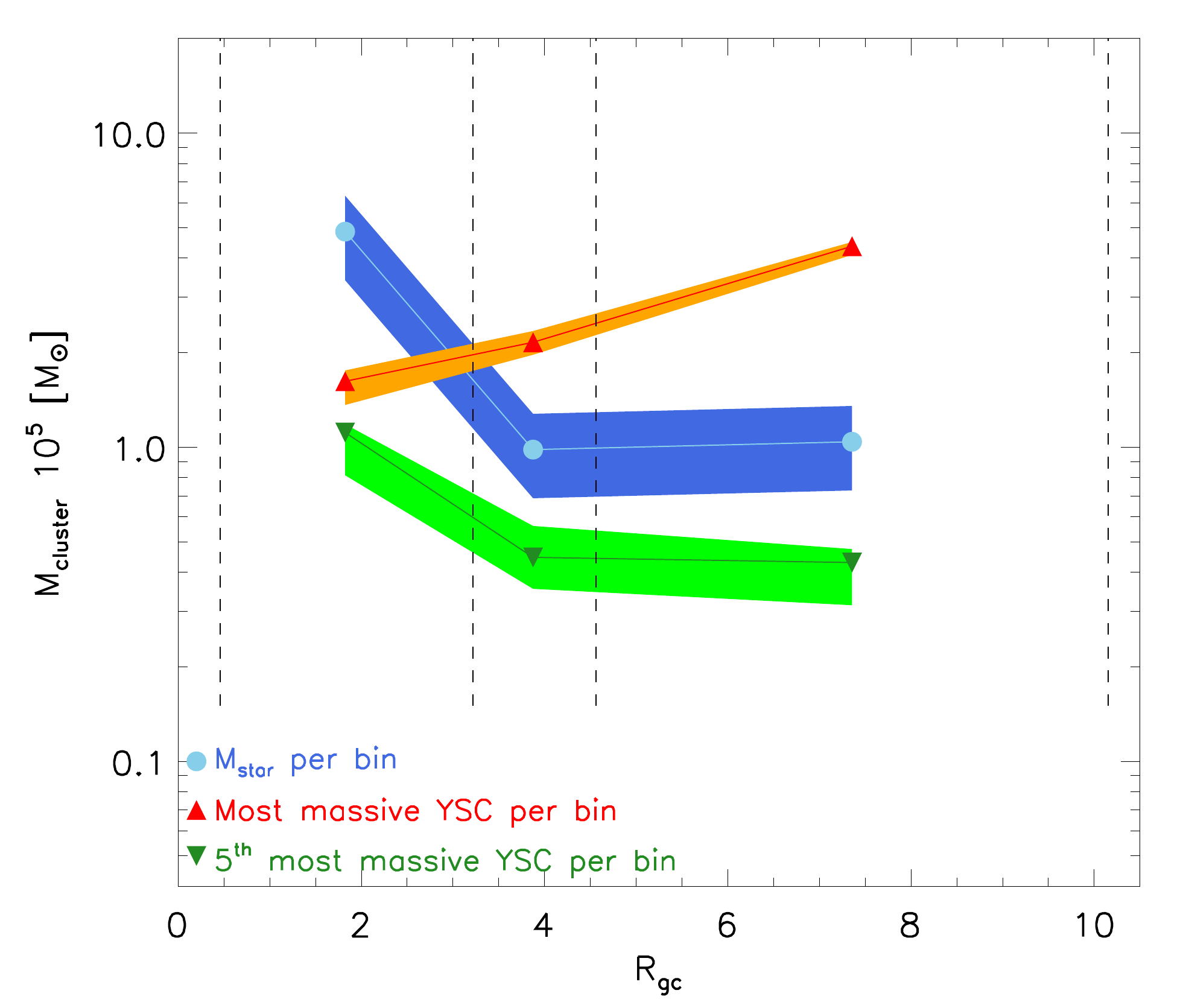}
    \caption{Comparison between the estimated truncation mass, \mstar (cyan dots), and the observed mass of the most massive cluster (red triangles) and fifth most massive cluster (green downward triangles) in bins of increasing galactocentric distance. The vertical dashed lines show the width of the bins, which were created to contain the same total number of class 1$+$2. The shadowed areas show the uncertainties associated with each derived \mstar\, and the errors on the mass estimates obtained with the SED fitting procedure described in Section 3.}
    \label{fig:fig16}
\end{figure}

We also test whether there is any variation in the CMF properties as a function of galactocentric distance performing a radial analysis of the CMF for the class 1 \& 2 clusters. In total our sample contains 320 objects more massive than 5000 \msun\, and younger than 200 Myr. We divide the sample in 3 radial bins of increasing distance from the centre of the galaxy, each containing the same number of objects so that we remove the size--of--sample effect. We then determine the \mstar, $N_0$, $\alpha_{SF}$, and $\alpha_{PLF}$ of the mass function of each bin using both a truncated and a more traditional power law function as described above. The recovered values are listed in Table~\ref{tab:massfunction}. In bin 1 the recovered \mstar is larger than the most massive cluster observed in the bin and the resulting $N_0$, including the uncertainties, is very close to 1. This means that the shape of the upper mass end of the CMF in the inner bin remains unconstrained and the solution with a single power law fit is as likely. In the second and third bins, between 3 and 10 kpc,  \mstar\, is consistently $1\times10^5$ \msun\, and does not decline significantly. The recovered $N_0$ is larger than one but the uncertainties are large too. In Figure~\ref{fig:fig16}, we plot the derived \mstar\, (cyan dots and blues bands including the uncertainties), the masses and uncertainties of the most massive, M$_{max}$ (red triangles and orange bands), and the fifth most massive cluster, M$_{max}^{5th}$ (green triangles and bands),  observed in each bin. The derived \mstar\, decreases significantly between the  innermost and the other two bins.  In the two outermost bins the observed masses of the most massive clusters are significantly more massive than the constrained \mstar, but the numbers of clusters close to the determined \mstar\, are small which makes the statistics (e.g. $N_0$) quite uncertain. 

Overall, we observe that if the recovered $N_0$ is larger than 1 the analysed cluster population has formed at least a few clusters with masses close to and larger than the truncation value, \mstar. This behaviour suggests that \mstar\, is not a sharp truncation and that the mass function is likely stochastically sampled. In the literature, it has been suggested that the CMF of young star cluster populations in local galaxies can be described with a Schechter function of slope close to $-2$ and a rapid exponential decline above a certain truncation mass \citep[][]{2006A&A...450..129G, 2009A&A...494..539L, 2012MNRAS.419.2606B}. The Schechter function differs from a pure power-law function only at the upper mass end of the distributions and could in principle explain the disagreement between the low numbers of observed massive clusters with respect to the expected one from the extrapolation of a pure power law function. The probability to form clusters more massive than \mstar\, declines exponentially but is not null.

We notice that the \mstar\, of class 1 \& 2 systems recovered for NGC628 is very similar to the one retrieved for M83 \citep{2015MNRAS.452..246A}, in agreement with the evidence presented in \cite{2009A&A...494..539L}, who suggested that \mstar\, in local spiral galaxies is about a few times $10^5$ \msun.

\begin{deluxetable*}{ccccccccccc}

  \tabletypesize{\scriptsize}

  \tablecaption{Parameters describing the cluster mass function in NGC\,628\tablenotemark{a} \label{tab:massfunction}}
  \tablewidth{0pt}

  \tablehead{
    \colhead{Region} &
    \colhead{Number} &
    \colhead{Radius} &
    \colhead{M$_{max}$} &
    \colhead{M$^{5th}_{max}$} &
    \colhead{ } &
    \colhead{$N_0$\tablenotemark{b}} & 
    \colhead{$\alpha_{SF}$\tablenotemark{b,c}} &
    \colhead{\mstar\tablenotemark{b}} & 
    \colhead{ } & 
    \colhead{$\alpha_{PLF}$\tablenotemark{c}} 
    \\
    \colhead{ } &
    \colhead{ } &
    \colhead{kpc} &
    \colhead{$10^5$\msun} &
    \colhead{$10^5$\msun} &
    \colhead{ } &
    \colhead{ } & 
    \colhead{ } &
   \colhead{$10^5$\msun} & 
    \colhead{ } & 
    \colhead{ }}
  
  \startdata 
& & & & & & & & & & \\

       bin 1  & 107 &    0.46--3.19 &  1.62$^{+0.13}_{-0.26}$ &  1.11$^{+0.07}_{-0.30}$ & &  2.29$\pm$3.40 &   $ -1.84 \pm 0.13 $ &  4.85 $\pm$   2.02 & &   $-1.90 \pm 0.10 $ \\

& & & & & & & & & & \\

       bin 2 & 107 &   3.22--4.53 & 2.16 $^{+0.18}_{-0.19}$ &  0.45$^{+0.11}_{-0.10}$ &  & 3.77 $\pm$ 2.80 & $-2.15 \pm$ 0.12  &   0.98 $\pm$ 0.42  & & $ -2.25 \pm  0.12 $ \\

& & & & & & & & & & \\

      bin 3  & 106 &  4.56--10.15 & 4.36$^{+0.14}_{-0.27}$  & 0.43$^{+0.04}_{-0.11}$ &  & 5.67$\pm$ 4.17 & $ -2.00 \pm$ 0.13 &    1.04 $\pm$  0.59 & &$ -2.13 \pm 0.09 $ \\

 & & & & & & & & & & \\
      \hline

& & & & & & & & & & \\

   all & 320 & 0.46--10.15 & 4.36$^{+0.14}_{-0.27}$ & 1.40$^{+0.09}_{-0.08}$& &7.58 $\pm$ 4.20 & $ -2.03 \pm$ 0.07 & 2.03$\pm$0.81 & & $ -2.09 \pm$ 0.06 \\
   
  \enddata
 \tablenotetext{a}{The table includes only observed and fitted values of class 1 \& 2 cluster population with ages smaller than 200 Myr and masses above 5000 \msun.}
 \tablenotetext{b}{Mass function parameter fits, computed via the maximum-likelihood method of Rosolowsky (2005). If $N_0 \gg 1$, a truncated CMF form is appropriate, while $N_0 \lesssim 1$ indicates a single power-law is more appropriate.}
 \tablenotetext{c}{$\alpha_{SF}$ is the slope derived assuming a truncated mass function, $\alpha_{PLF}$ has been derived assuming a pure power-law function. See text for details.} 
\end{deluxetable*}

\subsection{Using Monte Carlo simulations to link cluster mass and luminosity functions}

In the attempt to understand the uncertainties imposed by the low number statistics and, at the same time, link the observed CMF to the CLF we use simulated cluster populations. We stochastically sample a pure power law mass function (we will refer to this population as \emph{run A}) and a truncated one in the form of a Schechter function (\emph{run B}), with the same slopes and \mstar\, derived for the cluster (class 1 \& 2) population of NGC\,628. In both runs we sample the mass function from $2\times10^2$ to $100\times$M$_{max}$ \msun. We use $1.5 \times 10^4$ objects so that the resulting cluster population will approximatively have the same number of clusters more massive than 5000 \msun\,  as observed in NGC\,628(i.e. 320 class 1 \& 2 objects). For each cluster we stochastically assign an age between 1 and 200 Myr, i.e. we assume that star formation has been constant during this time range. Based on the evidence produced by the analysis of cluster dissolution time scales in Section 5.4 we also include mass dependent disruption using the formula:
\begin{equation}
M_i(t_i)=\left(M_i^{\gamma}-{\gamma}t_i/t_0\right)^{-\gamma}
\end{equation}
where we assume $\gamma =0.65$ and $t_0=4.91\times 10^5$ yr (values obtained from a maximum-likelihood fit to the data, see Section 5.4 for details).
 We then create 1000 realisations of each population. 

\begin{figure}
\centering
		\includegraphics[scale=0.45]{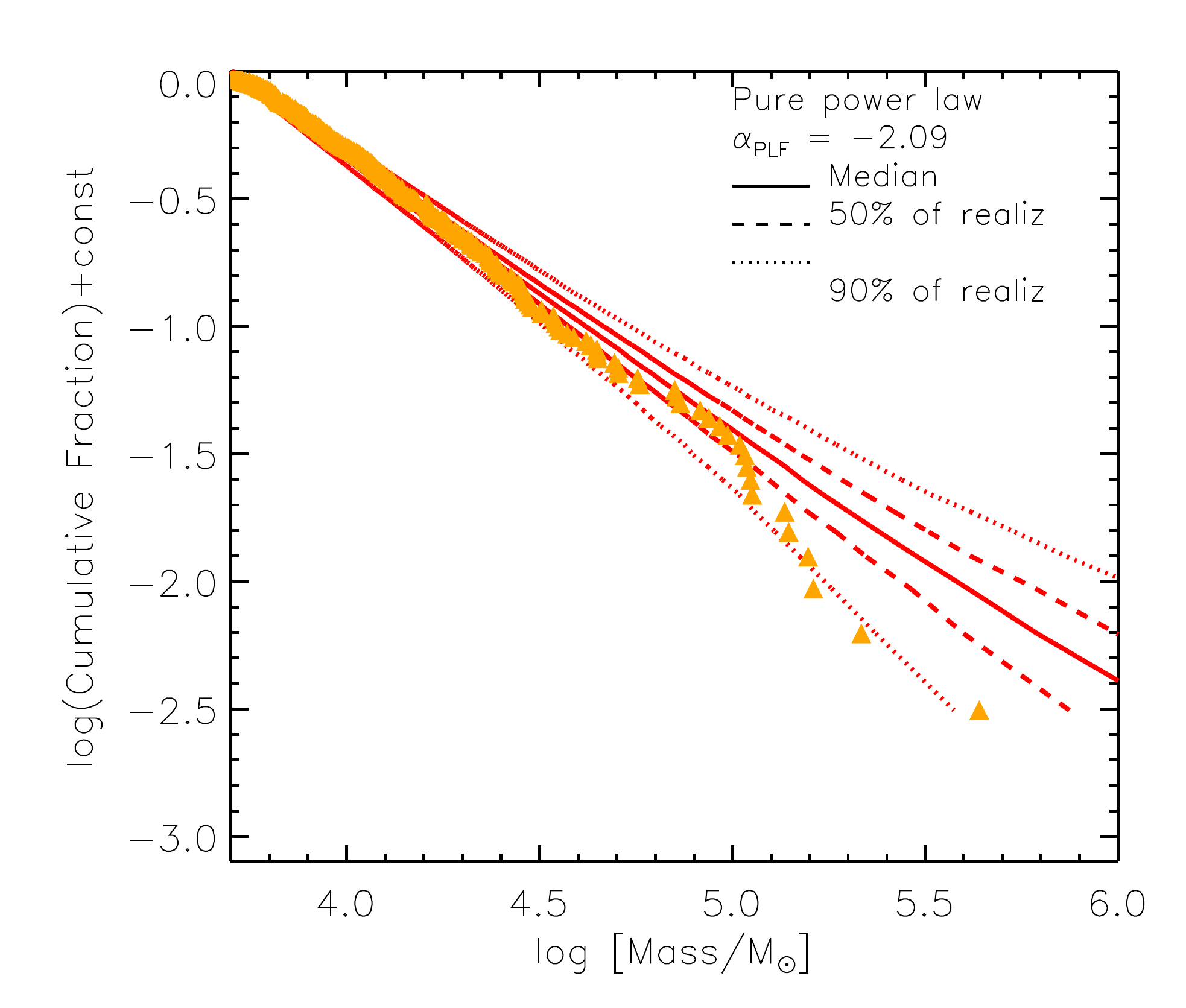}\\
		\includegraphics[scale=0.45]{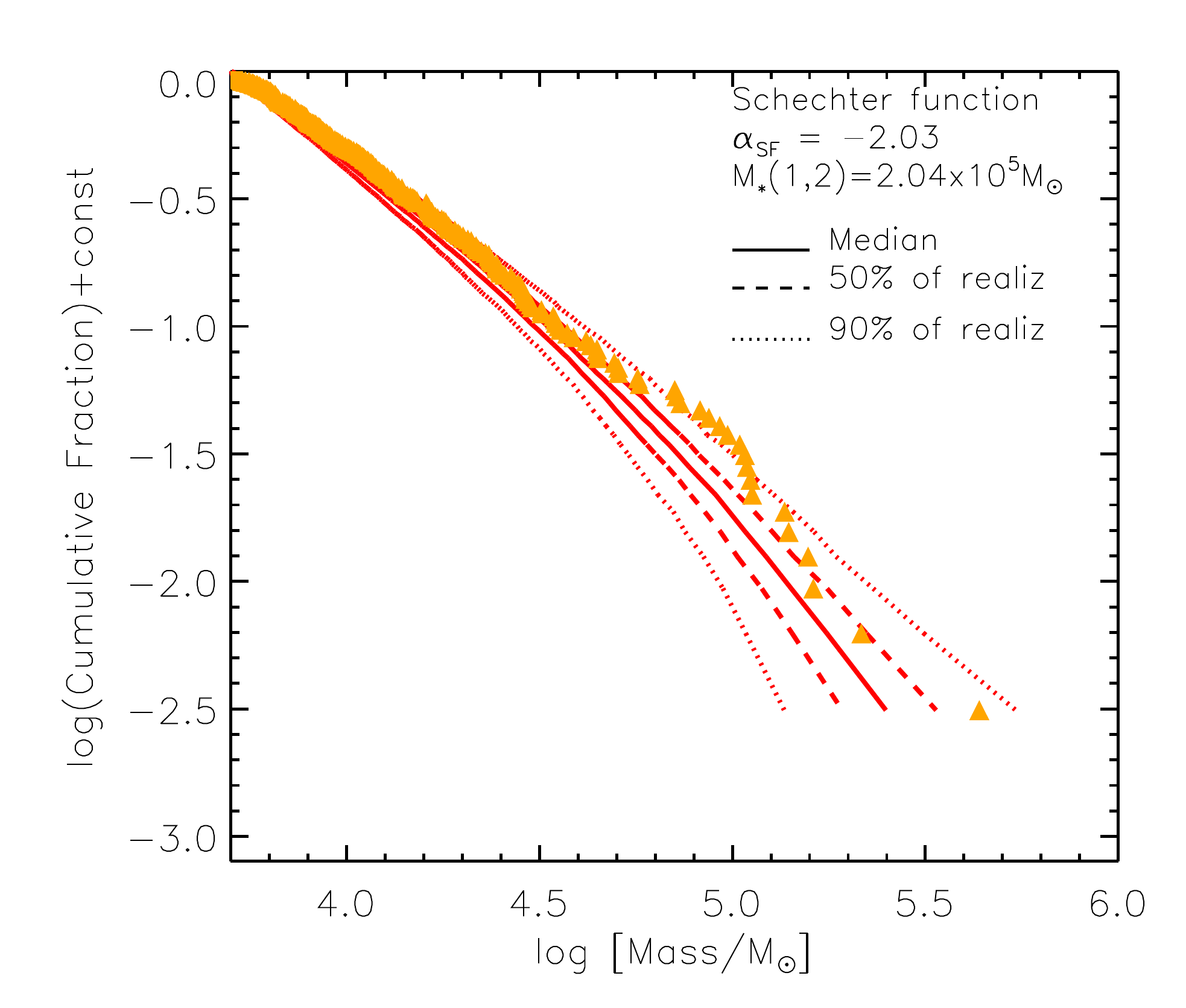}
    \caption{Monte Carlo simulations aim to reproduce the observed CMF for class 1 \& 2 (orange triangles). A 1000 realisations, containing a similar number of objects as in the real distributions, have been performed in each case. The median and the regions containing 50 and 90 \% realisations are shown as indicated in the inset. In the top panel, we assume a pure power-law shape (\emph{run A}). In the bottom panel, we assume a Schechter-type distribution(\emph{run B}). The slopes of the power law and the truncation masses used are reported in the top right inset of each panel. See text for a discussion of the results.}
    \label{fig:fig15}
\end{figure}

 In Figure~\ref{fig:fig15} we include the observed CMF of NGC\,628 (orange dots) and we overplot the median (red solid line) mass cumulative distribution and the distributions containing 50 \% (dashed red line) and 90 \% (dotted red line) of the 1000 Monte Carlo realisations sampled from an underlying pure power-law mass function (\emph{run A}, top panel) and Schechter truncated function (\emph{run B}, bottom panel). To build the cumulative distributions of the simulated populations we apply the same mass cut as in the observations at 5000 \msun. When comparing the loci occupied by the simulations with respect to the observed CMF we see that 90 \% of realisations of \emph{run A} overestimate the number of clusters more massive than $10^5$ \msun (the locations of the upper mass end of the observed CMF is off of the 90\% limits of realisations). We test the null hypothesis that the upper part of the median distribution of \emph{run A} and \emph{run B} realisations and of the observed CMF are drawn from the same parent population. Since the differences between the pure power-law function and the truncated function is at the upper mass end of the distributions we run both a Kolmogorov--Smirnov (KS) and an  Anderson-Darling (AD) test using clusters more massive than $10^4$ \msun. We recover similar probabilities from both statistics with p(AD) and p(KS) of $\sim$0.3 and $\sim$0.9 for \emph{run A} and \emph{run B}, respectively. This result does not discard any of the two functions, but it yields a marginal preference for a truncated mass function. 
 
 Next we analyse the resulting luminosity functions produced from the two \emph{run A} and \emph{run B} simulated cluster populations with the goal of understanding if the underlying CMF has a truncation. We use the median cluster population of the 1000 Monte Carlo realisations to yield the CLF. We do not apply any mass cut to the simulated populations but we use the age and mass of each mock object together with Yggdrasil models to estimate their fluxes in the $UBVI$ LEGUS bands. We then select only sources with  a luminosity brighter than $M_V \leq -6$ mag and brighter than the 90\% completeness values in the other two bands for which we require detection for a source to enter the catalogue (namely $B$ and $I$).  Extinction is not taken into account.  

\begin{figure*}
\centering
		\includegraphics[scale=0.43]{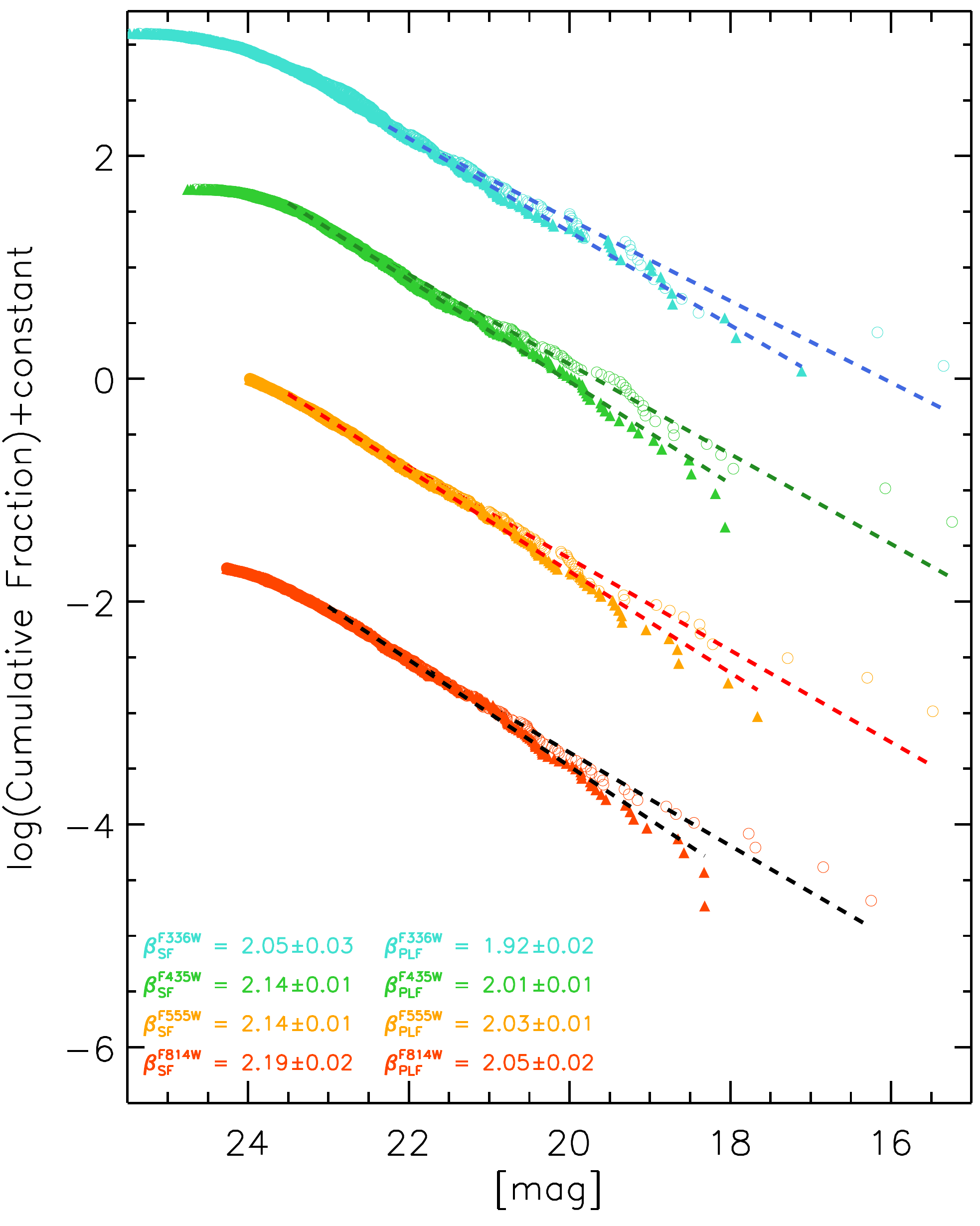}
		\includegraphics[scale=0.43]{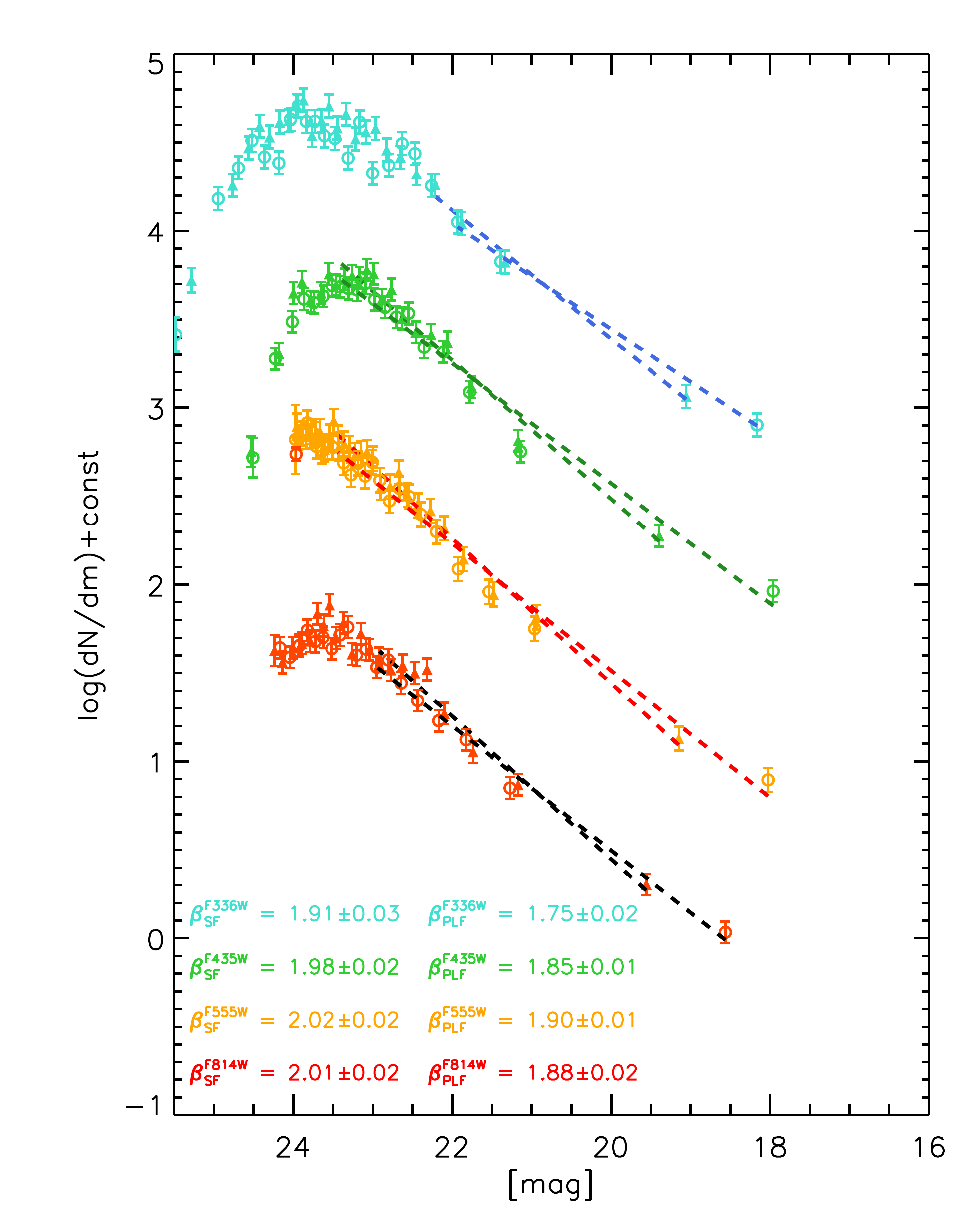}
    \caption{ Cumulative function (left) and binned distributions (right) of the luminosity produced with {run A} (pure power-law function, open circles) and {run B} (Schechter truncated function, solid triangles) simulated cluster populations. Dashed lines are the single power law fits performed to both distributions. The fit is carried out down to the same luminosity limits as in the observed CLFs to take into account the effect of detection limits imposed by the depth of the data. The resulting indexes for \emph{run A} ($\alpha^{\lambda}_{PLF}$) and \emph{run B} (($\alpha^{\lambda}_{SF}$)) are listed in the bottom left insets.}
    \label{fig:fig16b}
\end{figure*}

The resulting cumulative and binned CLFs as a function of waveband of the two simulated cluster populations are illustrated in Figure~\ref{fig:fig16b}. As done in Section 5.1 for the observed CLFs, we fit a single power law function to the simulated CLFs. To recreate the limiting detection depth of each waveband we use the same low luminosity limits as for the observed distributions. 

Since the underlying CMF of the simulated cluster populations is known we can directly verify the effect of using either binning or cumulative techniques on the resulting distributions. We notice that the recovered slopes $\beta^{\lambda}_{PLF}$ and $\beta^{\lambda}_{SF}$ listed in the left panel of Figure~\ref{fig:fig16b} are about 0.2 steeper then the ones listed in the right plot of the same figure. As discussed in Section 5.2, the way the equal-number object binning and cumulative distributions are built make them sensitive to different properties of the distributions. Since we are interested in understanding whether the CMF has a truncation at the high mass, we prefer to analyse cumulative distributions.

The cumulative distributions in the left panel of Figure~\ref{fig:fig16b} should be compared to the observed cumulative distributions of class 1 \& 2 objects (cumulative functions illustrated with triangles in the left plot in Figure~\ref{fig:fig12}). 
First, the resulting indexes of \emph{run A} ($\beta^{\lambda}_{PLF}$) and \emph{run B} ($\beta^{\lambda}_{SF}$) show that the effect of fading when comparing the slope of the $U$ band to the ones in the redder filters  is between 0.1 and 0.15.  The differences in the recovered slopes for the observed CLFs in the ultraviolet and optical are larger (about $0.2-0.3$). However, we note that our simulations do not include extinction, which preferentially affects the bluer bands and could explain the larger flattening of the observed ultraviolet slopes. 

Second, we notice that the recovered CLF slopes for a population of clusters sampled out of a pure power law CMF (\emph{run A}) are very close to the initial slope of the corresponding CMF (i.e. $\alpha_{PLF} = 2.09$).  The slopes obtained fitting a CLF sampled out of a CMF with a power-law shape (slope $-2$) and a truncation at the high mass end (\emph{run B}) are steeper than $-2$. We also notice a fundamental difference at the bright luminosity end between the cumulative distributions of the two runs (pure power law versus power law with a truncation). The presence of a truncation at the high mass end of the CMF (\emph{run B}) results in significant deviations at the bright end (i.e. solid triangles) of the CLF, with respect to the population sampled out from a pure power-law CMF (\emph{run A}, open circles). The drop at the bright luminosity observed for the \emph{run B} CLF (filled circles, left panel of Figure~\ref{fig:fig16b}) is very similar to the ones observed in the real CLF (class1 \& 2, left plot in Figure~\ref{fig:fig12}). This trend reinforces the evidence of a presence of a truncation in the underlying CMF of NGC628.


\subsection{The cluster and association age distributions and implications for evolution}

In this section we probe the evolution of the two classes of objects visually identified, class 1 \& 2 being likely clusters, and class 3 being a heterogeneous sample dominated by compact stellar associations.

We apply the same mass cut at 5000 \msun\, which ensures that we will be complete in their recovery rate up to 200 Myr, but to look now at the number densities of objects per unit time ($dN/dt$) as a function of increasing age. A decreasing rate of objects as a function of time $dN/dt \propto t^{\delta}$ is historically interpreted as an evidence for YSC dissolution within the galaxy \citep[see][for a short review]{2009Ap&SS.324..183L}.  In Figure~\ref{fig:fig17}, top panel, we plot the change in number densities of class 1 \& 2 (orange dots), class 3 (blue diamonds), and the whole sample (green triangles) using bins of the same width (left panel). The shadowed areas are the regions of the diagram where incompleteness mimics a rapid disappearance of objects. Therefore, these regions are excluded \citep[e.g. see][]{2009Ap&SS.324..183L}. Overall, we observe that the disruption rate of class 3 systems  is certainly more significant than class 1 \& 2. When the two classes are analysed together the resulting disruption rate is coincident with the one of class 1 \& 2, probably because this latter class is much more numerous. 

\subsubsection{Cluster and compact association evolution during the first 10 Myr}
We also observe variations depending on the age range. In all the plots of Figure~\ref{fig:fig17}, we clearly see that during the first 10 Myr the number densities of objects in both class 1 \& 2 and class 3 are significantly higher than in the range 10 to 100 Myr. The number densities per unit time declines roughly by factors of 4 and 3 for class 3 and class 1 \& 2 respectively, when comparing the rate between the range 1 to 10 Myr and 10 to 100 Myr. Two different explanations are typically advocated to describe such a downtrend, either ``infant mortality", in the classic terminology introduced by \citet{2003ARA&A..41...57L}, or contamination of our sample with systems that are not gravitationally bound \citep[e.g.][on the difficulty of distinguishing bound from unbound young star-forming regions]{2011MNRAS.410L...6G}. In the first scenario, it is assumed that all stars form in bound clusters that rapidly dissolve because gas evacuated by stellar feedback has destabilised the gravitational potential of the system. In the second scenario, the rapid disruption is only caused by the limits of the method we are employing. Dynamical imprints of very young star-forming regions in the Milky Way \citep[e.g.][]{2014MNRAS.438..639W} and Magellanic Clouds \citep{2014MNRAS.439.3775G}  suggest that even massive OB associations, like Cygnus OB2, are not the result of an evolution from a gravitationally bound status but are formed already unbound. Stellar feedback, therefore, does not appear to be a major agent of cluster disruption \citep[e.g.][for a review]{2014prpl.conf..291L}. In extragalactic studies, like the one performed on the LEGUS galaxies, we are not able to probe the boundness status of the objects we consider clusters. Objects older than 10 Myr and with effective radii of a few parsecs can be considered gravitationally bound since their crossing time is smaller than the age of the stars. However younger objects, often still nested within the large star-forming regions where they have formed, are more difficult to classify. Because of the complexity involved in characterising objects at these young ages, we restrict our $dN/dt$ analysis to the age range 10--200 Myr.

\subsubsection{The change in the number density of clusters and associations as a function of time, mass, galactocentric distance}
\begin{figure*}
\centering
		\includegraphics[scale=0.29]{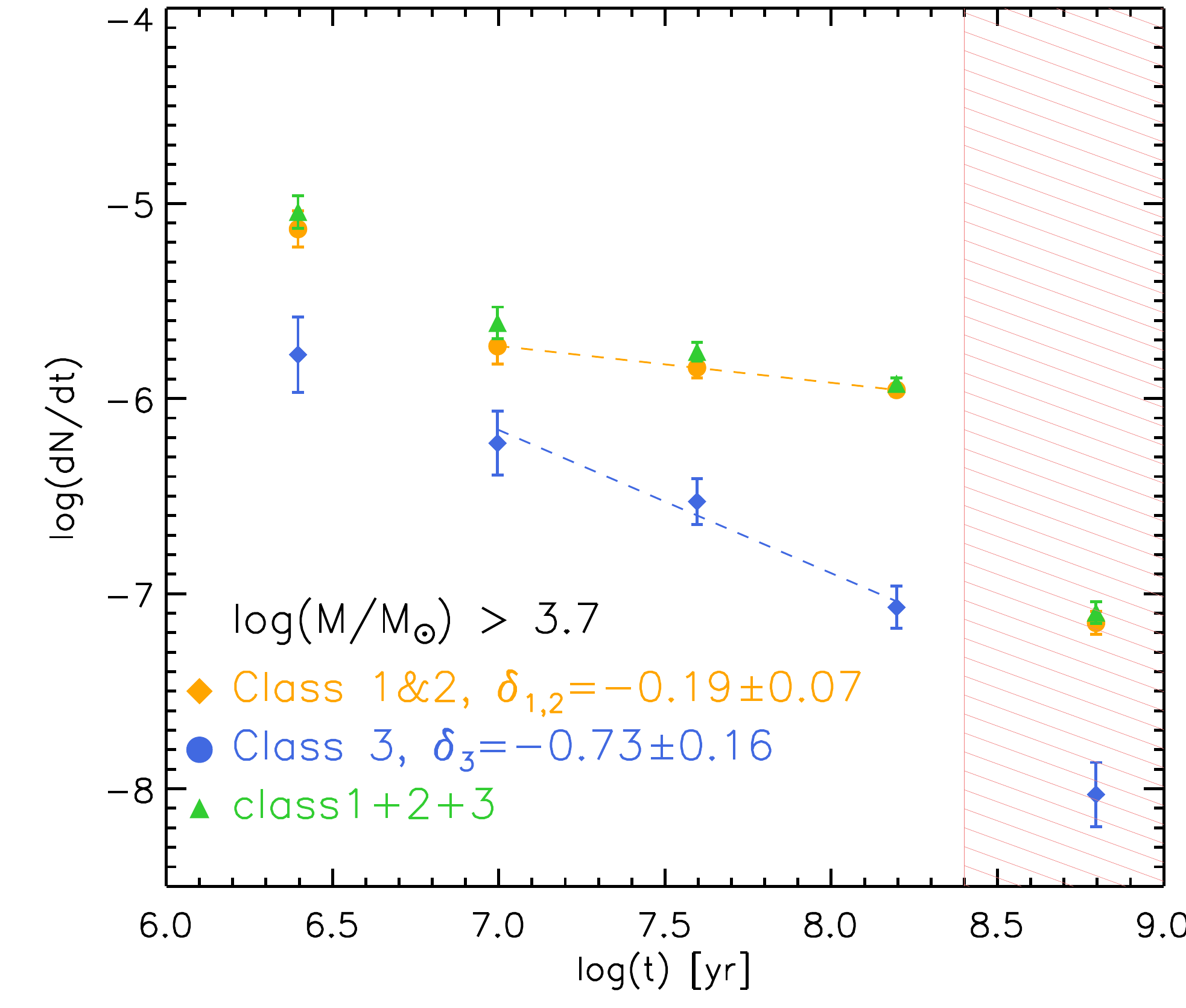}
		\includegraphics[scale=0.29]{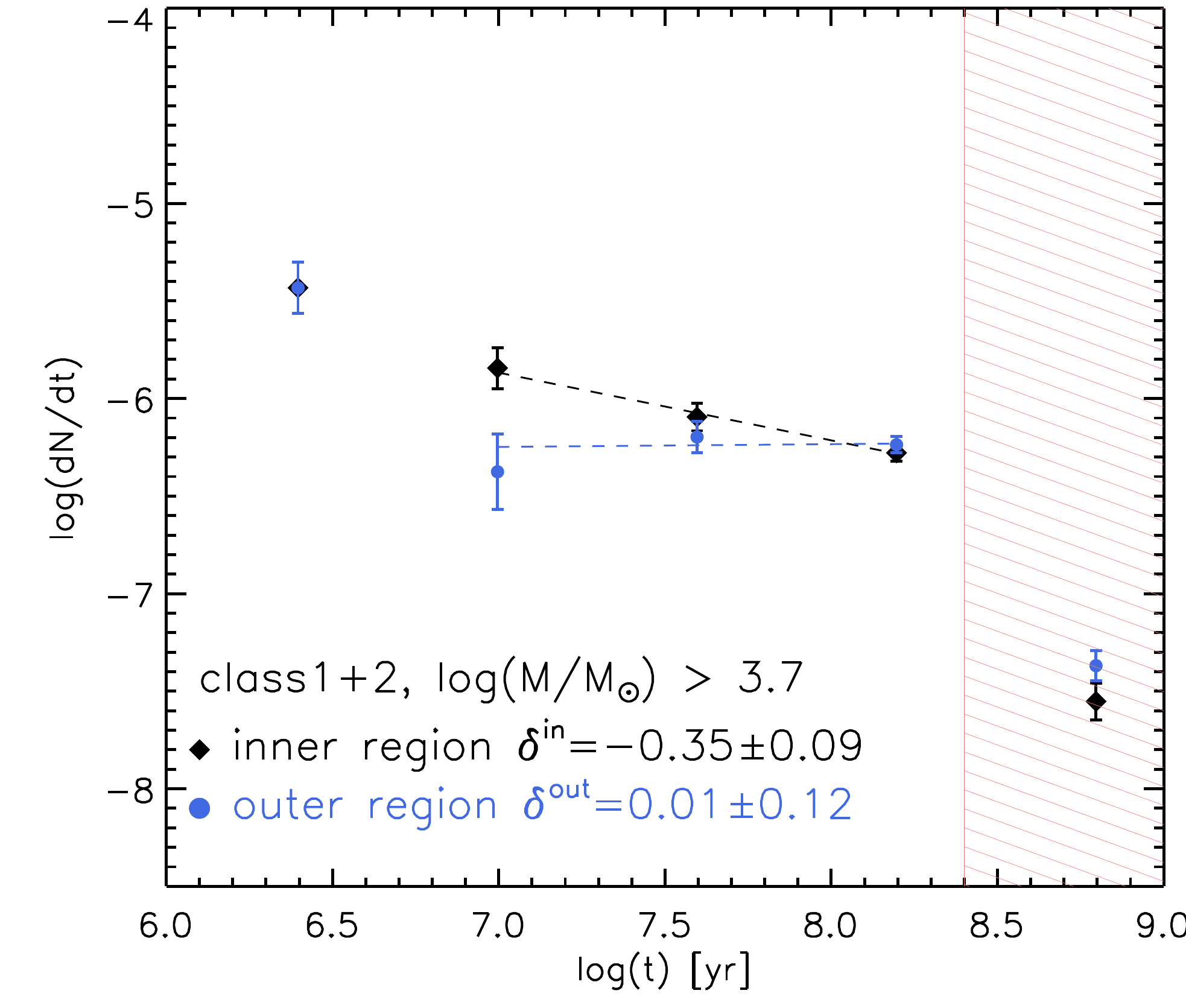}
		\includegraphics[scale=0.29]{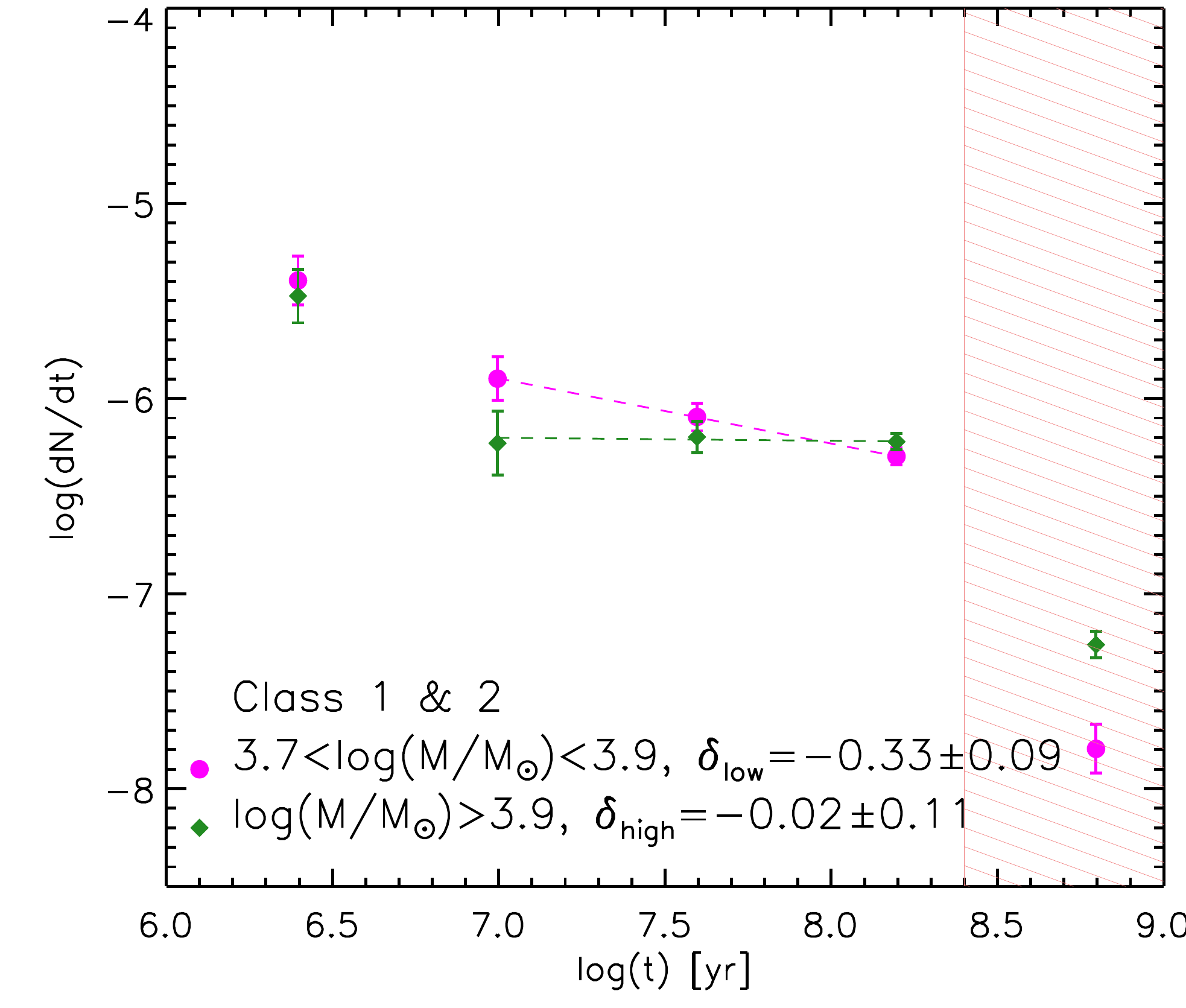}
    \caption{Number density of systems more massive than 5000 \msun\, per unit time as a function of age using equally spaced temporal bins (bin size is 0.6 dex). The shadowed areas show the regions of the diagrams that are affected by incompleteness and excluded from the analysis.  The fit to each distribution within the age range 10 to 200 Myr is illustrated with a dashed line. The recovered slopes are included in the corresponding insets.  The left panel illustrates the change in number density of the whole population (class 1, 2, \& 3, green triangles), cluster candidates (class 1 \& 2, orange dots), compact associations (class 3, blue diamonds). The central panel shows the number density of clusters as a function of age within an inner and outer region. The two regions contain the same number of clusters with mass above 5000 \msun. In the left panel we split the sample into low mass ($log(M) \leq 3.9$ \msun, magenta dots) and high mass ($log(M) > 3.9$ \msun, green diamonds) clusters. See text to follow the discussion of the results.}
    \label{fig:fig17}
\end{figure*}

Between 10 and 200 Myr the left plot of Figure~\ref{fig:fig17} shows that the change in the number density of class 1 \& 2 clusters is consistent with mild disruption. We do not observe any drastic dissolution of this class of objects up to 200 Myr. On the other hand the number density of class 3 systems keeps declining with a factor of 3 in the same age range, suggesting as already observed in the color-color diagrams analysis, their rapid disappearance within 100 Myr (the slope we recover is $\delta \sim -0.7$). These short time scales are in agreement with the clustering analysis performed by \citet{2015ApJ...KG} on the class 3 population. The lack of clustering after 40 Myr may be caused by the quick dissolution of these systems. On the other hand, since class 1 \& 2 survive longer, their lack of spatial clustering is possibly the result of the randomisation of their positions because they move away from their birthplaces. The clustering analysis of stellar structures performed by \citet{2015MNRAS.452.3508G} in another LEGUS galaxy, NGC6503, finds that hierarchical clustered stellar structure disappear and distribute into the stellar field within 60 Myr. This suggests that class 3 objects are well nested within the hierarchical properties of star formation, while stellar clusters, even though have formed within the same hierarchically structured ISM, as gravitationally bound systems, may follow a different fate.

Recent theoretical and numerical works \citep[e.g.][]{2010ApJ...712..604E, 2011MNRAS.414.1339K} point out that tidal forces exerted by GMC encounters
in a hierarchical ISM can reproduce the steady decrease in cluster numbers over time as a result of a decreasing ISM density in each cluster's environment as it drifts away from its birth place.  Higher gas densities towards the centres of
spiral galaxies or in starburst systems can increase the overall
dissolution rate of YSCs.

In Figure~\ref{fig:fig7}, the color-color diagrams of the inner and outer cluster population of \obj\, show interesting features suggesting that the outer cluster population is preferentially older.  We attempt here to investigate whether  we observe any change in the recovered dissolution time as a function of galactocentric distances. We split our sample of class 1 \& 2 objects into a central and outer bin containing the same number of objects. In the central panel of Figure~\ref{fig:fig17} we show the recovered disruption rates for the inner and outer bins. We observe that the number density of clusters in the outer bins is consistent with being constant between 10 and 200 Myr ($\delta \sim 0.0$), while in the inner bin we observe higher dissolution rates $\delta \sim 0.3$. The observed trends are consistent with theoretical expectations, suggesting a higher mass-independent disruption in denser galactic ISM, thus closer to the centre of the spiral galaxy.

In the right panel of Figure~\ref{fig:fig17}, we probe the disruption rate of low (magenta dots) and high (green diamonds) mass cluster candidates. We observe that low mass clusters show evidence of mild disruption ($\delta \sim -0.3$) while the number densities of clusters with mass similar to or larger than $10^4$ \msun\, are consistent with being constant ($\delta \sim 0.0$). This result suggests the presence of a mass dependency in the disruption rate of clusters. 

\begin{figure}
\centering
		\includegraphics[scale=0.45]{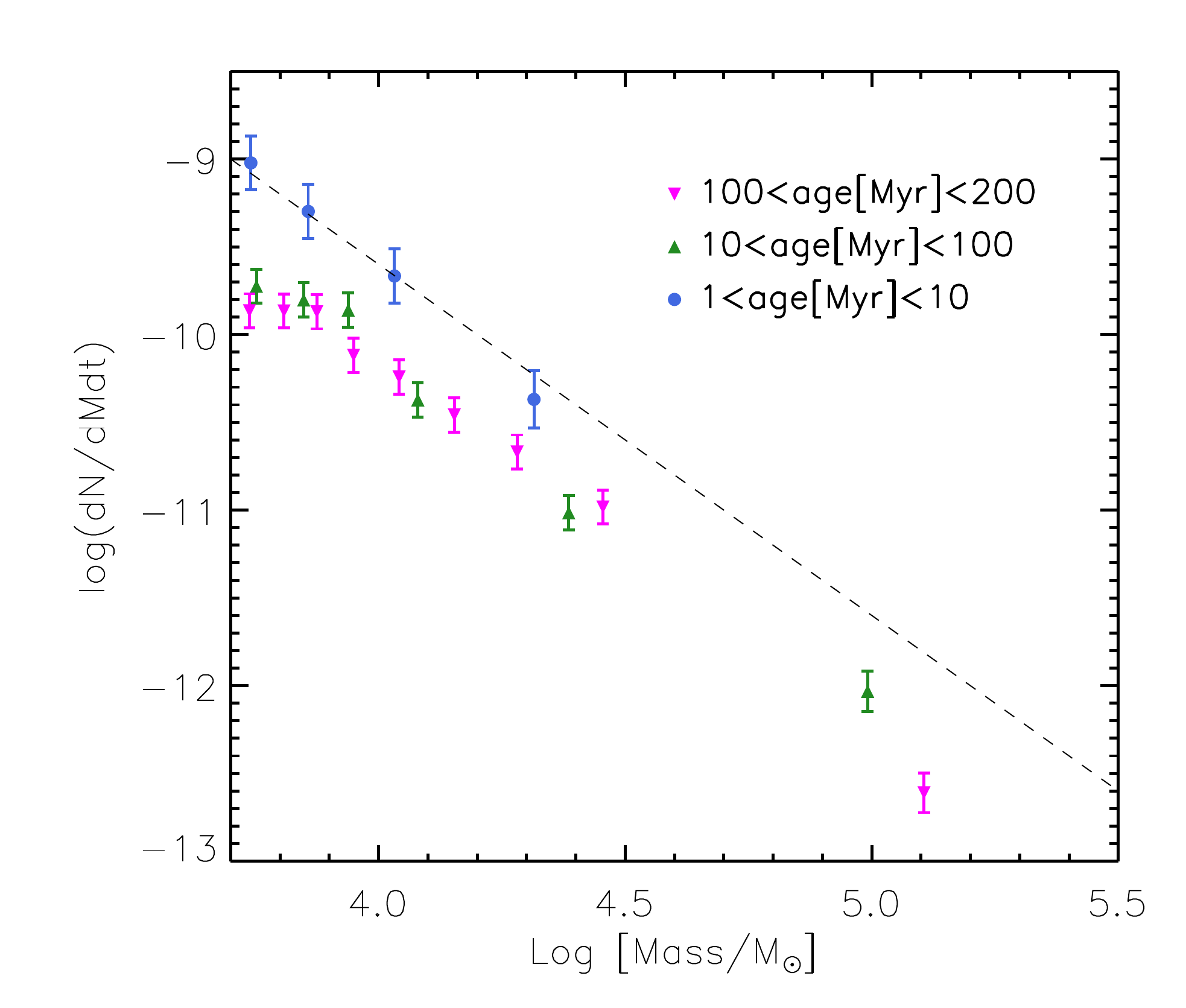}
		
\caption{Number of clusters per mass and time units ($dN/dMdt$). Only clusters more massive than 5000 \msun\, and classified as class 1\&2 are included. Blue dots show the CMF for ages below 10 Myr. Green and magenta triangles show the CMF between 10 and 100 Myr and between 100 and 200 Myr, respectively. The dashed line is included to provide a reference for a power-law CMF with slope $-2$. See text for details. }
    \label{fig:fig18}
\end{figure}

Another way to investigate cluster evolution as a function of the cluster mass is to encode in a bivariate distribution, $g(M, t)$ the mass function and time dependence for formation and evolution of clusters \citep[see][]{2009ApJ...704..453F, 2009MNRAS.394.2113G}. The function $g(M, t)$ is thus the number of clusters observed as a function of time and mass expressed as 

\begin{equation}
g(M, t)=\frac{\dr^2 N}{\dr M\dr t}
\end{equation}

Integrating $g(M, t)$ over the mass provides the $dN/dt$ distributions, while integrating $g(M, t)$ over a time range provides the observed CMF. 
\begin{figure*}
\centering
		\includegraphics[scale=0.65]{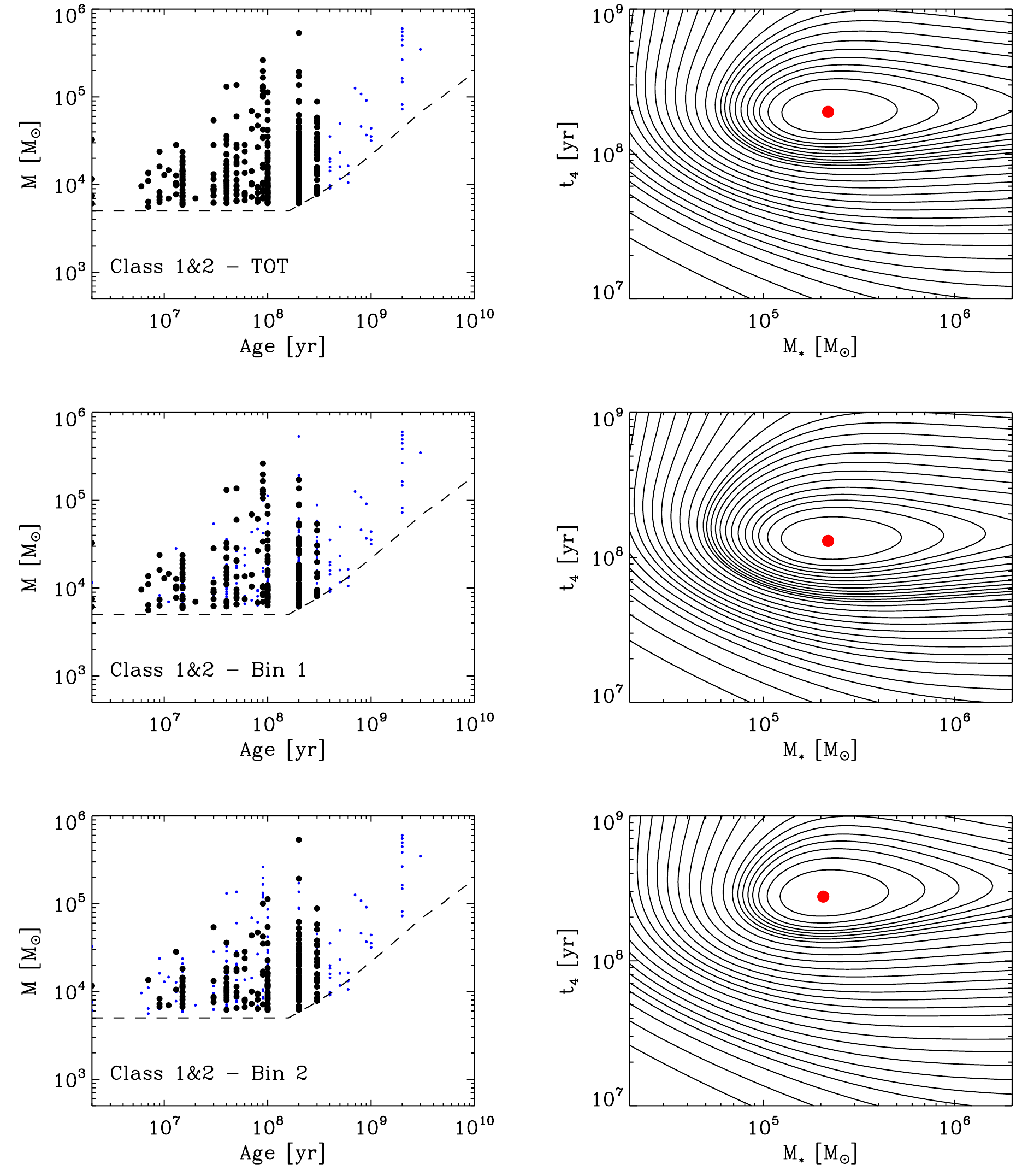}
		
\caption{Maximum-likelihood fit of the YSC population of \obj. Only class 1 \& 2 clusters more massive than 5000 \msun and younger than 300 Myr are included. A magnitude limit corresponding to the $M_V=-6$ magnitude cut applied to select cluster candidates is also taken into account in the fitting. The left panels show the age-mass diagram of the sources included in the fit (black dots) overplotted on the whole class 1 \& 2 sample (blue dots). The dashed line shows how the mass and magnitude limits select clusters at different age ranges. At young ages the mass cut is limiting the sample (horizontal dashed line), at older ages the luminosity cut is more important (transversal dashed line at ages larger than 200 Myr). The right panels show the values of $t_4$ and \mstar\, that provide the maximum likelihood (red dots). In the {\bf  top panels} the total YSC population of \obj\, has been fitted. The {\bf  central} and {\bf  bottom panels} show the results of the fit performed on the inner and outer bin population as showed in Figure 18. See text for details.}
    \label{fig:fig19}
\end{figure*} 

In Figure~\ref{fig:fig18} we show the recovered mass distributions as a function of cluster age normalised by the corresponding age interval (i.e. $dN/dMdt$ diagnostic). The distribution is built to contain the same number of objects in each bin.
The $dN/dMdt$ diagnostic is sensitive to the evolution of the mass function with time. If the number of clusters rapidly declines independently of the cluster mass then the shape of the mass function will be unchanged but the distribution at each age interval will be shifted because the number of clusters is diminishing. On the other hand, if the cluster disruption time-scale depends on cluster mass then the ageing CMFs should overlap at the mass ranges untouched by disruption and deviate where disruption of low mass clusters is significant. The plot in Figure~\ref{fig:fig18} shows a clear offset of the CMF for clusters younger than 10 Myr. As already seen in the $dN/dt$ distribution the number of clusters at this age range is significantly higher. The $dN/dMdt$ analysis provides further insights showing that the number of clusters younger than 10 Myr is higher at any mass range in units of time. The CMFs of the other two age ranges (10 to 100 and 100 to 200 Myr) overlap at masses larger than $10^4$ \msun, while deviations become significant at lower masses. These deviations suggest that the number of clusters with masses below $10^4$ \msun\, are decreasing, flattening the CMF. This trend is consistent with both the effect of detection limits causing the loss of increasing number of low mass clusters at older ages and higher disruption rates of clusters below $10^4$ \msun\, (consistent with the trends observed in the right plot of Figure~\ref{fig:fig17}). As pointed out in Section 5.2, our mass cut of 5000 \msun\, is very close to the completeness limits at 200 Myr (see Figure~\ref{fig:fig13}). Incompleteness can mimic a mass dependent cluster dissolution, and we cannot exclude that the observed flattening is in part caused by incompleteness.

To understand whether the trends observed in the bivariate distribution of the CMF as a function of time are compatible with mass dependent disruption we perform a maximum likelihood fit to our data. We refer to \cite{2009MNRAS.394.2113G} for the formalism behind the fit and \cite{2012MNRAS.419.2606B} for an application to the YSC population in the M83 galaxy. The fit is performed assuming a Schechter function that describes the probability to form a cluster of mass $M_i$ in the time interval $t$ and $t+\dr t$ as
\begin{equation}
\frac{\dr^2 N}{\dr M_i\dr t} = A\,M_i^{-2}\,\exp\left(-\frac{M_i}{M_{\star}}\right)
\end{equation}
where $M_i$ depends of time. If only cluster mass-dependent disruption is taken into account, the disruption time can be described as $t_{dis}=t_0M^{\gamma}$ with $t_0$ and $\gamma$ depending on the galactic environment \citep[see e.g.][]{2005A&A...441..117L} and the mass evolution is described by $M=\left(M_i^{\gamma}-\gamma t/t_0\right)^{1/\gamma}$ \citep{2005A&A...441..117L, 2009MNRAS.394.2113G}. In the fit $\gamma$ is fixed to the average value found in local spirals ($\gamma=0.65$). The fitting algorithm thus finds the values of \mstar, $t_0$, and $t_4$ that maximise the likelihood. The $t_4$ is the time scale necessary to dissolve a cluster of $M_4=10^4$ \msun\, and it depends on $t_0$ as $t_4=(t_0/\gamma)M_4^{\gamma}$. The results of maximum-likelihood fits performed on the YSC population of \obj\, are shown in Figure~\ref{fig:fig19}. The fit is performed on class 1 \& 2 clusters more massive than 5000 Msun, brighter than the magnitude cut $M_V=-6$ ($V=23.98$ mag) mag and younger than 300 Myr. The age-mass diagram on the left side panels shows the YSC population taken into account in the fit. The mass cut removes low mass objects up to an age of 200 Myr. The magnitude cut becomes important at older ages because it removes clusters more massive than 5000 \msun\, at ages older than 200 Myr. The fit is done for the entire cluster population (top), and the inner (centre) and outer (bottom) regions of the galaxy corresponding to two bins containing the same number of clusters. 
The maximum-likelihood fit performed on the entire cluster population finds \mstar$= 2.2\times 10^5$ \msun and a time scale for dissolution of a $10^4$ \msun\, cluster of $t_4=190$ Myr.  These values are in agreement with the \mstar\, found in Section 5.2 and the evidence of slow disruption in the $dN/dt$ analysis performed above. With the same technique we derive \mstar\, and $t_4$ for clusters of the inner and outer regions containing the same number of objects, as defined above. The cluster population located within the inner bin has the same \mstar\, value found for the entire galaxy (\mstar$= 2.2\times 10^5$ \msun) and a time scale for dissolution of a $10^4$ \msun\, cluster of $t_4=130$ Myr. The clusters in the outer region have a factor of 2 longer disruption time scales with $t_4=270$ Myr and \mstar$= 2.0\times 10^5$ \msun. The maximum-likelihood fitting analysis confirms the results and trends obtained from the analyses of the CMF and the $dN/dt$ distributions. The average time scale for cluster disruption in the galaxy are long enough to produce a shallow $dN/dt$ distribution. However, the age interval to which our analysis is sensitive too is long enough that we should start to see the effect of disruption on low mass clusters. Indeed, in this respect the $dN/dMdt$ analysis is a more sensitive diagnostic than the simple $dN/dt$ one. The longer time scales for the disruption of clusters in the outer part of the galaxy are also in agreement with the expectation from theoretical works \citep[e.g.][]{2010ApJ...712..604E, 2011MNRAS.414.1339K} and the results obtained for the M\,83 spiral galaxy \citep{2012MNRAS.419.2606B}.

To verify the effect of incompleteness at the low mass end as a function of age, we repeat the maximum-likelihood fit using the same mass cut at 5000 \msun\, together with a more conservative luminous cut, i.e. $M_v = 22.93$ mag, i.e. one magnitude brighter than previously done. The mass and magnitude cuts result in a selection of clusters more massive than 5000 \msun\, up to 100 Myr and objects brighter than 22.93 mag at older ages. We obtain values of \mstar, and $t^4$, within a factor of two from the analysis performed above with a less conservative magnitude cut. With a limiting brightness of $M_v = -22.93$ mag, we retrieve \mstar$=1.7\times 10^5$ \msun\, and $t^4 = 170$ Myr for the entire sample. \mstar$=2.2 \times 10^5$ and $1.4 \times 10^5$\msun, and $t^4 = 100$ and 530 Myr for the central and out region of \obj, respectively. This further test reinforces the evidence, found above, of a mass-dependent component necessary to describe cluster disruption in \obj.
 
\section{Discussion}

\subsection{The shape of the CMF at the high-mass end, constraints on the cluster formation process}
CLF and CMF are powerful tools to investigate the formation of YSCs. 

In the spiral galaxy NGC628, using the LEGUS dataset, we recover the properties of the CLF from the UV to the NIR. The luminosity distributions of the cluster (class 1 \& 2) population is close to the slope $-2$, while the compact stellar associations (class 3) appear to have steeper slopes. The analysis of increasing size of stellar aggregates \citep{2006ApJ...644..879E} finds also slopes close to $-2$. Hence, star formation is consistent with a hierarchical process, driven by turbulence, from the largest scales, i.e. star-forming complexes, down to the densest and smallest physical scales, i.e. star clusters. Our analysis reveals also some interesting variations in the CLF for both clusters and associations.  At each waveband  we observe a steepening at the brightest end of the cumulative distributions. Extinction cannot explain the steepening at the brightest end of the CLF. If more luminous clusters are more extinguished than faint ones it would imply preferential extinction as a function of luminosity, a trend not observed in studies of local galaxies. 

The cause of the steepening of the CLF could be connected to the nature of the cluster formation process. The CLF is a direct observable tracer of the underlying CMF integrated over time. If the CMF is a pure power-law function of slope $-2$ at any age and cluster dissolution is independent of the cluster mass, the resulting CLF should consistently be a function with the same slope. The steepening that we are observing at the bright end of the distributions suggests that we find fewer luminous (massive) clusters than expected for a pure power-law CMF. 

 Increasing evidence in the literature suggests that the YSC mass function is better described by a truncated power law. \citet{2017arXiv170310312J} studying the YSC population of M\,31 find that the CMF in this galaxy has a Schechter type form, with \mstar\, $\sim1\times10^4$ \msun. Cosmological simulations of Milky Way type of galaxies, where YSCs are implemented as star formation units, show that the resulting YSC populations have CMF that are better described by a Schechter function and \mstar\, scales as a function of SFR \citep{2017ApJ...834...69L}. 

In the case of NGC\,628, both a steeper power law ($\alpha \sim 2.1$) or a Schechter function ($\alpha \sim 2.0$ and \mstar$=2.0\times 10^5$ \msun) can equally reproduce the observed CMF. We investigate whether the steepening observed in the mass function, which suggests a dearth of very massive systems (if a pure power-law function is assumed), is a sign of an underlying soft truncation. The difficulty to constrain the upper-mass distribution of the CMF in local spirals has already being pointed out by \citet{2006astro.ph..6625L}. The combination of cluster formation being a stochastic process \citep[e.g.][]{aa..nb..2015Spr} and relatively low SFR combined with detection limits and rapidly fading luminosities above a few hundreds of Myr, make it very challenging to put strong constraints on the shape of the cluster upper mass distribution, because of low number statistics. 

\subsection{Time scales for cluster disruption in NGC628; probing cluster evolution}
Another fundamental aspect which gives important constraints on the formation and evolution of YSCs is understanding how they disrupt. Star-forming regions and stellar complexes are hierarchical in space and time \citep{1998MNRAS.299..588E}, thus their crossing times are comparable to their ages  and appear to dissolve on time scales below 60 Myr \citep[e.g.][]{2012AJ....144..182P,2015MNRAS.452.3508G, 2015ApJ...808...76C}. 
On the other hand, the fate of YSCs, forming in the densest peaks of these very regions, is not yet well understood. 

In our analysis, we look at the number density of stellar systems between 1 and $\sim$200 Myr. We are not able to derive the dynamical status of the stars within our objects, but since the average size of our class 1 \& 2 systems peaks at 3 pc (Ryon et al. 2017, submitted), their crossing times, for those older than 10 Myr, are much shorter than the stellar ages so we can consider them likely gravitationally bound. We see a clear decline between the number of objects during the first 10 Myr and the next age bin suggesting a significant loss of both class 1 \& 2 and class 3 systems. The recovered number densities of YSCs versus associations in the age range 10 to 200 Myr are significantly different. Compact stellar associations (class 3) rapidly decline in number and disappear on time scales ($\sim$ 50 Myr) comparable to those of hierarchically structured star-forming regions. Objects that we have identified as potentially bound clusters (class 1 \& 2) show close to constant number densities ($\delta \sim -0.2$) in the $dN/dt$ analysis. 

Small disruption rates within the first 100--200 Myr have also been reported in the literature for other local galaxies, e.g., in M31 \citep{2014ApJ...786..117F}, LMC \citep{2013MNRAS.430..676B}, M83 \citep{2014MNRAS.440L.116S}. However, we notice here, that these results are not ubiquitous. A compilation of previous works \citep{aa..nb..2015Spr} clearly shows that disruption rates of YSCs may change significantly from galaxy to galaxy and become very high for YSCs in hostile environments like the Antennae system \citep{2010AJ....140...75W}. 

Our results suggest a steeper disruption slope for YSCs located in the inner portion of the galaxy ($\delta \sim -0.3$).  This finding is observationally supported by the difference in the number densities of YSCs in regions of the color--color diagrams (Figure~\ref{fig:fig7}) corresponding to young ages. In the inner pointing the number of YSCs is larger than in the outer pointing at young ages, while similar number densities are observed at older ages between the populations of the two pointings. If star formation has been constant in the last few hundreds of Myr then these differences suggest a longer survival time in the outer regions. A larger coverage of the galaxy would certainly improve the results, as recently shown for M83 \citep[e.g.][]{2010ApJ...719..966C, 2014ApJ...787...17C, 2012MNRAS.419.2606B, 2014MNRAS.440L.116S}.

\section{Conclusions}

We present the methods and pipelines developed and applied to build uniform YSC catalogues of the LEGUS galaxies. Our method consists of a mixture of automated and visually optimised procedures, which take into account, differences in the quality of the data, distance of the galaxy, coverage, varying local background. We implement a quality-flag system, based on human and machine learning approaches under development, which describe the morphology of sources that are detected in at least 4 bands, and have luminosities in the visual band brighter than $-6$ mag.
We provide final YSC catalogues which include positions, CI of the source, photometry in the 5 LEGUS bands; ages, masses, extinctions, and uncertainties; $\chi^2$ analysis including residuals; visual classification flags. Cluster photometry is produced with the two most used methods in the literature which consist of fixed aperture photometry corrected by 1. an average aperture correction derived from observed clusters in each band,  or 2. a CI based aperture correction as a function of wavelength. The SED of each source is then fitted with both deterministic and stochastic SSP models which include a treatment for nebular emission. We used both Padova and Geneva stellar libraries and 3 different recipes for internal extinction.

We provide, as a proof of concept, a detailed description of all the steps necessary to produce the final cluster candidate catalogues, with parameters and assumptions optimised for the LEGUS target NGC628. Two regions of the galaxy have been targeted by LEGUS, an inner and an outer one. 

In the attempt to probe YSC formation and evolution, we analyse the luminosity and physical properties of the YSC population in NGC628, using as reference catalogue the one obtained with Padova evolutionary tracks, Kroupa IMF, solar metallicity, Cardelli extinction law. A comparison between the cluster properties obtained with the two photometric methods (AV\_APCOR and CI\_BASED) shows some differences in the color-color distributions, and recovered cluster physical properties. We conclude that significant deviations affect about 20\% of the sources in common between the two catalogues. We perform the analysis for both photometric catalogues, but results are not affected by the choice of the catalogue.

The color-color diagrams of class 1 \& 2 systems (which, according to our visual classification scheme, are YSC candidates) and class 3 objects (likely compact stellar associations) show interesting differences in their density distributions. YSC candidates follow closely the SSP models at all ages, while stellar associations are more concentrated around the regions of the tracks younger than 50-60 Myr. We also compare the color properties of the YSCs and stellar associations in the inner and outer regions of NGC628. We do not observe any significant difference between the number of stellar association distributed along the evolutionary models, suggesting similar age ranges for the associations in these two pointings. On the other hand, in regions of the color-color diagram corresponding to young ages the number density of YSCs changes between the inner and outer pointing. On average, the population in the outer field is older, probably reflecting a slower disruption rate in the outer part of the galaxy. 

Thanks to the LEGUS multiband coverage, we produce a complete luminosity function analysis from the UV to the NIR. A power-law fit to the luminosity distributions yields slopes which are consistently close to $-2$. We also observe, for both YSC candidates and stellar associations, that the recovered slopes show a steepening from the shorter to the longer wavebands possibly consistent with wavelength-dependent fading effects. At the bright end of the CLF, at all wavelengths, we see a clear deviation from the power-law shape, suggesting a dearth of luminous clusters. The analysis of the CMF and Monte Carlo simulations of cluster populations suggest that the CMF can be described by a power law function with a slope close to $-2$ with a truncation at \mstar$\sim 2 \times 10^5$ \msun. However, due to the low number statistics the solution of a pure power-law function with a slope of $\alpha \sim -2.1$ cannot be discarded. 

The analysis of the number densities of objects as a function of age shows different trends for YCS candidates and stellar associations.  The numbers of stellar associations decline rapidly and they tend to disappear on short time scales as already observed in the color-color diagrams.  On the other hand, bound systems do not show any drastic decline in their numbers between 10 and $\sim$200 Myr. We find evidence of a more significant cluster disruption rate in the inner region of NGC628, in agreement with expectations of higher chances of encounters with GMCs and consistent with theoretical predictions of an environmental dependency in cluster disruption. We estimate that the time scale to disrupt a cluster with M$=10^4$ \msun\, is shorter (130 Myr) closer to the centre and significantly longer in the outer part of the galaxy (270 Myr). These time scales should be considered as lower limits, because we cannot exclude that incompleteness at older ages is affecting our results.

Since we do not observe significant disruption for cluster with masses above $10^4$ \msun\, we conclude that the observed \mstar\, values are linked to the formation mechanism of the YSC population and not to their evolution.

The analysis performed on the YSC population of NGC628 shows important evidence for our understanding of the formation and evolution of YSC and stellar associations within their host galaxy. Our morphologically based classification provides, for the first time, insights in the properties of our stellar systems within the framework of a hierarchically driven star formation process. YSCs and stellar associations form within star-forming regions and inherit the imprints of the turbulent status of the ISM as proved by their power-law slope $-2$. However, evidence suggests that at some physical scales other physical parameters may play an important role and affect the shape of the CMF. After their formation YSCs and associations seem to follow different evolution paths, with the formers surviving untouched for a longer timeframe, while associations disappear on time scales comparable to hierarchically organised star-forming regions.

These initial results show interesting trends, but only the sampling provided by a survey like LEGUS, can allow us to address fundamental open questions related to YSC formation and evolution.
A large pool of different galactic environments will provide us with the means to investigate whether the CMF is universal, what physical properties affect the upper-mass end of the mass function, what mechanism dominates cluster disruption, and whether cluster formation efficiency is a relevant quantity to describe cluster formation in the local universe.

\acknowledgments

We are thankful to the anonymous referee for comments and suggestions that have improved the manuscript. AA thanks Nate Bastian for sharing the codes used to make density contours of the color-color diagrams and the maximum-likelihood analysis. Mark Gieles and Nate Bastian are thanked for comments on an early draft of the manuscript.  AA acknowledges partial support from the Swedish Royal Academy. G.A. acknowledges support from the Science and Technology Facilities Council (ST/L00075X/1 and ST/M503472/1). CD acknowledges funding from the FP7 ERC starting grant LOCALSTAR (no. 280104). M.F. acknowledges support by the Science and Technology Facilities Council (grant number ST/L00075X/1). D.A.G. kindly acknowledges financial support by the German Research Foundation (DFG) through program GO\,1659/3-2. AH thanks the Spanish MINECO for grant AYA2015-68012-c2-1. These observations are associated with program \# 13364.
Support for program \# 13364 was provided by NASA through a grant from the Space Telescope Science Institute. Based on observations obtained with the NASA/ESA Hubble Space Telescope, at the Space Telescope Science Institute, which is operated by the Association of Universities for Research in Astronomy, Inc., under NASA contract NAS 5-26555.





{\it Facilities:}  \facility{HST (WFC3, ACS)}.



\appendix




\clearpage




\clearpage

\end{document}